%% file: main.tex
\documentclass[fleqn,usenatbib]{mnras}

\usepackage{newtxtext,newtxmath}

\usepackage[T1]{fontenc}
\usepackage{ae,aecompl}

\usepackage{subfig}
\usepackage{tikz}
\usepackage{standalone}

\usepackage{graphicx}	
\usepackage{amsmath}	
\usepackage{amssymb}	
\usepackage{booktabs}

\usepackage{hyperref}

\def\beq{\begin{eqnarray}}
\def\eeq{\end{eqnarray}}
\newcommand{\Mpch}{h^{-1}\mathrm{Mpc}}
\newcommand{\hMpc}{h\,\mathrm{Mpc}^{-1}}

\newcommand{\av}[1]{\left\langle{#1}\right\rangle} 
\let\vec\mathbf
\newcommand{\bias}[1]{b^{({#1})}}
\newcommand{\deltaRx}[1]{\delta_R(\vec{x}_{{#1}})}
\newcommand{\deltaRxn}[2]{\delta_R^{{#2}}(\vec{x}_{{#1}})}

\newcommand{\thex}{\Theta_\mathrm{ex}}
\newcommand{\sex}{S_\mathrm{ex}}
\newcommand{\vex}{V_\mathrm{ex}}

\newcommand{\resub}[1]{#1}

\numberwithin{equation}{section}

\title[The Effective Halo Model]{The Effective Halo Model: \\\Large Creating a Physical and Accurate Model of the Matter Power Spectrum and Cluster Counts}

\author[O.\,H.\,E. Philcox et al.]{
Oliver H.\,E. Philcox$^{1}$\thanks{E-mail: \href{mailto:ohep2@cantab.ac.uk}{ohep2@alumni.cam.ac.uk}},
David N. Spergel$^{1,2}$ and
Francisco Villaescusa-Navarro$^{1,2}$
\\
$^{1}$Department of Astrophysical Sciences, Princeton University, Princeton, NJ 08544, USA\\
$^{2}$Center for Computational Astrophysics, Flatiron Institute, 162 Fifth Avenue, New York, NY 10010, USA\\
}

\pubyear{2020}

\begin{document}
\label{firstpage}
\pagerange{\pageref{firstpage}--\pageref{lastpage}}
\maketitle

\begin{abstract}
We introduce a physically-motivated model of the matter power spectrum, based on the halo model and perturbation theory.  This model achieves 1\% accuracy on all $k-$scales between $k=0.02\hMpc$ to $k=1\hMpc$. Our key ansatz is that the number density of halos depends on the {\it non-linear} density contrast filtered on some unknown scale $R$. Using the Effective Field Theory of Large Scale Structure to evaluate the two-halo term, we obtain a model for the power spectrum with only two fitting parameters:  $R$ and the effective `sound speed', which encapsulates small-scale physics. This is tested with two suites of cosmological simulations across a broad range of cosmologies and found to be highly accurate. Due to its physical motivation, the statistics can be easily extended beyond the power spectrum; we additionally derive the {one-loop covariance matrices of cluster counts and their combination with the matter power spectrum}. This yields a significantly better fit to  simulations than previous models, and includes a new model for super-sample effects, which is rigorously tested with separate universe simulations. At low redshift, we find a significant ($\sim 10\%$) exclusion covariance from accounting for the finite size of halos which has not previously been modeled. Such power spectrum and covariance models will enable joint analysis of upcoming large-scale structure surveys, gravitational lensing surveys and cosmic microwave background maps on scales down to the non-linear scale. We provide a publicly released Python code.
\end{abstract}

\begin{keywords}
Cosmology: theory, large-scale structure of Universe, dark matter -- methods: analytical, numerical -- software: simulations
\end{keywords}



\section{Introduction}\label{sec: intro}
Canonical, yet inexact. Such would be a worthy description of the `Halo Model'.  Since its inception a number of decades ago, this venerable model has provided a useful phenomenological description of the clustering of matter   \citep[e.g.][]{1952ApJ...116..144N,1980lssu.book.....P,1991ApJ...381..349S,2000ApJ...543..503M,2000MNRAS.318.1144P,2000MNRAS.318..203S,2001ApJ...554...56C,2002PhR...372....1C,2003MNRAS.341.1311S}. At heart, this rests upon the simple assumption that all matter in the universe lies within halos of some size. For the matter power spectrum, this delineates a separation between two regimes corresponding to particles within the same halo clustered according to some density profile (i.e. `one-halo' contributions), and those in different halos, which trace large scale density fluctuations (i.e. `two-halo' contributions). The combination of the two terms yields a useful model for the power spectrum across a broad range of scales. There are rich sets of  applications of the halo model beyond the power spectrum; a nonexhaustive list includes the bispectrum \citep[e.g.][]{2002PhR...372....1C,2008PhRvD..78b3523S,2011A&A...532A...4V}, weak lensing \citep{2000ApJ...535L...9C,2011PhRvD..84f3004K,2017MNRAS.470.3574G}, covariance matrices \citep{2001ApJ...554...56C,2018A&A...615A...1L}, neutral hydrogen modeling \citep{2017ApJ...846...21F,2017MNRAS.469.2323P,2017MNRAS.471.1788C} and Sunyaev-Zel`dovich effects \citep[e.g.][]{2002MNRAS.336.1256K,2012PhRvD..85b3007F,2013PhRvD..88f3526H,2019PhRvD..99j3511T}. The flaws of the model are however well known, particularly regarding the transition between the one- and two-halo regime, where the model is accurate to less than 20\%, {due to a lack of inclusion of non-linear physics}. 

On an entirely different plane lives cosmological perturbation theory, which is able to provide accurate predictions for matter clustering, yet is fundamentally limited to scales above the non-linear threshold $k_\mathrm{NL}^{-1}$. In this regime, the Universe can be accurately described by an ideal fluid, and solving the standard fluid equations leads to a perturbative description of the matter field, extensively reviewed in \citet{2002PhR...367....1B}. Solving these na\"ively leads to Standard Perturbation Theory, which is known to be inadequate treatment of short-scale contributions. Over the past decade, a new method has been developed, dubbed the `Effective Field Theory of Large Scale Structure' (EFT), which accounts for \textit{all} physical effects on the large-scale clustering of matter via a perturbative expansion combined with counterterms to parametrize short-scale physics and proper treatment of ultraviolet and infrared modes \citep{2012JHEP...09..082C,2012JCAP...07..051B,2014JCAP...07..057C,2015JCAP...02..013S,2015PhRvD..92l3007B,2018JCAP...05..019S}. However, whilst the model predicts power spectra (and higher order statistics) accurately on large scales (corresponding to small wavenumbers $k$), the radius of convergence of the perturbative expansion is finite, and its applicability is unlikely to extend beyond $k\sim 0.5\hMpc$ \citep{2019JCAP...11..027K}. This places fundamental limitations on the model for applications such as weak lensing analyses, which must consider integrals of power spectra across a wide range of scales.

Given the limitations of the above models, it is natural to consider their modification and combination. An interesting example is that of \citet{2011A&A...527A..87V,2011A&A...532A...4V}, which aimed to insert perturbative modeling in the halo model framework, making use of a Lagrangian description of halo exclusion. Whilst this had some success, it was limited by an incorrect description of perturbative physics and achieved only $\sim 10\%$ accuracy for the one- to two-halo transition. In \citet{2014MNRAS.445.3382M} and \citet{2015PhRvD..91l3516S}, a different approach was adopted, combining a Zel'dovich model for the two-halo power spectrum with a purely empirical (Pade-resummed) model for the one-halo term. Whilst the latter work was able to achieve a model with 1\% accuracy in the matter power spectrum down to $k = 1\hMpc$, it requires a complex `compensation function' and a number of free parameters that do not have clear physical motivations, thus the model is difficult to extend to more involved contexts, for example the bispectrum and other higher-order statistics. Recent contributions to this effort include \citet{2013PhRvD..88h3507B}, \citet{2015PhRvD..91l3516S}, \citet{2017JCAP...10..009H} and  \citet{2019OJAp....2E...4C}, which consider low-$k$ modifications to the halo model to compensate for a known ultra-large scale deficiency with a varying number of fitting parameters. Furthermore, \citet{2020arXiv200306411V} showed that the inclusion of a voids could reduce the modeling deficiency in the non-linear transition region. Finally, a number of works have included simulation-based calibration of the halo model, including \citet{2020PhRvD.101j3522C} and \citet{2015MNRAS.454.1958M,2016MNRAS.459.1468M}, with the latter works including the effects of non-standard cosmologies and baryonic feedback.

\begin{figure}%
    \centering
    \subfloat[Matter Power Spectrum (Sec.\,\ref{sec: Pk_derivation})]{{\includegraphics[width=0.48\textwidth]{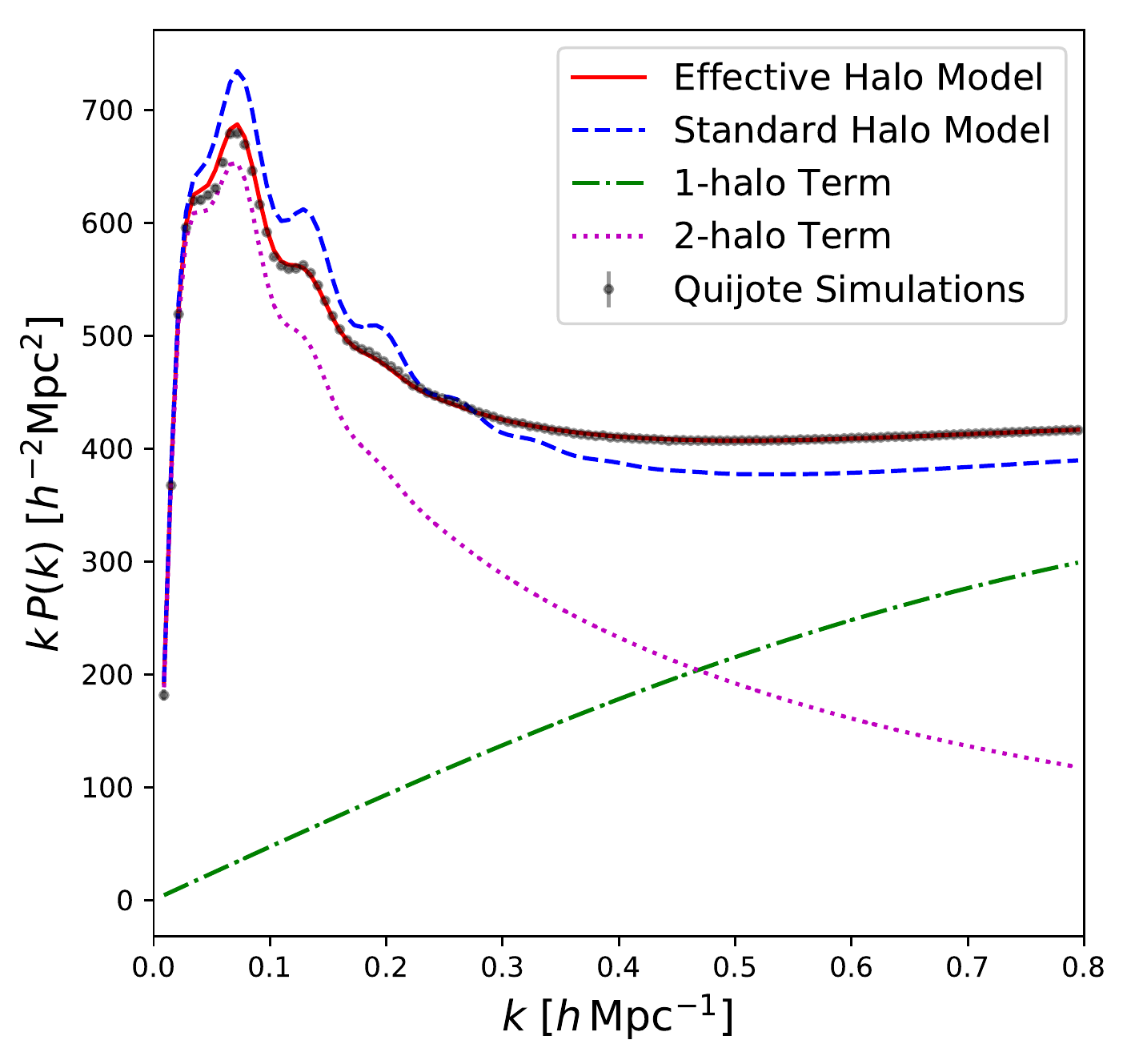} }}%
    \qquad
    \subfloat[Covariance of halo counts and the matter power spectrum (Sec.\,\ref{sec: cov_N_Pk_derivation})]{{\includegraphics[width=0.48\textwidth]{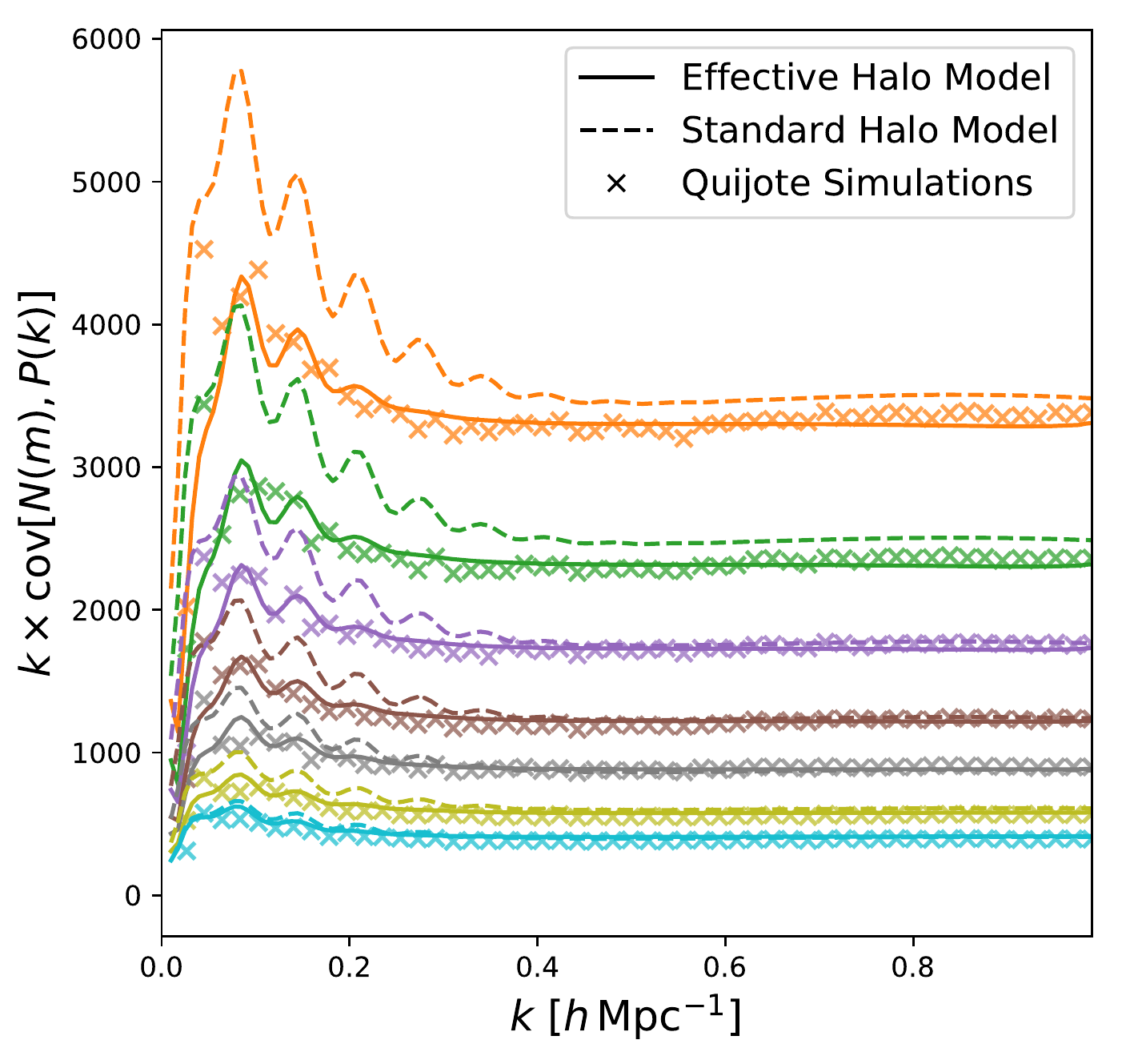} }}%
    \caption{Summary of the key results of this work. In both panels, we compare the analytical models introduced in this work with those from the \texttt{Quijote} suite of $N$-body simulations. In both cases, we work at redshift zero and plot the predictions from the standard halo model \citep[e.g.][]{2002PhR...372....1C,2007NJPh....9..446T} for reference. {In the power spectrum plot, we additionally include the contributions from the one- and two-halo terms in the effective halo model, with the one-halo term matching that of the standard halo model.} For the covariance figure, data are plotted for halo counts in mass bins with $\Delta \log_{10}\left(M/h^{-1}M_\odot\right) = 0.2$, from central mass $10^{13.2}h^{-1}M_\odot$ (top line) to $10^{14.4}h^{-1}M_\odot$ (bottom line). Note that these plots use data from Figs.\,\ref{fig: pk_comparison}\,\&\,\ref{fig: subbox-cov} respectively.}%
    \label{fig: introductory-plots}%
\end{figure}

In this paper, we construct a variant of the model, {dubbed `The Effective Halo Model'}, which is both physically motivated and produces a percent-level accurate matter power spectrum across a broad range of scales. 
This rests on two key assumptions, beyond that of the standard halo model: (1) that the positions of dark matter halos are a function of the underlying \textit{non-linear} density field smoothed on an unknown scale $R$ and (2) that the long-wavelength density field can be described by EFT at one-loop order. Since the model is derivable from a minimal set of assumptions, it can be simply extended to observables beyond dark matter spectra. Here, we apply it to the covariances {of halo counts, both alone and in combination with the dark matter power spectra}, noting that these are key observables for weak lensing and thermal Sunyaev-Zel'dovich cosmology \citep{2007NJPh....9..446T,2014PhRvD..90l3523S,2014MNRAS.441.2456T,2016JCAP...08..005L,2017A&A...604A..71H}. These are particularly interesting since, despite an array of previous work, {when compared to simulations}, current models show severe deficiencies at low redshift 
due to lack of consideration of halo exclusion.

The power spectrum formalism introduced herein is similar to that of \resub{\citet{2011PhRvD..83d3526S} and} \citet{2016PhRvD..93f3512S}, both of which include quasi-linear effects via perturbative modeling, improving the agreement in the non-linear transition regime. \resub{The former work adds in perturbative corrections to the usual two-halo term, additionally including density-field smoothing and a prescription for halo exclusion based on \citet{2007PhRvD..75f3512S}. This can reduce the aforementioned spurious power at low-$k$, though has strong dependence on short-scale modeling.} In \citet{2016PhRvD..93f3512S} (and to an extent \citealt{2017PhRvD..96h3528G}), the non-linear halo model terms are sourced naturally from perturbation theory, by way of a semi-empirical stochasticity field, whose behavior is well understood on large and small scales. In a sense, this is complementary to this work, with our philosophy being to incorporate perturbation theory into the halo model rather than the inverse. \citet{2016PhRvD..93f3512S} also includes a discussion of a number of additional subdominant features such as halo triaxiality and environment-dependent halo concentration, which are omitted here to obtain an easily computable model. \resub{Both works still fail in the mildly non-linear regime however; this is solved in this work by the inclusion of a free smoothing scale parameter and EFT counterterms.}


Given that the derivations in this paper are somewhat lengthy, we provide a brief summary of the power spectrum model for the casual reader below. Our main results are also shown in Fig.\,\ref{fig: introductory-plots}. In the standard halo model, one assumes that all the mass in the Universe is embedded in Poisson-distributed halos with masses drawn from some halo mass function $n(m)$, modulated by the local linear overdensity, $\delta_\mathrm{L}$, ensuring that overdense regions are more likely to form halos. This leads to a power spectrum separable into one- and two-halo terms, with the two halo term scaling as the linear power spectrum, $P_\mathrm{L}(k)$, on large scales. Our model is a natural extension of this; we make the ansatz that the halo mass function depends on the \textit{non-linear} overdensity $\delta_R$, smoothed 
on an unknown scale $R$. Whilst {we assume this not to modify the mass function or one-halo term}, the two-halo term now scales as the $W^2(kR)P_\mathrm{NL}(k)$ where $W$ is {the Fourier transform of the smoothing window.}
Evaluating the non-linear power spectrum using effective field theory (EFT) with long-wavelength modes resummed leads to a power spectrum model that can be written as
\beq\label{eq: pk-intro}
    P_\mathrm{HM}(k) &\equiv& P^{2h}(k)+P^{1h}(k)\\\nonumber
    P^{2h}(k) &=& \left[\int dm\,\frac{m}{\bar\rho}n(m)\bias{1}(m)u(k|m)\right]^2W^2(kR)P_\mathrm{EFT}(k; c_s^2)\\\nonumber
    P^{1h}(k) &=& \int dm\,\frac{m^2}{\bar\rho^2}n(m)u^2(k|m),
\eeq
where the smoothing scale $R$ and the effective squared sound speed $c_s^2$ are the only free parameters. This uses the linear bias $\bias{1}(m)$ (that can be predicted from theory or measured in simulations), and the normalized halo profile $u(k|m)$, usually assumed to take an NFW form \citep{1996ApJ...462..563N}. This is described in detail in Sec.\,\ref{sec: Pk_derivation} and leads to a power spectrum model which comparison to $N$-body simulations shows to be percent-level accurate up to $k = 1\hMpc$, as demonstrated in Figs.\,\ref{fig: introductory-plots}\,\&\,\ref{fig: pk_comparison}. {Since we do not modify $P^{1h}(k)$, we expect also good agreement at larger $k$, where the one-halo term is known to be a good fit.}


The structure of this work is as follows. We begin with a discussion of our power spectrum model in Sec.\,\ref{sec: Pk_derivation}, including a full derivation and discussion of the {effective} halo model assumptions, before assessing its accuracy in Sec.\,\ref{sec: sims} by comparison with $N$-body simulations. In Sec.\,\ref{sec: cov_N_Pk_derivation} a model for the covariance of {cluster counts and their combination with the matter power spectrum} is derived, additionally incorporating super-sample and halo exclusion effects. This is compared to simulations in Sec.\,\ref{sec: sims-cov}, before we present a summary in Sec.\,\ref{sec: conclusion} alongside discussions of extensions of the work including the incorporation of baryon physics. Appendix \ref{appen: cov-Pk} presents a brief derivation of the covariance of our power spectrum model, {whilst its dependence on the choice of perturbation theory is discussed in Appendix \ref{appen: zel-Pk}}. Finally, useful results concerning the loop-corrections of the power spectrum of objects in boxes of finite size are given in Appendix \ref{appen: Pk-trunc}. A Python package implementing our power spectrum model is publicly available with extensive documentation.\footnote{\href{https://EffectiveHalos.rtfd.io}{EffectiveHalos.readthedocs.io}}

\section{The Matter Power Spectrum: Theoretical Model}\label{sec: Pk_derivation}

We begin by describing the {effective halo model}, and its application to the matter power spectrum. This is similar to the standard derivation \citep[e.g.][]{1980lssu.book.....P,2000ApJ...543..503M,2000MNRAS.318.1144P,2000MNRAS.318..203S,2001ApJ...554...56C,2002PhR...372....1C}, though with a significantly modified two-halo term, {similar to that of \citet{2016PhRvD..93f3512S}}. Note that we do not include baryonic effects in this derivation, though this is possible, for example via the approaches of \citet{2019JCAP...03..020S} and \citet{2019OJAp....2E...4C}.

\subsection{Effective Halo Model Phenomenology}\label{subsec: halo-phenom}
The standard halo model makes a simple assumption; that all matter is contained within halos of some size. This implies that the matter overdensity, $\hat{\delta}_\mathrm{HM}(\vec r)$ may be written as a sum over all halos
\beq
    1+\hat{\delta}_\mathrm{HM}(\vec r) &=& \frac{1}{\bar{\rho}}\sum_i \rho_h(\vec r-\vec x_i|m_i)\\\nonumber
    &=& \frac{1}{\bar\rho}\int d\vec x\,dm\,\rho_h(\vec r-\vec x|m)\left[\sum_i\delta_D(\vec x-\vec x_i)\delta_D(m-m_i)\right]\\\nonumber
    &\equiv&  \frac{1}{\bar\rho}\int d\vec x\,dm\,\rho_h(\vec r-\vec x|m)\hat{n}(m|\vec x),
\eeq
where $\rho_h(\vec r -\vec x_i|m_i)$ is the (assumed universal) density profile of a halo of mass $m_i$ centered at $\vec x_i$ and $\bar\rho$ is the mean matter density. The term in parentheses may be identified with the stochastic number density per unit mass of halos at $\vec x$ of mass $m$, $\hat{n}(m|\vec x)$. In expectation, $\av{\hat{n}(m|\vec x)}\equiv n(m)$, independent of position, thus we obtain
\beq
    1+\av{\hat{\delta}_\mathrm{HM}(\vec r)} &=& \frac{1}{\bar\rho}\int dm\,n(m)\left[\int d\vec x\,\rho_h(\vec r-\vec x|m)\right]\\\nonumber
    &=& \frac{1}{\bar\rho}\int dm\,n(m)m = 1\\\nonumber
    \Rightarrow \hat{\delta}_\mathrm{HM}(\vec r) &=& \frac{1}{\bar\rho}\int  d\vec x\,dm\,\rho_h(\vec r-\vec x|m)\left[\hat{n}(m|\vec x)-n(m)\right],
\eeq
where we note that the integral of a halo density profile is simply its mass. The penultimate line follows from the definition of $\bar\rho$ as the mean density of the volume, and this implies that $\hat{\delta}_\mathrm{HM}$ is correctly normalized. It is pertinent to note that for a simulation containing a fixed total mass of particles, $\av{\hat{\delta}_\mathrm{HM}}$ is equivalent to a volume average over the box. Whilst this is in general true in the infinite volume limit via ergodicity, it is not true for general statistics in finite volumes, as will be important for halo number count statistics. Conventionally, the halo density profile $\rho_h$ is written in dimensionless units as $\rho_h(m|\vec x) = mu(m|\vec x)$ leading to the final form for the matter density field
\beq\label{eq: deltah-normed}
    \hat{\delta}_\mathrm{HM}(\vec r) &=& \int d\vec x\,dm\,\frac{m}{\bar\rho}u(\vec r-\vec x|m)\,\delta\hat{n}(m|\vec x),
\eeq
where we have additionally defined $\delta\hat{n}(m|\vec x) \equiv \hat{n}(m|\vec x) - n(m)$.

To compute cosmological observables containing $\hat{\delta}_\mathrm{HM}$, we must have knowledge of the statistical properties of the random field $\hat{n}(m|\vec x)$. Firstly, since it is fundamentally a sum over Dirac delta functions at the halo positions, $\hat{n}$ is expected to obey Poisson statistics (ignoring halo exclusion effects), for example
\beq\label{eq: poisson_expansion}
    \av{\hat n(m|\vec x)}_P &=& n(m|\vec x)\\\nonumber
    \av{\hat n(m_1|\vec x_1)\hat{n}(m_2|\vec x_2)}_P &=& n(m_1|\vec x_1)n(m_2|\vec x_2) + \delta_D(\vec x_1-\vec x_2)\delta_D(m_1-m_2)n(m_1|\vec x_1),
\eeq
where $\av{...}_P$ indicates a Poissonian average (without averaging over realizations of the Universe). Henceforth, we will use the notation that $\hat{f}$ refers to $f$ before Poisson averaging.

Throughout this work we make a simple ansatz: the number density is a function of the \textit{non-linear} local overdensity smoothed on an unknown scale $R$. This underpins our model, and is noticeably different to standard approaches in which $n(m|\vec x)$ is assumed to depend only on the linear overdensity field, \resub{though it is similar to that of \citet{2011PhRvD..83d3526S}.} Using this assumption, we can expand $n(m|\vec x)$ perturbatively in the smoothed field $\delta_R(\vec x)$,\footnote{An alternative approach would be to expand $n(m|\vec x)$ fully in terms of all possible density and tidal field operators up to third order \textit{viz.} the EFT of biased tracers \citep{2015JCAP...09..029A,2020JCAP...01..009F}. This would result in the standard biased tracer EFT spectrum, but with each term multiplied by a $k$-dependent prefactor, integrated over mass. Whilst this would be the most general approach, it has limitations since the associated bias parameters are not calculable within EFT, thus the mass integrals cannot be simply performed. In addition, given the bias consistency relations of Eq.\,\ref{eq: bias_consistency_relation}, the resulting spectrum will be identical to our approach for small wavenumber $k$ (where $u(k|m)\approx 1$.} 
\beq\label{eq: n_m_expansion}
    n(m|\vec x) = n(m)\left\{1 + \bias{1}(m)\delta_R(\vec x) + \frac{1}{2!}\bias{2}(m)\left[\delta_R^2(\vec x)-\av{\delta_R^2}\right]+\frac{1}{3!}\bias{3}(m)\left[\delta_R^3(\vec x)-\av{\delta_R^3}\right]+...\right\},
\eeq
\resub{\citep[cf.\,][]{2007PhRvD..75f3512S}} where the biases $\bias{i}(m)$ are defined as Taylor expansion coefficients by 
\beq\label{eq: bias-definitions}
    \bias{i}(m)&=& \frac{1}{n(m)}\left.\frac{\partial^i n(m|\vec x)}{\partial\delta_R(\vec x)^i}\right|_{\delta_R(\vec x) = 0},
\eeq
and $\delta_R$ is defined by a convolution of the matter density field $\delta$ with the window function $W_R$, here chosen as a top-hat with radius $R$;\footnote{{The results in this paper are highly insensitive to the choice of smoothing function; our model has the same accuracy when a Gaussian window is used instead of a top-hat.}}
\beq
    \delta_R(\vec x) = \int d\vec y\,W_R(\vec x-\vec y)\delta(\vec y).
\eeq
Models for arbitrary cosmological statistics can then be computed by applying perturbation theory to the various terms arising from the bias expansion of Eq.\,\ref{eq: bias-definitions}.

{Before continuing, it is instructive to discuss the physical motivation for the above ansatz, which assumes an Eulerian and non-linear halo distribution function. This differs significantly from the standard halo model, in which one assumes that the halo formation process is local in \textit{Lagrangian} space, with evolution only via the linear growth factor. Our justification is twofold; firstly, we note that most of the mass in halos has been recently accreted, thus the non-linear field is expected to be a better predictor of halo properties. In essence, this approach takes the (assumed Gaussian) smoothed non-linear field and associates halos to points above some critical density. Secondly, this formalism arises naturally from perturbation theory. In the large-scale limit, one expects that individual halos can be ignored, thus the matter power spectrum can be described by some perturbative theory which ignores virial collapse. Effective Field Theory (hereafter EFT) is an excellent candidate for this (since it arises from the smoothed fluid equations), and it is therefore important that our halo model should asymptote to such a description on large scales. Practically, this is impossible without using the non-linear density field. Furthermore, the halo density field $\hat{n}(m|\vec x)$ is empirically known to be well described by EFT \citep[e.g.,][]{2015JCAP...11..007S,2015JCAP...07..030M,2015JCAP...09..029A,2020JCAP...01..009F}, which utilizes a set of bias parameters (integrated over the halo formation history). In a sense, our model is equivalent to adding in the effects of finite halo profiles to a biased EFT model, with bias integrals controlled by enforcing agreement with the standard results for a matter power spectrum on large scales.}

\subsection{Deriving the Matter Power Spectrum Model}
We now proceed to apply the above model to the real-space matter power spectrum. It is useful to begin in configuration space, defining the two-point correlation function (hereafter 2PCF) as
\beq\label{eq: 2PCF-definition}
    \hat{\xi}_\mathrm{HM}(\vec r) &=& \frac{1}{V}\int d\vec y\,\av{\hat{\delta}_\mathrm{HM}(\vec y)\hat{\delta}_\mathrm{HM}(\vec y+\vec r)}.
\eeq
Inserting the definition of $\hat{\delta}_\mathrm{HM}$ (Eq.\,\ref{eq: deltah-normed} and taking the expectation, we obtain
\beq
    \xi_\mathrm{HM}(\vec r) &=& \frac{1}{V}\int d\vec yd\vec x_1d\vec x_2dm_1dm_2\,\frac{m_1m_2}{\bar\rho^2}u(\vec y-\vec x_1|m_1)u(\vec y+\vec r-\vec x_2|m_2)\av{\delta\hat{n}(m_1|\vec x_1)\delta\hat{n}(m_2|\vec x_2)}\\\nonumber
    &=& \frac{1}{V}\int d\vec yd\vec x_1d\vec x_2dm_1dm_2\,\frac{m_1m_2}{\bar\rho^2}u(\vec y-\vec x_1|m_1)u(\vec y+\vec r-\vec x_2|m_2)\av{\delta n(m_1|\vec x_1)\delta n(m_2|\vec x_2)}\\\nonumber
    &&\,+ \frac{1}{V}\int d\vec yd\vec x_1dm_1\,\frac{m_1^2}{\bar\rho^2}u(\vec y-\vec x_1|m_1)u(\vec y+\vec r-\vec x_1|m_1)\av{n(m_1|\vec x_1)}
\eeq
where we perform Poissonian averaging (Eq.\,\ref{eq: poisson_expansion}) to split the expression into one- and two-halo terms in the final line. For ease of notation we will write $n(m|\vec x)$ in terms of the fractional number density fluctuation field $\eta$ via
\beq\label{eq: eta_definition}
    n(m|\vec x) \equiv n(m)\left[1+\eta(m|\vec x)\right],
\eeq
satisfying $\av{\eta(m|\vec x)}=0$ such that $\av{n(m|\vec x)}=n(m)$. Rewriting the one- and two-halo terms of Eq.\,\ref{eq: 2PCF-definition} in terms of $\eta$, we obtain
\beq
    \xi_\mathrm{HM}(\vec r) &\equiv& \xi^{2h}(\vec r)+\xi^{1h}(\vec r)\\\nonumber
    \xi^{2h}(\vec r) &=& \frac{1}{V}\int d\vec yd\vec x_1d\vec x_2dm_1dm_2 \frac{m_1m_2}{\bar\rho^2}n(m_1)n(m_2)u(\vec y-\vec x_1|m_1)u(\vec y+\vec r-\vec x_2|m_2)\av{\eta(m_1|\vec x_1)\eta(m_2|\vec x_2)}\\\nonumber
    &=& \int dm_1dm_2\frac{m_1m_2}{\bar\rho^2}n(m_1)n(m_2)\left[u\ast u\ast \av{\eta(m_1)\eta(m_2)}\right](\vec r)\\\nonumber
    \xi^{1h}(\vec r) &=& \frac{1}{V}\int d\vec yd\vec xdm\,\frac{m^2}{\bar\rho^2}n(m)u(\vec y-\vec x|m)u(\vec y+\vec r-\vec x|m)\\\nonumber
    &=& \int dm\,\frac{m^2}{\bar\rho^2}n(m)\left[u\ast u\right](\vec r),
\eeq
where $\left[f\ast g\right]$ represents the convolution of $f$ and $g$ and we have noted that $\av{\eta(m_1|\vec x_1)\eta(m_2|\vec x_2)}$ can only depend on $\vec x_1-\vec x_2$ via statistical homogeneity. These are more easily represented in Fourier space, via a trivial application of the convolution theorem;
\beq\label{eq: P2h-temp}
    P^{2h}(\vec k) &\equiv& \mathcal{F}\left[\xi^{2h}\right](\vec k) = \int dm_1 dm_2 \,\frac{m_1m_2}{\bar\rho^2}n(m_1)n(m_2)u(\vec k|m_1)u(\vec k|m_2)\av{\eta(m_1)\eta(m_2)}(\vec k)\\\nonumber
    P^{1h}(\vec k) &\equiv& \mathcal{F}\left[\xi^{2h}\right](\vec k) = \int dm\,\frac{m^2}{\bar\rho^2}n(m)u^2(\vec k|m)
\eeq
where $\mathcal{F}$ is the Fourier operator. Note that the one-halo term agrees with the standard approach \citep[e.g.][]{2002PhR...372....1C}. 

To proceed, we require the statistical properties of $\eta$, which may be found using Eq.\,\ref{eq: n_m_expansion};
\beq\label{eq: eta-statistics}
    \av{\eta(m|\vec x)} &=& 0\\\nonumber
    \av{\eta(m_1|\vec x_1)\eta(m_2|\vec x_2)} &=& \bias{1}_1\bias{1}_2\av{\deltaRx{1}\deltaRx{2}}\\\nonumber
        &+&\frac{1}{2}\bias{1}_1\bias{2}_2\av{\delta_R(\vec x_1)\deltaRxn{2}{2}}+\frac{1}{2}\bias{2}_1\bias{1}_2\av{\deltaRxn{1}{2}\deltaRx{2}}\\\nonumber
    &+&\frac{1}{3!}\bias{1}_1\bias{3}_2\av{\delta_R(\vec x_1)\delta_R^3(\vec x_2)}+\frac{1}{3!}\bias{3}_1\bias{1}_2\av{\delta_R^3(\vec x_1)\delta_R(\vec x_2)}\\\nonumber
    &+&\frac{1}{2}\bias{2}_1\bias{2}_2\left[\av{\delta^2(\vec x_1)\delta^2(\vec x_2)}-\av{\delta^2_R}^2\right]\\\nonumber
    &+&\mathcal{O}(\delta_R^5),
\eeq
where $\bias{i}_j\equiv \bias{i}(m_j)$, and we have used $\av{\delta_R}=0$. For a Gaussian density field, the expectation of any odd number of fields vanishes; this is not true here since $\delta_R$ is not the linear density field, and is hence non-Gaussian. 

Whilst Eq.\,\ref{eq: eta-statistics} seems to contain a lot of terms, our expressions may be greatly simplified by considering the large scale regime where ($u(\vec k|m)\approx 1$) and applying the bias consistency relation
\beq\label{eq: bias_consistency_relation}
    \int dm\,\frac{m}{\bar\rho}n(m)\bias{i}(m) = \begin{cases}1 \quad i=1\\0 \quad i>1\end{cases},
\eeq
which follows from enforcing that the halo model power spectrum tends to the perturbative result in the linear regime.\footnote{\resub{At low $k$, where the one-halo term can be neglected, we require $P^{2h}(\vec k) \rightarrow P_\mathrm{L}(\vec k)$ for linear power spectrum $P_\mathrm{L}(\vec k)$. Only terms in Eq.\,\ref{eq: eta-statistics} proportional to $\bias{1}_1\bias{1}_2$ have this $k$ dependency, requiring the higher order terms to vanish at $k\rightarrow 0$. Since all biases appear in $P^{2h}(\vec k)$ (Eq.\,\ref{eq: P2h-temp}) multiplied by $m/\bar\rho n(m)u(\vec k|m)$ and integrated over mass, we arrive at Eq.\,\ref{eq: bias_consistency_relation}, since $u(\vec k|m) = 1$ on these scales.}}
In our context, this implies that only terms involving linear biases $\bias{1}$ can survive at low $k$. At high $k$, we expect the power spectrum to be dominated by the one-halo term, thus we are justified in neglecting all higher-order bias terms in the analysis of the two-point correlator. This is found to be an excellent approximation in practice.\footnote{{The analysis of \citet{2016PhRvD..93f3512S} included these terms and found them to be subdominant on all scales.}} 

The expectation of the linearly biased term is given in terms of the perturbation theory 2PCF as
\beq
    \av{\delta_R(\vec x_1)\delta_R(\vec x_2)} &=& \int d\vec y_1d\vec y_2W_R(\vec x_1-\vec y_1)W_R(\vec x_2-\vec y_2)\av{\delta(\vec y_1)\delta(\vec y_2)}\\\nonumber
    &=& \int d\vec y_1d\vec y_2W_R(\vec x_1-\vec y_1)W_R(\vec x_2-\vec y_2)\xi_\mathrm{NL}(\vec y_1-\vec y_2) = \left[W_R\ast W_R\ast\xi_\mathrm{NL}\right](\vec x_1-\vec x_2).
\eeq
In Fourier space, this simply translates to $W^2(kR)P_\mathrm{NL}(k)$ is the Fourier transform of the window function (equal to $3j_1(kR)/kR$ for spherical Bessel function $j_1$) and $k = |\vec k|$. Note that $P_\mathrm{NL}(k)$ and $\xi_\mathrm{NL}(\vec r)$ are \textit{non-linear} quantities unlike in the standard halo model; this is a result of our initial ansatz, that $n(m|\vec x)$ should depend on the smoothed non-linear density field. The modeling of this is discussed below.

Collecting results, we have final expressions for the {effective} halo model power spectra (ignoring higher-order bias terms suppressed by the consistency relation);
\beq\label{eq: pk-summary}
    P^{2h}(\vec k) &=& \left[\int dm\,\frac{m}{\bar\rho}n(m)\bias{1}(m)u(\vec k|m)\right]^2W^2(kR)P_\mathrm{NL}(\vec k)\\\nonumber
    P^{1h}(\vec k) &=& \int dm\,\frac{m^2}{\bar\rho^2}n(m)u^2(\vec k|m).
\eeq
In similar notation to \citet{2001ApJ...554...56C}, we introduce the compact notation
\beq\label{eq: integral-notation}
    I_p^{q}(\vec k_1,...,\vec k_p) = \int dm\,\left(\frac{m}{\bar\rho}\right)^p n(m)\bias{q}(m)\prod_{i=1}^p u(\vec k_i|m),
\eeq
giving
\beq
    P^{2h}(\vec k) &=& \left[I^1_1(\vec k)\right]^2W^2(kR)P_\mathrm{NL}(\vec k)\\\nonumber
    P^{1h}(\vec k) &=& I^0_2(\vec k,\vec k),
\eeq
with $\bias{0}(m)=1$ for all $m$. Note that, on very large scales ($k\lesssim 10^{-2}\hMpc$), our model {(and the standard halo model)} is sub-optimal, since our one-halo term tends to a constant that can exceed the two-halo term, implying that the model does \textit{not} match linear theory. {This is a consequence of over-counting, since the perturbative two-halo term naturally includes contributions from particle pairs within the same halo, which are also captured by the one-halo term.} As a consequence of mass and momentum conservation, the leading contribution of the one-halo term should in fact scale as $k^4$ \citep[e.g.][]{2002PhR...372....1C,2003MNRAS.341.1311S,2016JCAP...03..007B}, which would give the correct asymptotic behavior. Whilst there a number of approaches which will reduce the low-$k$ power (e.g., via halo exclusion \citep{2011A&A...527A..87V}, compensation functions \citep{2014MNRAS.445.3382M,2015PhRvD..91l3516S} or compensated halo profiles \citep{2020PhRvD.101j3522C}), {on the scales considered in this work, we have been able to achieve percent-level accuracy without consideration of this effect. Further discussion of this is found in Sec.\,\ref{subsec: model-assumptions}.}

Via similar arguments, one may derive the full covariance matrix of the model power spectrum $P_\mathrm{HM}$ using the same set of assumptions, arriving at a form that depends density field correlators as well as a set of $I_p^q$ functions. A brief derivation of this at fourth order in $\delta_R$ is presented in Appendix \ref{appen: cov-Pk}.

\subsection{Predicting $P_\mathrm{NL}(k)$}\label{subsec: quasi-linear-power-model}
A key component of the two-halo term in Eq.\,\ref{eq: pk-summary} (and the principal difference between our model and canonical approaches) is the non-linear power spectrum. In this work, it is modeled using EFT \citep{2012JCAP...07..051B,2012JHEP...09..082C}), the ingredients of which are discussed below. {Note that this is not the only possible choice; the results of using the Zel`dovich approximation to compute $P_\mathrm{NL}(k)$ instead of EFT are shown in Appendix \ref{appen: zel-Pk}.}

On quasi-linear scales, the real-space matter power spectrum may be written at one-loop order as 
\beq\label{eq: Pnl-def}
    P_\mathrm{NL}(k) = P_\mathrm{L}(k)+P_\mathrm{SPT}(k)+P_\mathrm{ct}(k),
\eeq
where $P_\mathrm{L}(k)$ is the usual linear power spectrum, easily generated by \texttt{CAMB} \citep{2011ascl.soft02026L} or \texttt{CLASS} \citep{2011JCAP...07..034B}. The one-loop power spectrum is given by standard (Eulerian) perturbation theory (herefter SPT) as
\beq\label{eq: 1-loop-SPT}
    P_\mathrm{SPT}(k) &=& P_\mathrm{22}(k)+2P_\mathrm{13}(k)\\\nonumber
    P_\mathrm{22}(k) &=& 2\int \frac{d\vec q}{(2\pi)^3}P_\mathrm{L}(q)P_\mathrm{L}(|\vec k-\vec q|)|F_2(\vec q,\vec k-\vec q)|^2\\\nonumber
    P_\mathrm{13}(k) &=& 3P_\mathrm{L}(k)\int \frac{d\vec q}{(2\pi)^3}P_\mathrm{L}(q)F_3(\vec k,\vec q,-\vec q),
\eeq
where $F_2$ and $F_3$ are the standard coupling kernels (see \cite{2002PhR...367....1B} for a comprehensive review) and $|\vec p|\equiv p$ in general. The can be computed efficiently by codes such as \texttt{FAST-PT} \citep{2016JCAP...09..015M} or via FFT-log decompositions \citep{2018JCAP...04..030S}.

At one-loop order SPT is known to over-predict the true power spectrum, even on mildly non-linear scales, and is thus not a complete model. Its inadequacies are due to a failure to account for the effects of small-scale (non-perturbative) displacements on large-scale modes, with the integrals in Eq.\,\ref{eq: 1-loop-SPT} extending into the high-$k$ ultraviolet (UV) regime where perturbation theory is known to be invalid. To account for this, an \textit{effective} technique is used, smoothing the equations of motion and accounting for the UV divergences of the above kernels. At one-loop order, this practically leads to a \textit{counterterm};
\beq\label{eq: counterterm_initial}
    P_\mathrm{ct}(k) = -c_s^2k^2P_\mathrm{L}(k),
\eeq
where the effective squared speed-of-sound $c_s^2$ is a free parameter whose magnitude (or sign) cannot be predicted from theory.\footnote{Note that the relevant quantity is the \textit{squared} sound speed, rather than the sound speed itself. Since this is purely an effective quantity, $c_s^2<0$ is possible.} In this work, we are interested in modeling the power spectrum well into the non-linear regime, where the counterterm is ill-behaved due to its $k^2$ scaling that becomes unbounded as $k$ becomes large. We therefore adopt a Pade approximant of Eq.\,\ref{eq: counterterm_initial};
\beq\label{eq: counterterm_final}
    \tilde{P}_\mathrm{ct}(k) = -c_s^2\frac{k^2}{1+(k/\hat{k})^2}P_\mathrm{L}(k),
\eeq
for $\hat{k} = 1\hMpc$, which has the correct asymptotic behavior at small $k$ and remains finite in the non-linear regime. This is a valid assumption since the leading order difference between the two counterterms scales as $k^4P_\mathrm{L}(k)$, and would hence be absorbed into the two-loop counterterm. Whilst $\hat{k}$ should be chosen such that it damps the counterterm on the characteristic scales of halos, its exact value is not found to be important (and it is highly degenerate with $c_s^2$ and $R$), thus we assume it to be fixed henceforth.

One further ingredient is needed to accurately model the quasi-linear matter power spectrum; the resummation of long-wavelength (infrared, hereafter IR) modes. In canonical perturbation theory, all displacements are considered to be small (and thus can be perturbatively expanded), which introduces a non-negligible error for the IR modes, causing excess sharpening of the Baryon Acoustic Oscillation (BAO) peak. Whilst a number of exact formalisms exist for ameliorating this \citep{2015JCAP...02..013S,2020JCAP...03..018L}, we here adopt the approximate (and commonly used) method proposed in \citet{2015PhRvD..92d3514B} and rigorously developed in \citet{2016JCAP...07..028B} and \citet{2018JCAP...07..053I}, using time-sliced perturbation theory \citep{2016JCAP...07..052B}. This damps the oscillatory part of the input power spectrum, resulting in power spectra of the form
\beq\label{eq: one-loop-IR}
    P_\mathrm{L,IR}(k) &=& P_\mathrm{L}^{nw}(k)+e^{-k^2\Sigma^2}P_\mathrm{L}^{w}(k)\\\nonumber
    P_\mathrm{NL,IR}(k) &=& P_\mathrm{L}^{nw}(k) + P_\mathrm{1-loop}^{nw}(k)+e^{-k^2\Sigma^2}\left[\left(1+k^2\Sigma^2\right)P_\mathrm{L}^{w}(k)+P_\mathrm{1-loop}^{w}(k)\right],
\eeq
for $P_\mathrm{1-loop}(k) = P_\mathrm{SPT}(k)+P_\mathrm{ct}(k)$ where 
\beq\label{eq: Sigma2-def}
    \Sigma^2 \equiv \frac{4\pi^2}{3}\int_0^\Lambda dq\,P_\mathrm{NL}^{nw}(q)\left[1-j_0(q\ell_\mathrm{BAO})+2j_2(q\ell_\mathrm{BAO})\right],
\eeq
for BAO scale $\ell_\mathrm{BAO}\sim 110h^{-1}\mathrm{Mpc}$ and spherical Bessel function $j_n(x)$ integrating up to $\Lambda = 0.2\hMpc$, following \citet{2016JCAP...07..028B}. Here, the superscripts `$w$' and `$nw$' refer to the wiggle and no-wiggle parts of the power spectrum, with $P_\mathrm{L}^{nw}(k)$ found from $P_\mathrm{L}^w(k)$ using the fourth-order wiggly-smooth decomposition algorithm of \citet{2010JCAP...07..022H}. From this, the other components are defined via
\beq
    P_\mathrm{L}^w(k) &\equiv& P_\mathrm{L}(k) - P_\mathrm{L}^{nw}(k)\\\nonumber
    P_\mathrm{1-loop}^{nw}(k) &\equiv& P_\mathrm{1-loop}\left[P_\mathrm{L}^{nw}\right](k)\\\nonumber
    P_\mathrm{1-loop}^w(k) &\equiv& P_\mathrm{1-loop}\left[P_\mathrm{L}\right](k)-P_\mathrm{1-loop}\left[P_\mathrm{L}^{nw}\right](k),
\eeq
where $P_\mathrm{1-loop}\left[P\right]$ is taken as a functional of the input power spectrum $P$. Note that the IR resummation is \textit{fully deterministic} given input power spectra, and as such, carries no free parameters. Following IR resummation, one-loop EFT claims to be percent-level accurate up to $k\approx 0.3\hMpc$ for matter in real-space {at $z = 0$} 
\citep{2015JCAP...02..013S}. With the Pade resummation of the counterterm and the smoothing function (with its additional free parameter), we expect this to extend to larger $k$, into the one-halo dominated regime.

\section{The Matter Power Spectrum: Comparison to Simulations}\label{sec: sims}

We are now ready to test the {effective} halo model formalism developed above. In this section, we first consider a number of practicalities relating to the choice of mass functions and bias parameters (Sec.\,\ref{subsec: practical-evaluation}) before comparing our model with two suites of $N$-body simulations (Sec.\,\ref{subsec: quijote-and-abacus-sims}), using both a large number of realizations with fixed cosmology (Sec.\,\ref{subsec: sim-comparison}) and a set encompassing a broad range of cosmologies (Sec.\,\ref{subsec: parameters-on-cosmology}).

\subsection{Practical Evaluation of $P_\mathrm{HM}(k)$}\label{subsec: practical-evaluation}
\subsubsection{{Halo Mass Function}}
The first required ingredient to calculate the halo model integrals (Eqs.\,\ref{eq: pk-summary}) is the mass function $n(m)$, usually defined defined in terms of the universal form \citep{1974ApJ...187..425P}
\beq\label{eq: universal-mass-function}
    \frac{m}{\bar\rho}n(m)dm = f(\nu)\frac{d\nu}{\nu},
\eeq
where $\nu \equiv \delta_c(z)/\sigma(m)$, $\delta_c(z)$ is the spherical collapse threshold at redshift $z$ and $\sigma(m)$ is the r.m.s. mass overdensity in a sphere whose Lagrangian radius contains mass $m$. In the $N$-body simulations discussed below (Sec.\,\ref{subsec: quijote-and-abacus-sims}), halos are identified by way of the Friends-of-Friends (FoF) algorithm \citep[e.g.][]{1982ApJ...257..423H} with a linking length of 0.2; for this reason we adopt the {mass function of \citet[Eq.\,12]{2011ApJ...732..122B} (similar to the mass functions proposed in \citealt{2006ApJ...646..881W} and \citealt{2010ApJ...724..878T})}. This takes the form
\beq\label{eq: mass-fn=bhatt}
    f_\mathrm{Bhattacharya}(\nu)=A\sqrt{\frac{2}{\pi}}e^{-a\nu^2/2}\left[1+\left(a\nu^2\right)^{-p}\right]\left(a\nu^2\right)^{q/2}.
\eeq
This is a simple generalization of the canonical \citet{2002MNRAS.329...61S} mass function, and is preferred to other recent formalisms since it respects the normalization condition $\int_0^\infty d\nu\,f(\nu)/\nu = 1$ without divergence (unlike the \citet{2010MNRAS.403.1353C} mass function, which has $f(\nu)/\nu \propto 1/\nu$ for $\nu\rightarrow0$). For this work, we recalibrate the model parameters using a set of measured $z = 0$ halo counts at high mass, giving $a = 0.774$, $p = 0.637$, $q = 1.663$, with $A$ set by normalization. We further compute $\nu$ assuming $\delta_c = 1.686$ (as expected from spherical collapse), with $\sigma$ computed from the \texttt{CLASS} cosmology package \citep{2011JCAP...07..034B}.\footnote{\href{https://lesgourg.github.io/class\_public/class.html}{lesgourg.github.io/class\_public/class.html}} \resub{Note that the primary dependence of $n(m)$ (and the one-halo term) on cosmology is sourced by $\sigma(m)$.}

\subsubsection{{Halo Bias}}
{The second important consideration is the choice of halo biases, which relate the halo density field $n(m|\vec x)$ to the dark matter overdensity $\delta(\vec x)$, as in Eq.\,\ref{eq: n_m_expansion}.}
Numerous published choices exist for the linear bias $\bias{1}(m)$ (see \citet{2018PhR...733....1D} for a comprehensive review), many of which are calibrated from simulations \citep[e.g.][]{2010ApJ...724..878T}. For consistency with the halo mass function, we instead derive the biases from the peak-background-split (PBS) formalism \citep{1999MNRAS.308..119S}, which gives a \textit{Lagrangian} bias of
\beq\label{eq: b1L-def}
    \bias{1}_L(m) = \frac{1}{n(m)}\frac{\partial n(m)}{\partial\delta_l} \equiv -\frac{1}{\delta_c}\frac{d\log f(\nu)}{d\log\nu},
\eeq
where $\delta_l$ is a long wavelength mode. This is related to the linear Eulerian bias by $\bias{1}(m) = 1+\bias{1}_L(m)$. For later use, it is convenient also to define the quadratic bias $\bias{2}(m)$, via 
\beq
    \bias{2}(m) &=& \frac{8}{21}\bias{1}_L(m) + \bias{2}_L(m)\\\nonumber
    \bias{2}_L(m) &=& \frac{1}{n(m)}\frac{\partial^2n(m)}{d\delta_l^2} \equiv \frac{1}{\delta_c^2}\frac{1}{f(\nu)}\frac{\partial}{\partial\log\nu}\left[\nu^2\frac{\partial f(\nu)}{\partial\log\nu}\right].
\eeq
Computing biases in this manner ensures that they automatically satisfy the consistency relation (Eq.\,\ref{eq: bias_consistency_relation}). With the mass function of Eq.\,\ref{eq: mass-fn=bhatt}, we obtain the Lagrangian biases
\beq
    \bias{1}_\mathrm{L,PBS}(m) &=&  -\frac{1}{\delta_c}\left[a \nu ^2-\frac{2 p}{\left(a \nu ^2\right)^p+1}+q\right]\\\nonumber
    \bias{2}_\mathrm{L,PBS}(m) &=& \frac{1}{\delta_c^2}\left[\frac{4p^2-4 p\left(a\nu^2+q\right)}{\left(a \nu^2\right)^p+1} + a \nu ^2 \left(a \nu ^2+2\right)+2 a \nu ^2 q+q^2\right],
\eeq
requiring no further free parameters. Whilst this approach is theoretically appealing, we note that peak-background-split is known to be imprecise and we do not expect the bias models above to exactly match those found in simulations. Whilst it is possible to measure biases from halo-matter power spectra and bispectra \citep[e.g.][]{2012PhRvD..86h3540B}, this is in tension with our overarching goal; to have a power spectrum model that does not require heavy calibration from simulations.\footnote{Whilst we do calibrate the mass function parameters from simulations, these can also be set from established models or determined robustly from a single simulation.} We therefore opt to use PBS biases throughout, and note that, for the power spectrum model, the choice of bias model is of limited importance, since the bias only appears in integrals over mass which are heavily constrained by the consistency condition.

\subsubsection{{Halo Structure}}
For the halo profile, we use the standard NFW parametrization, truncated at the virial radius $r_\mathrm{vir}$ (recently shown to be a fair approximation on a wide range of scales in \citealt{2019arXiv191109720W});
\beq
    u_\mathrm{NFW}(r|m) = \frac{1}{m}\frac{\rho_s}{(r/r_s)(1+r/r_s)^2}
\eeq
\citep{1996ApJ...462..563N} with Fourier transform
\beq
    u_\mathrm{NFW}(k|m) = \frac{4\pi\rho_s r_s^3}{m}\left\{\sin(kr_s)\left[\operatorname{Si}([1+c]kr_s)-\operatorname{Si}(kr_s)\right]-\frac{\sin(ckr_s)}{(1+c)kr_s}+\cos(kr_s)\left[\operatorname{Ci}([1+c]kr_s)-\operatorname{Ci}(kr_s)\right]\right\},
\eeq
where $\operatorname{Si}$ and $\operatorname{Ci}$ are the Sine and Cosine integrals with $\rho_s$ constrained by enforcing $\int_0^{r_\mathrm{vir}} r^2dr\,u(r|m)\equiv 1$. The scale radius $r_s$ is related to the virial radius via $r_s \equiv r_\mathrm{vir}/c$ for halo concentration parameter $c(m,z)$. {Whilst it is possible for $c(m,z)$ to have dependence on the underlying density field $\delta_R$, this effect is expected to be small \citep{2016PhRvD..93f3512S} and is thus ignored in this work.} We parametrize $c$ using the form of \citet{2010MNRAS.405.2161D}:
\beq
    c(m,z) = A\left(\frac{m}{10^{12}h^{-1}M_\odot}\right)^\alpha (1+z)^\beta,
\eeq
for $A = 7.85$, $\alpha = -0.081$, $\beta = -0.71$ (with no refitting performed here). Note that the results in this work are largely insensitive to the exact concentration parametrization {and halo profile truncation} adopted, though this may become important on smaller scales.

\subsubsection{{Consistency Relations}}
Given the above ingredients in addition to the power spectrum model discussed in Sec.\,\ref{subsec: quasi-linear-power-model}, one may compute the mass integrals in Eqs.\,\ref{eq: pk-summary}. For the two-halo prefactor integral, $I_1^1(k)$, convergence is difficult to achieve due to large contributions from low-mass halos that cannot be probed in simulations. To this end, it is standard practice to use the consistency relation (Eq.\,\ref{eq: bias_consistency_relation}) to approximate the integral as
\beq
    I^1_1(k) \equiv \int_0^\infty dm\,\frac{m}{\bar\rho}n(m)\bias{1}(m)u(k|m) \approx \int_{m_\mathrm{min}}^{m_\mathrm{max}}dm\,\frac{m}{\bar\rho}n(m)\bias{1}(m)u(k|m) + A\frac{u(k|m_\mathrm{min})}{u(0|m_\mathrm{min})},
\eeq
(equivalent to \citealt[Appendix A]{2016PhRvD..93f3512S}) where $A = 1 - \int_{m_\mathrm{min}}^{m_\mathrm{max}} dm\,(m/\bar\rho)n(m)\bias{1}(m)u(0|m)$, which reproduces the consistency relation by construction and has little dependence on the mass limits. In this paper we adopt $\{m_\mathrm{min} = 10^6h^{-1}M_\odot, m_\mathrm{max} = 10^{17}h^{-1}M_\odot\}$. {As noted in \citet{2016PhRvD..93f3512S}, we could also opt to truncate at the minimum mass for which $n(m)$ is probed by simulations. This incurs an error scaling as $(kR_\mathrm{min})^2$ where $R_\mathrm{min}$ is the typical halo radius at $m_\mathrm{min}$, and is small except at very large $k$.} 

\subsection{$N$-body Simulations}\label{subsec: quijote-and-abacus-sims}
To test the validity of our power spectrum model, we turn to $N$-body simulations, first from the \texttt{Quijote} project \citep{2019arXiv190905273V},\footnote{\href{https://github.com/franciscovillaescusa/Quijote-simulations}{github.com/franciscovillaescusa/Quijote-simulations}} consisting of 43,100 simulations spanning over 7000 cosmologies using a total of above 8.5 trillion particles, run using the \texttt{GADGET-III} TreePM + SP code \citep{2005MNRAS.364.1105S}. Each simulation has the form of a periodic $1h^{-3}\mathrm{Gpc}^3$ simulation box containing a large number of cold dark matter (CDM) particles that are evolved forward from $z = 127$, with initial conditions generated from second-order Lagrangian perturbation theory (2LPT). In this work we make extensive use of four types of simulations;
\begin{enumerate}
    \item \textbf{Fiducial High-Resolution}: These comprise 100 simulations each containing $1024^3$ particles generated using the fixed cosmology $\{\Omega_m = 0.3175, \Omega_b = 0.049, h = 0.6711, n_s = 0.9624, \sigma_8 = 0.834, M_\nu = 0\,\mathrm{eV}, w = -1\}$. {These have mass resolution $M_\mathrm{min} = 8.2\times 10^{10}h^{-1}M_\odot$} and are expected to produce percent-level accurate power spectrum up to $k\approx 0.8\hMpc$ \citep[Fig.\,15]{2019arXiv190905273V}.
    \item \textbf{Fiducial Standard-Resolution}: We use 15,000 simulations at lower resolution ($512^3$ particles, {$M_\mathrm{min} = 6.6\times 10^{11}h^{-1}M_\odot$}) to test our model for the covariance between the halo counts and the matter power spectra (Secs.\,\ref{sec: cov_N_Pk_derivation}\,\&\,\ref{sec: sims-cov}).
    \item \textbf{Latin Hypercube High-Resolution}: These are a set of 2000 simulations (containing $1024^3$ particles) spanning a latin hypercube of $\Omega_{m,0} \in [0.1,0.5]$, $\Omega_{b,0}\in [0.03,0.07]$, $h\in [0.5,0.9]$, $n_s\in [0.8,1.2]$, $\sigma_{8,0}\in [0.6,1.0]$, again assuming a flat universe, $w=-1$, and no massive neutrinos. These are used to test the dependence of our model on cosmology.
    \item \textbf{Separate Universe}: \texttt{Quijote} includes {standard}-resolution 
    simulations which emulate the effect of a large-scale overdensity in the box by way of altered cosmological parameters (as discussed in \citealt{2014PhRvD..89h3519L}). Here we use 100 simulations with background overdensity $\delta_b = -0.035$ and a further 100 with $\delta_b = 0.035$. These can be compared to the fiducial simulations.
\end{enumerate}
For each simulation, power spectra have been computed from 3D density fields created using triangle-in-cell interpolation with $N_\mathrm{grid}=2048$, with no redshift-space distortions included. 
We principally work at $z = 0$ since this is the redshift at which the conventional halo model is least accurate. For each snapshot, halos are identified using the Friends-of-Friends algorithm with a linking length $b = 0.2$ \citep[e.g.][]{1982ApJ...257..423H}; these are required for testing of our covariance model.

As an important cross-check, we additionally test our power spectrum model on simulations from the \texttt{AbacusCosmos} suite, which is computed using an entirely separate $N$-body code: \texttt{Abacus} \citep{2018ApJS..236...43G,2019MNRAS.485.3370G}. Each contains $1440^3$ CDM particles in a periodic box of volume $(1.1h^{-1}\mathrm{Gpc})^3$ based on the Planck 2015 cosmology \citep{2016A&A...594A..13P}. These use 2LPT initial conditions at $z = 49$, with power spectra computed as for the \texttt{Quijote} simulations. These boast a mass resolution of $4\times 10^{10}h^{-1}M_\odot$, and are thus expected to be more accurate then the \texttt{Quijote} simulations at high $k$. Due to their small number, the \texttt{AbacusCosmos} simulations will not be used to probe the covariance model introduced in Sec.\,\ref{sec: cov_N_Pk_derivation}.

\subsection{Comparing the $P(k)$ Model to Simulations with Fixed Cosmology}\label{subsec: sim-comparison}
We begin by testing the {effective halo model} prediction for $P_\mathrm{HM}(k)$ using the 100 fiducial high-resolution simulations described above. All power spectrum estimates are computed using our public \texttt{EffectiveHalos} code,\footnote{\href{EffectiveHalos.rtfd.io}{EffectiveHalos.readthedocs.io}} which computes halo-model spectra from scratch in a few seconds.
Following the method of Sec.\,\ref{sec: Pk_derivation}, we compute the one- and two-halo power spectra using the following models;
\begin{enumerate}
    \item `Full Model': The complete power spectrum model given in Eq.\,\ref{eq: pk-summary}, including one-loop EFT corrections, density field filtering (via the smoothing function $W(kR)$) and IR resummation.
    \item `No IR': As above, but without any IR resummation (effectively setting $\Sigma^2 = 0$ in Eq.\,\ref{eq: one-loop-IR}).
    \item `No Truncation': As above, but without any density field smoothing (effectively setting $R=0$).
    \item `No Counterterm': As above, but without the UV counterterm of Eq.\,\ref{eq: counterterm_final} (effectively setting $c_s^2=0$). This is equivalent to using an (IR-resummed) SPT power spectrum instead of EFT.
    \item `No Pade': As above, but using the EFT counterterm of Eq.\,\ref{eq: counterterm_initial}, without Pade resummation.
    \item `Vanilla': The standard halo model, which uses a linear power spectrum with no IR resummation, counterterms or halo truncation.
\end{enumerate}
For each model, we fit the halo truncation radius ($R$) and speed-of-sound parameter ($c_s^2$) to the mean of all 100 simulations, fitting for 125 $k$-bins up to $k_\mathrm{max} = 0.8\hMpc$ (beyond which the simulations are not expected to be 1\% accurate).\footnote{{The \texttt{AbacusCosmos} simulations used below have greater small-scale accuracy, justifying our claim of percent-level accuracy up to $k = 1\hMpc$.}} For simplicity, we assume a diagonal covariance matrix to perform this fit, such that
\beq
    \operatorname{var}\left[P(k)\right] = \frac{2}{N_\mathrm{modes}(k)}P^2_\mathrm{HM}(k)
\eeq
where $N_\mathrm{modes}$ is the number of modes in the $k$-space bin centered at $k$.\footnote{This is simply the Gaussian part of the full covariance matrix of Eq.\,\ref{eq: full-covariance-Pk}, excluding the sub-leading thrice-contracted term.} Including off-diagonal elements in the covariance was not found to make an appreciable difference to the fit.

\begin{figure}
    \centering
    \subfloat[100 \texttt{Quijote} simulations at $z = 0$]{\includegraphics[width=0.5\linewidth]{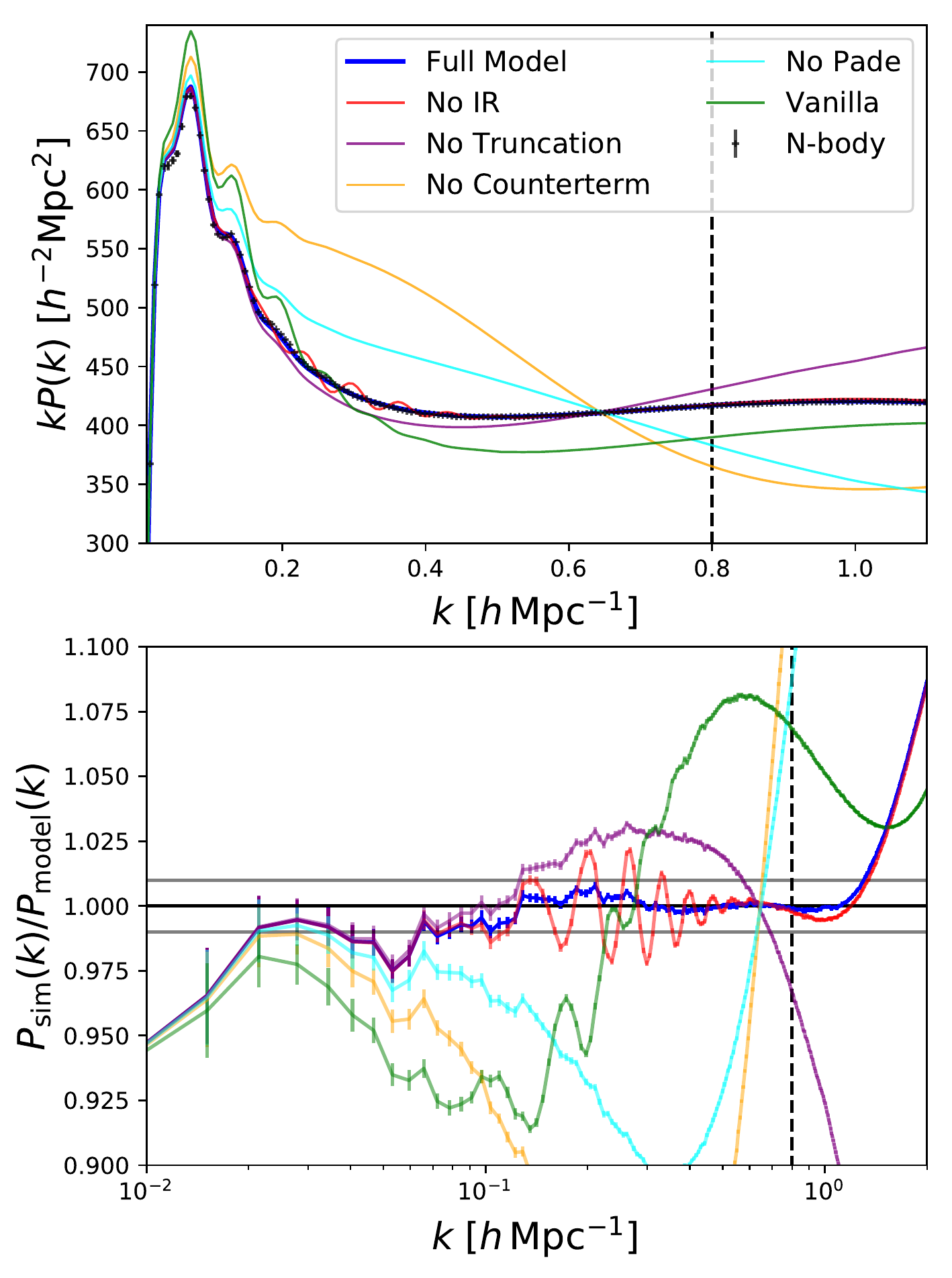}}
    \subfloat[20 \texttt{AbacusCosmos} simulations at $z = 0.3$]{\includegraphics[width=0.5\linewidth]{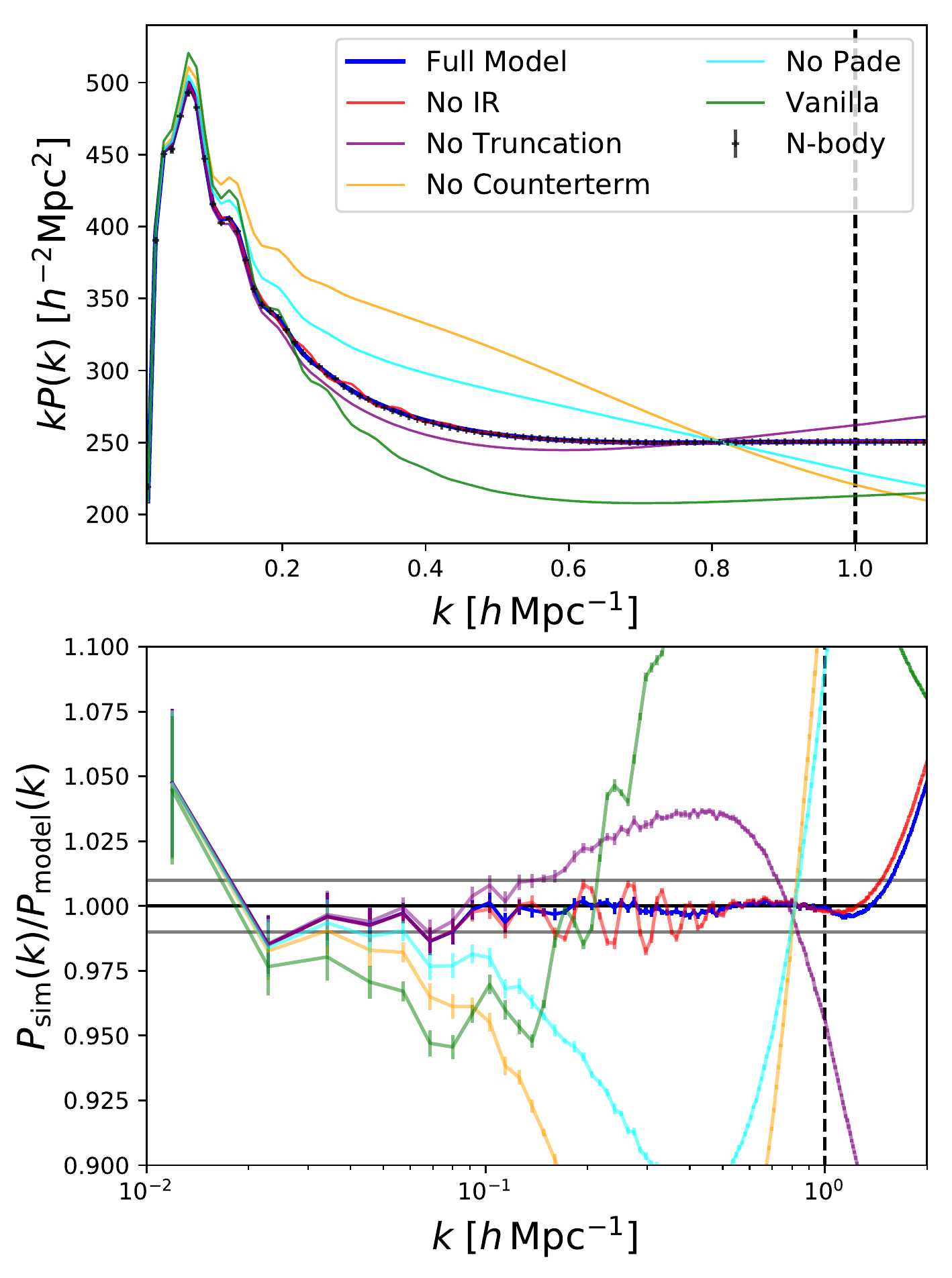}}\\
    \caption{Comparison of simulated and model power spectra using a variety of power spectrum models, as discussed in Sec.\,\ref{subsec: sim-comparison}. In the upper plots, we show the mean spectrum from $N$-body simulations as black points, with the fitted models shown as colored curves. Parameters were optimized using all data-points to the left of the dashed line. In the lower sections, we plot the ratio of simulated to model power, with the gray horizontal lines indicating 1\% errors. For both plots, error bars are computed using the standard error {in the mean of all simulations}. We note that the full {effective halo model}, which is the main result of this paper, achieves 1\% accuracy (within statistical error) across the fitting range in both cases. {The slight under- (over-)prediction of power on the largest scales in the left (right) panel is attributed to cosmic variance. At large $k$, deviations will arise from the lack of precision in the simulations.}}
    \label{fig: pk_comparison}
\end{figure}

The results for this are shown in Fig.\,\ref{fig: pk_comparison}a, plotting the simulated and model power spectra as well as their ratio. {For the $k$-range probed herein}, we first note that the canonical halo model is clearly deficient on all but the largest scales (where it is identical to other power spectrum models), with a particularly large power deficit noted between one- and two-halo scales. {(The vanilla model is additionally expected to be an accurate predictor of the \resub{dark matter power spectrum} at large $k$, though these scales are not accurately predicted by \texttt{Quijote}.} In contrast, the full power spectrum model considered in this paper achieves 1\% accuracy (within statistical error) on all $k$-scales from $k = 0.02\hMpc$ to the fitting radius of $k = 0.8\hMpc$. This uses optimal parameters of $\widehat{c}_s^2=9.26h^{-2}\mathrm{Mpc}^2$ and $\widehat{R} = 1.71\Mpch$, suggesting that the density field should be smoothed on scales $\sim 1.5\Mpch$. We note that this is comparable to the physical scale of halos of mass $\sim 10^{14}h^{-1}M_\odot$, where the mass function $n(m)$ peaks, as one may expect from a na\"ive estimate. {We note that our $c_s^2$ estimate is somewhat higher than that found in canonical EFT papers (e.g., $c_s^2\sim 3h^{-2}\mathrm{Mpc}^2$  in \citealt{2014JCAP...07..056C}). We expect this to arise from (a) the smoothing window, which reduces the counterterm amplitude, (b) the broad fitting range applied and (c) the one-halo term, which contributes additional power on perturbative scales due to the lack of halo compensation.}

By considering the additional models described above, we can see the effects of including various components in our power spectrum model.\footnote{Note that the parameters $R$ and $c_s^2$ are fitted for each model separately, and are not directly comparable, e.g., the halo truncation radius will be affected by the counterterm, since both give corrections $\sim k^2$ at low $k$.} The inclusion of IR resummation (which does not carry additional free parameters) is not seen to affect the broadband spectrum (as expected), but significantly reduces spurious wiggles in the power spectrum model at mildly non-linear $k$. These have an amplitude of $\sim 2\%$ at $z = 0$, demonstrating that IR resummation is a necessary procedure if we wish to obtain percent-level accurate spectra. Setting the halo truncation radius (or density field smoothing scale) to zero is seen to have only minor effects on the power spectrum on large scales (as expected since $R\sim 1\Mpch$), yet without its effect we have a large excess of power on non-linear scales where the two-halo term remains large. In some ways, the smoothing term is analogous to the exclusion model of \citet{2011A&A...527A..87V}, which similarly reduces the two-halo contributions at high-$k$. The action of the EFT counterterm is also clear; without its inclusion the power spectrum is grossly overestimated on even mildly non-linear scales. This is inline with the general finding that one-loop SPT overestimates the true dark matter power spectrum. In addition, we note that our optimizer is unable to produce a good fit of model to data if the Pade-resummation is not applied to the EFT counterterm; this is because the na\"{i}ve counterterm scales as $k^2P_\mathrm{L}(k)$ which is unphysically large on non-linear scales. {An additional benefit of our model is that the free parameters $c_s^2$ and $R$ naturally correct any minor inaccuracies in the one-halo term close to the transition scale.}

To check for consistency, we show analogous results using the \texttt{AbacusCosmos} simulations in Fig.\,\ref{fig: pk_comparison}b. These were computed from 20 simulations at $z = 0.3$, as described in Sec.\,\ref{subsec: quijote-and-abacus-sims}, and, due to their higher resolution, we choose to fit up to $k = 1\hMpc$. Here, this gives $\widehat{c}_s^2=5.75h^{-2}\mathrm{Mpc}^2$ and $\widehat{R} = 1.29\Mpch$, {with the difference with respect to \texttt{Quijote} mainly attributed to the higher redshift.}
Our conclusions are qualitatively similar to those for \texttt{Quijote}; our power spectrum model is 1\% accurate for $k \in [0.02,1]\hMpc$, and we note the importance of including a resummed EFT counterterm and density field smoothing. Note that the excess wiggles in the model spectra without IR-resummation are smaller in this case; this is due to the higher redshift, which reduces structure formation (and additionally $c_s^2$, {whose time-dependence is expected to scale as the growth factor squared}). We further note that it is difficult to probe $k\lesssim 0.02\hMpc$ with these simulations due to cosmic variance, but linear theory is expected to work well in this regime.

It is important to consider how well our model performs when fit to only a single simulation (as would be done in analysis of observational data). For this, we perform the above fitting procedure on each of the 100 fiducial \texttt{Quijote} simulations separately (at $z =0$ and $1$), to obtain the optimal parameters $\{R, c_s^2\}$ in each case. The {mean}
parameters are $\widehat{R} = \left(1.71\pm 0.03\right)\Mpch$, $\widehat{c_s}^2 = \left(9.26\pm 0.11\right)h^{-2}\mathrm{Mpc}^2$ at $z = 0$ and $\widehat{R} = \left(1.27\pm 0.02\right)\Mpch$, $\widehat{c_s}^2 = \left(2.36\pm 0.04\right)h^{-2}\mathrm{Mpc}^2$ at $z = 1$, with the standard deviations indicating the variation between mocks. Note that $R$ roughly scales with $(1+z)$ and $c_s^2$ reduces at higher $z$ (since non-linearities become less important). We also note a significant anticorrelation between the parameters, with correlation coefficient of $-0.31$ ($-0.78$) at $z = 0$ ($z=1$). 
This may be rationalized by noting that, at leading order in $k$, both effects modify the low-$k$ two-halo power spectrum by the same factor, $k^2P_\mathrm{L}(k)$. {At high redshift, we expect $c_s^2$ and $R$ to be fully degenerate, thus our formalism reduces to a one-parameter model.}

\begin{figure}
    \centering
    \subfloat[Allowing parameters to vary between simulations]{\includegraphics[width=0.5\linewidth]{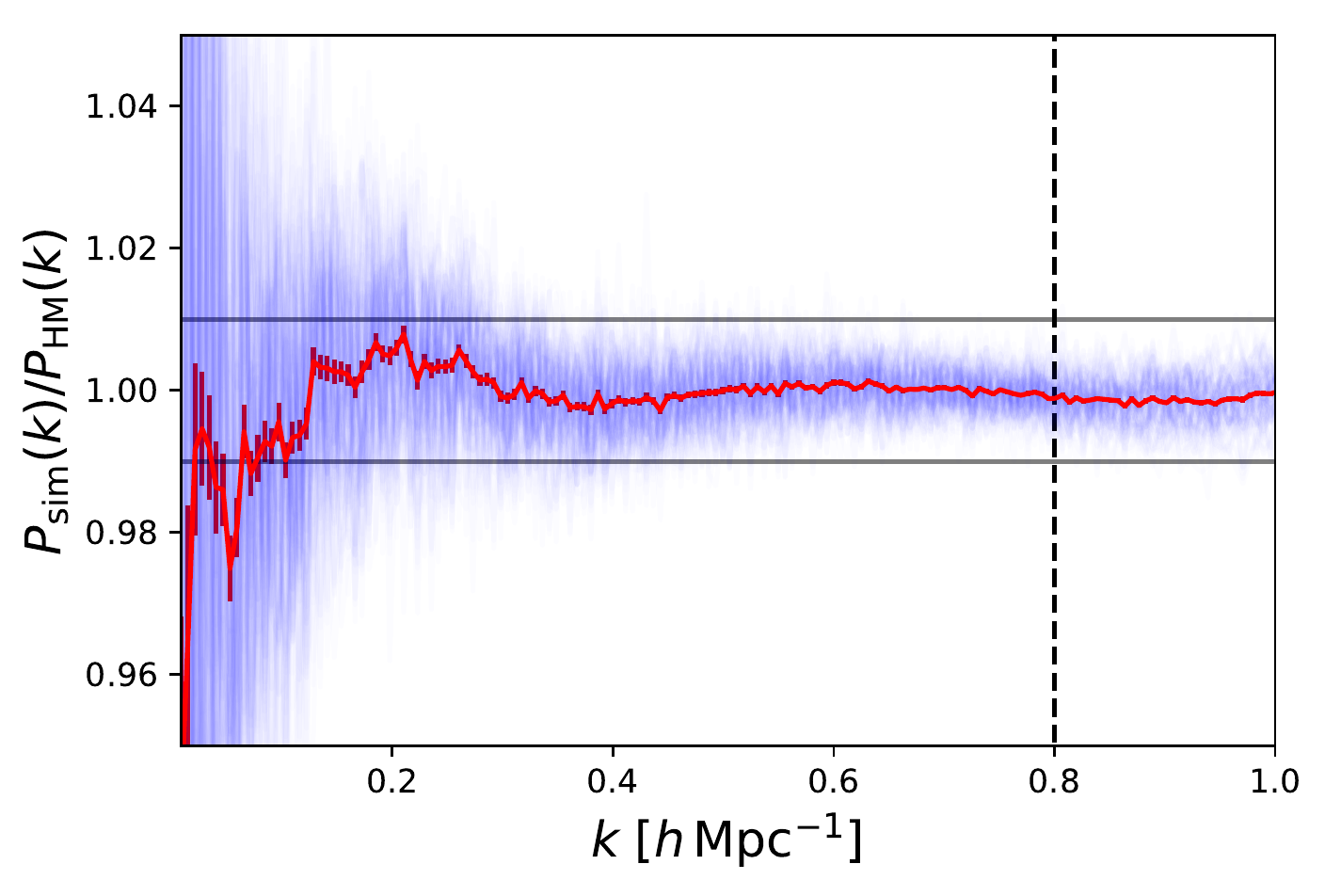}}
    \subfloat[Fixing parameters to their mean values]{\includegraphics[width=0.5\linewidth]{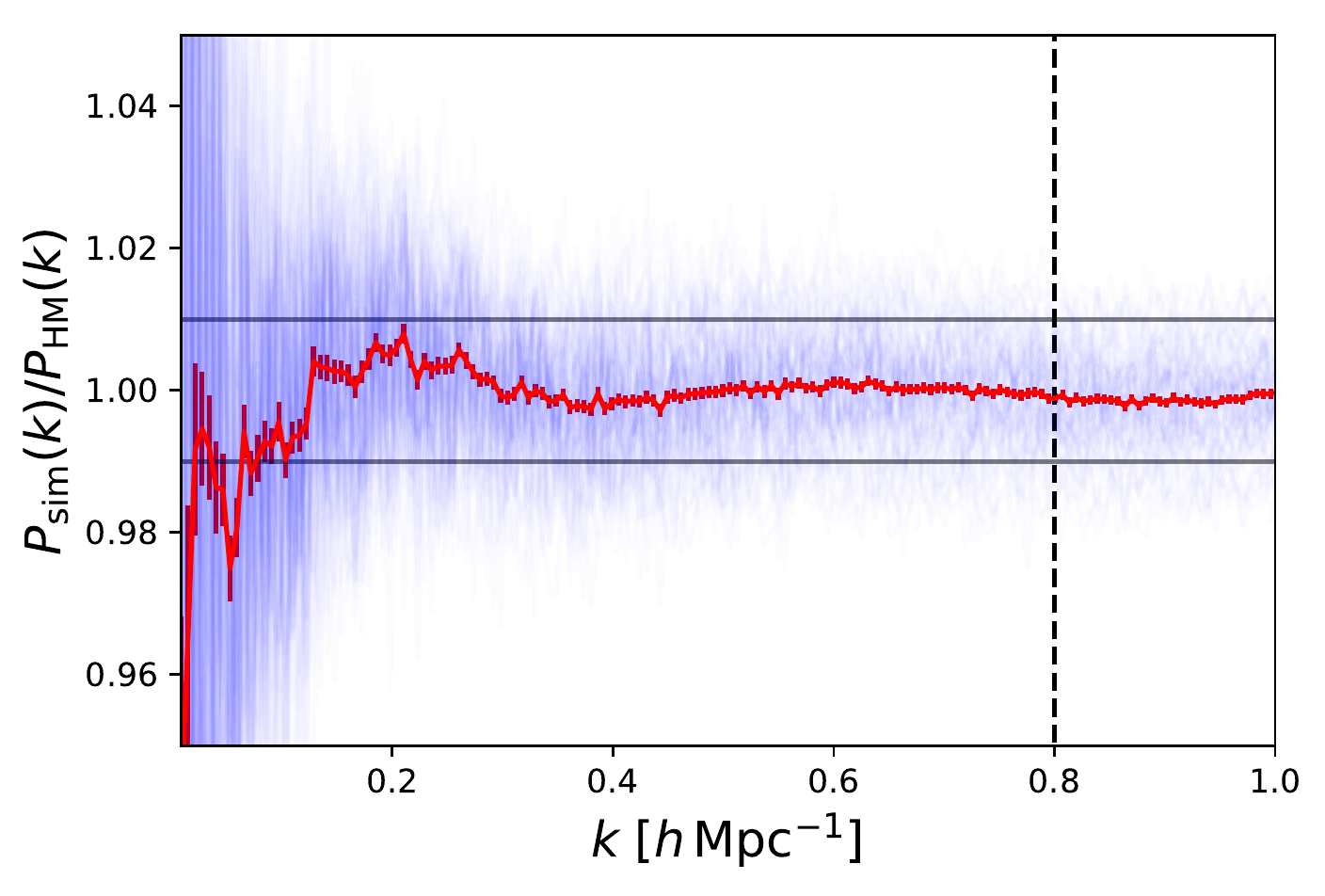}}\\
    \caption{Ratio of $N$-body to model power spectrum for 100 high-resolution \texttt{Quijote} simulations with fiducial cosmology. Individual ratios are plotted in blue with the mean in red, and each uses the full $P_\mathrm{HM}(k)$ model defined in Sec.\,\ref{subsec: sim-comparison}. A vertical line gives the maximum $k$ for which the power spectra are fit. In the left panel the model parameters $R$ and $c_s^2$ are allowed to vary between simulations, whereas in the right, they are fixed to the values obtained from the mean of 100 simulations, giving a somewhat worse fit.}\label{fig: fid_pk}
\end{figure}

The ratio of the model and simulated power spectrum for each simulation is shown in Fig.\,\ref{fig: fid_pk}a; we note excellent agreement between the two in all cases, with percent-level agreement obtained for $k\in[0.3,0.8]\hMpc$ in individual simulations.\footnote{At lower $k$, the intrinsic variance of the simulations makes this difficult to test, but from Fig.\,\ref{fig: pk_comparison} we may be confident in our model down to $k\sim 0.02\hMpc$.} The mean ratio is also shown, which is again seen to be in excellent agreement. When performing these fits, it is important to allow for the intrinsic scatter between different mocks; to demonstrate this, we plot the ratio of simulated to model $P(k)$ in Fig.\,\ref{fig: fid_pk}b, where each model is computed using the \textit{mean} parameters obtained from the mean-of-100-mocks analysis above. Here, we observe substantially larger variation between mocks, and percent-level accuracy is not always achieved. {This difference arises due to cosmic variance, indicating that the parameters depend on the specific realizations of small-scale physics. By marginalizing over them in full analyses, we can ensure that we fully account for the simulation-specific non-linear physical processes.}

\subsection{Dependence of Model Parameters on Cosmology}\label{subsec: parameters-on-cosmology}

Given the power spectrum parametrization of Eq.\,\ref{eq: pk-summary}, it is interesting to consider the dependence of the free parameters $R$ and $c_s^2$ on cosmology. Whilst the intrinsic parameter scatter found above indicates that we will not be able to obtain percent-level agreement from a model where the parameters are taken from deterministic cosmology-dependent relation, it is important to understand these relations in order to place sensible priors on $R$ and $c_s^2$. In particular, we expect there to be strongest dependence on $\Omega_m$ and $\sigma_8$ which describe the mean and variance of the universe's matter density respectively.

For this purpose, we utilize the 2000 latin-hypercube \texttt{Quijote} simulations described in Sec.\,\ref{subsec: quijote-and-abacus-sims}. To ensure that our simulations are physically reasonable, we additionally restrict to $\Omega_{m,0}\in[0.2,0.4]$, $\sigma_{8,0} \in [0.7,0.9]$. For each simulation, we compute the power spectrum model of Eq.\,\ref{eq: pk-summary} (including all model components) and fit for the parameters $R$ and $c_s^2$ as in Sec.\,\ref{subsec: sim-comparison}. To account for the redshift evolution of our parameters, the analysis is performed for three redshifts ($z\in\{0,0.5,1\}$) and we additionally store the value of $\sigma_8(z)$ at each redshift. Whilst these simulations are not phase-matched (and thus have different cosmic variance effects), we do not expect this to affect our fit, which depends predominantly on mildly non-linear scales.

Given the set of model parameters and cosmologies across three redshifts (treated as independent samples), we find the optimal parameters are well fit by the relations
\beq
    \overline{R} &\approx& \left(1.94\,\mathrm{Mpc}\right)\, \left(\frac{\Omega_{m,0}}{0.3}\right)^{1.142}\left(\frac{\sigma_8(z)}{0.8}\right)^{0.911} \left(\frac{n_s}{0.96}\right)^{2.167}\\\nonumber
    \overline{c}_s^2 &\approx& \left(7.34\,\mathrm{Mpc}^2\right)\, \left(\frac{\Omega_{m,0}}{0.3}\right)^{-0.139}\left(\frac{\sigma_8(z)}{0.8}\right)^{2.487} \left(\frac{n_s}{0.96}\right)^{-0.56}.
\eeq
Note that $R$ depends strongly on both $\Omega_{m,0}$ and $\sigma_8(z)$, whilst $c_s^2$ principally depends on the amplitude of clustering, $\sigma_8(z)$. In both cases, we found a need to include dependence on an additional parameter, with $n_s$ found to give best results. Furthermore, the dependence on redshift appears to be entirely encapsulated by the $\sigma_8(z)$ parameter. Whilst we caution that these relations should not be used to give fixed values to the model parameters {(since they are approximate and do not capture the variations in small-scale physics between simulations)}, they should be useful in setting priors.

\begin{figure}
    \centering
    \includegraphics[width=0.5\textwidth]{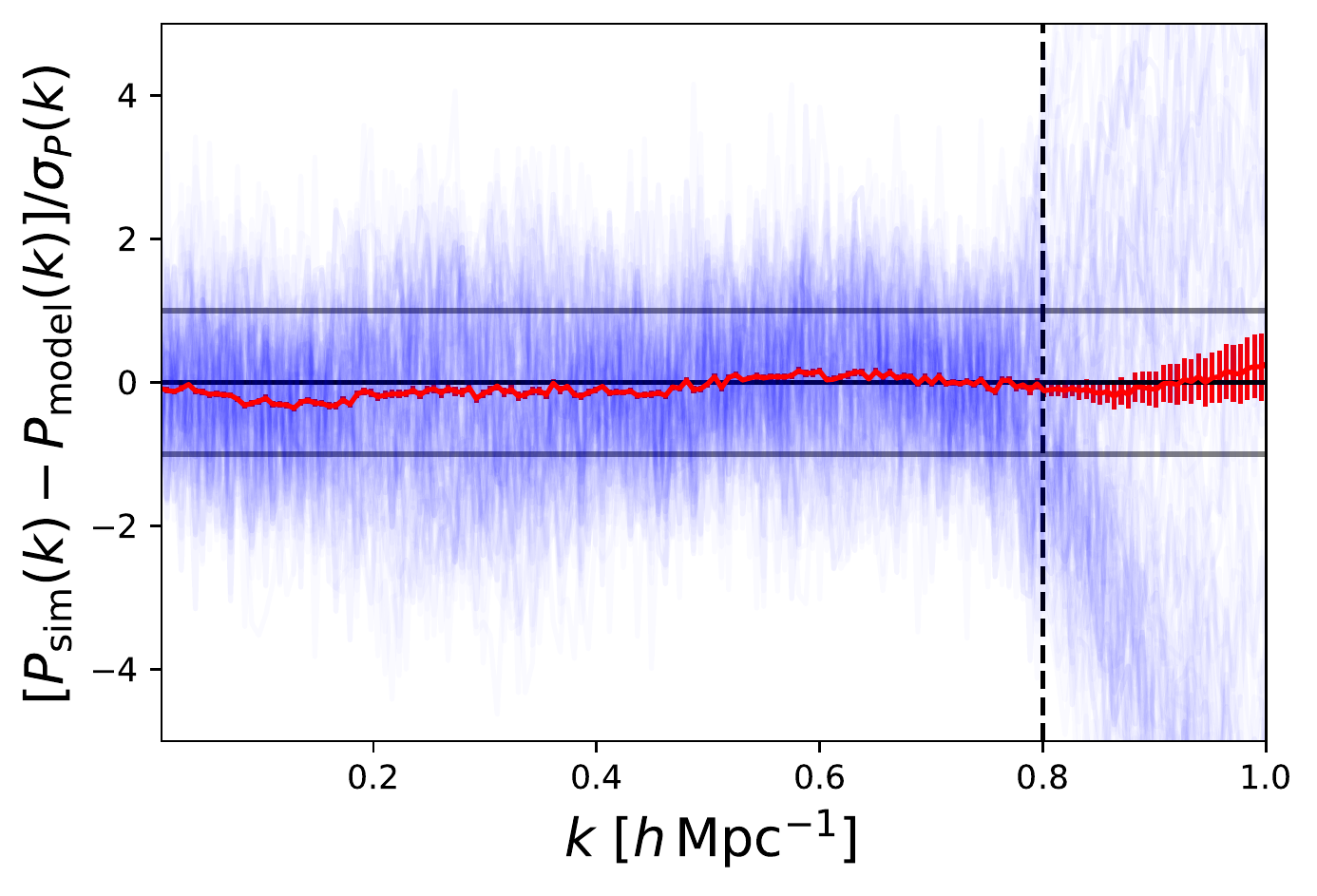}
    \caption{Comparison of the model- to statistical-error in the matter power spectrum using a large suite of high-resolution \texttt{Quijote} `latin-hypercube' simulations. {These contain a variety of $\Lambda$CDM cosmologies, with parameters drawn from broad distributions.} The blue lines show the ratios obtained for 100 randomly selected simulations, whilst the red shows the mean (and variance) of the set. Solid black lines show $0$ and $1\sigma$ deviations, and $\sigma$ is estimated from the model, assuming a diagonal covariance. We fit the two parameter power spectrum of Sec.\,\ref{sec: Pk_derivation} to each simulation, and note excellent agreement up to the fitting $k$. {This implies that our model remains appropriate for non-standard cosmologies.}}
    \label{fig: all_cosmology_ratio}
\end{figure}

Using the above set of simulations, we may additionally ask the question: is our model still accurate for non-standard cosmologies? To answer this, we plot the ratio of the difference between simulated- and model-spectrum and the statistical error {(assuming a Gaussian covariance)} 
in Fig.\,\ref{fig: all_cosmology_ratio}, with the parameters allowed to vary freely in each realization. From this it is clear that the deviations between model and true $P(k)$ are {compatible with} 
that expected from statistical error for all $k$ up to $k_\mathrm{fit}$ for a large variety of cosmologies; our model is thus applicable to many scenarios. It may also be shown that the model remains percent level accurate up to $k=0.8\hMpc$. {Whilst this has been performed only for $\Lambda$CDM cosmologies, it is likely to extend to those including massive neutrinos, simply by replacing $P_\mathrm{NL}$ with the non-linear power spectrum of CDM and baryons, computed in the presence of neutrinos \citep{2014JCAP...12..053M}, \resub{and assuming linear spectra for the neutrino and cross terms}. Whilst this results in non-linear spectra that are not separable in space and time, it is an acceptable assumption to use the standard Einstein de-Sitter growth factors, since leading-order corrections are absorbed into the EFT counterterm \citep{2019JCAP...11..034C}}.

\subsection{Validity of the Model Assumptions}\label{subsec: model-assumptions}
{Before continuing, it is important to discuss the validity of our physical model in light of the accurate predictions made above. Firstly, we consider its behavior on ultra-large scales, i.e. $k\ll k_\mathrm{NL}$ for non-linear scale $k_\mathrm{NL}$. In this limit, $u(k|m)\rightarrow 1$, and $P_\mathrm{EFT}(k)\rightarrow P_\mathrm{L}(k)$, thus 
\beq\label{eq: ultra-low-k}
    \lim_{k\rightarrow 0}P_\mathrm{HM}(k) = P_\mathrm{L}(k) + \int dm\, \frac{m^2}{\bar\rho^2}n(m) \neq P_\mathrm{L}(k)
\eeq
Our model therefore violates linear theory on the largest scales, due to an additional shot-noise term. This has not been obvious in the previous sections, since we restrict our analysis to $k\gtrsim 0.02\hMpc$ (due to the finite simulation volume). This effect arises from over-counting, since we do not remove the two-halo contributions from pairs of particles in the same halo. This phenomenon is discussed in detail in \citet{2016JCAP...03..007B}, who suggest that $P^{1h}$ should be altered to use the \textit{difference} of true and perturbation-theory one-halo profiles; this scales as $k^4$ at low-$k$ (as expected on physical grounds), nulling this effect. Alternative approaches include invoking halo exclusion \citep{2011A&A...527A..87V}, a halo compensation function \citep{2015PhRvD..91l3516S,2020PhRvD.101j3522C}, \resub{or an explicit restriction of the two-halo term to scales outside the virial radius \citep{2007PhRvD..75f3512S,2011PhRvD..83d3526S}}. {Of particular interest is the method proposed by \citet{2016PhRvD..93f3512S}; this includes an additional stochastic field $\epsilon$ in the bias expansion of Eq.\,\ref{eq: n_m_expansion}, which sources the one-halo term at high-$k$ and is required to have zero intercept by the consistency relations. A phenomenological interpolation function is used to unite the regimes, smoothly removing the extra term in Eq.\,\ref{eq: ultra-low-k}.} We note that any of these methods will only affect large scales; in \citet{2015PhRvD..91l3516S}, the compensated one-halo term has a dominant $k^0$ shot-noise component beyond $k\sim 0.1\hMpc$ matching the standard one-halo term, and the canonical one-halo term is found to be a good fit to simulations in \citet{2019ApJ...884...29N}.}

{Related to this is the behavior close to the non-linear scale. As shown in Fig.\,\ref{fig: introductory-plots}a, our model receives large contributions from both one- and two-halo terms at mildly non-linear $k$, with equal power by $k\sim 0.5\hMpc$. This may seem unusual, since EFT is known be percent-level accurate up to $k\sim 0.2\hMpc$ at $z = 0$ \citep{2012JHEP...09..082C}, yet, on its own, our two-halo term (which is simply one-loop EFT for $I_1^1(k)\approx 1$) severely underpredicts the power spectrum. This is not a mistake in our implementation of EFT; instead the presence of the significant one-halo term (which is verified by simulations, as discussed above) is compensated for by the free parameters $c_s^2$ and $R$, which act to reduce the two-halo contribution. This additionally explains the large value of $c_s^2$ found in Sec.\,\ref{subsec: sim-comparison}. (This can be verified by fitting the model using only the two-halo term; restricting to $k \lesssim 0.3\hMpc$, one still obtains percent-accurate predictions with only $c_s^2$ varied). Close to $k_\mathrm{NL}$, our model thus receives significant contributions from both EFT and the one-halo term. Indeed, the combination of both effects also allows us to provide an excellent model of the power between $k \approx 0.2\hMpc$ and $k \approx 0.4\hMpc$, despite the lack of inclusion of the two-loop EFT terms.}

\section{Halo Count Covariances: Theoretical Model}\label{sec: cov_N_Pk_derivation}
Given the success of the {effective} halo model in describing the power spectrum $P(k)$, we now apply the architecture of Sec.\,\ref{sec: Pk_derivation} to additional statistics: halo number counts, {considering both their auto-covariance and cross-covariance with $P(k)$. These are useful observables} for thermal Sunyaev-Zel'dovich and weak lensing analyses \citep[e.g.][]{2007NJPh....9..446T,2014MNRAS.441.2456T,2014PhRvD..90l3523S,2017A&A...604A..71H}. To begin with, note that the number of clusters in a region of volume $V$ with mass in $\left[m,m+\delta m\right]$ can be written
\beq\label{eq: N-def}
    \delta \hat{N} \equiv \hat{N}(m)\delta m = \int d\vec x\,\hat{n}(m|\vec x)\times \delta m,
\eeq
with expectation $Vn(m)\delta m$. {(Throughout this work, $N(m)$ will refer to number counts and $n(m)$ to number densities.)}\footnote{This is an example of a scenario in which it is \textit{not} correct to invoke ergodicity and equate the integral of $\hat{n}(m|\vec x)$ across the survey with the expectation $n(m)$. This would imply that $N(m)$ was everywhere fixed, nulling the statistic.} It is simpler to first use only the differential form $N(m)$ and consider the effects of finite mass bins later. Working in configuration space, the covariance between $\hat{N}(m)$ and the 2PCF $\xi(\vec r)$ is defined as 
\beq
    \operatorname{cov}(N(m),\xi(\vec r)) &\equiv& \av{\hat{N}(m)\hat{\xi}(\vec r)}-\av{\hat{N}(m)}\av{\hat{\xi}(\vec r)}.
\eeq
{Likewise, the covariance between cluster counts of two different masses (labelled $m_1$ and $m_2)$ is given by}
\beq
    \operatorname{cov}(N(m_1),N(m_2)) &\equiv& \av{\hat{N}(m_1)\hat{N}(m_2)}-\av{\hat{N}(m_1)}\av{\hat{N}(m_2)}.
\eeq

In practice, there are three principal contributors to the covariance:
\begin{enumerate}
    \item \textbf{Intrinsic covariance}, i.e. the standard covariance expected that one would obtain for a survey of infinite volume. This arises from $\hat{N}(m)$ tracing the underlying matter density field $\delta(\vec x)$ as well as from Poissonian contractions of the field $\delta\hat{n}(m|\vec x)$;
    \item \textbf{Halo exclusion covariance}, from the condition that multiple halos cannot co-exist in the same spatial location, \resub{both physically, and due to the halo-finder algorithm}. {Whilst previous works have considered this for the matter power spectrum \citep[e.g.,][]{2011PhRvD..83d3526S,2013MNRAS.430..725V,2013PhRvD..88h3507B},} it has not been accounted for in the halo covariance, and we find it to be an important piece of the puzzle;
    \item \textbf{Super-sample covariance}, arising from density field fluctuations on scales comparable to the survey width, which modulate the background number density of the spatial region in question. Whilst important cosmologically, these effects are not usually included in $N$-body simulations due to the condition of fixed mass in the box.
\end{enumerate}
Each contribution will be discussed in detail in the following sections.

\subsection{Intrinsic Covariance}\label{subsec: no-ssc-deriv}
\subsubsection{Cross-Covariance of $N(m)$ and $P(\vec k)$}
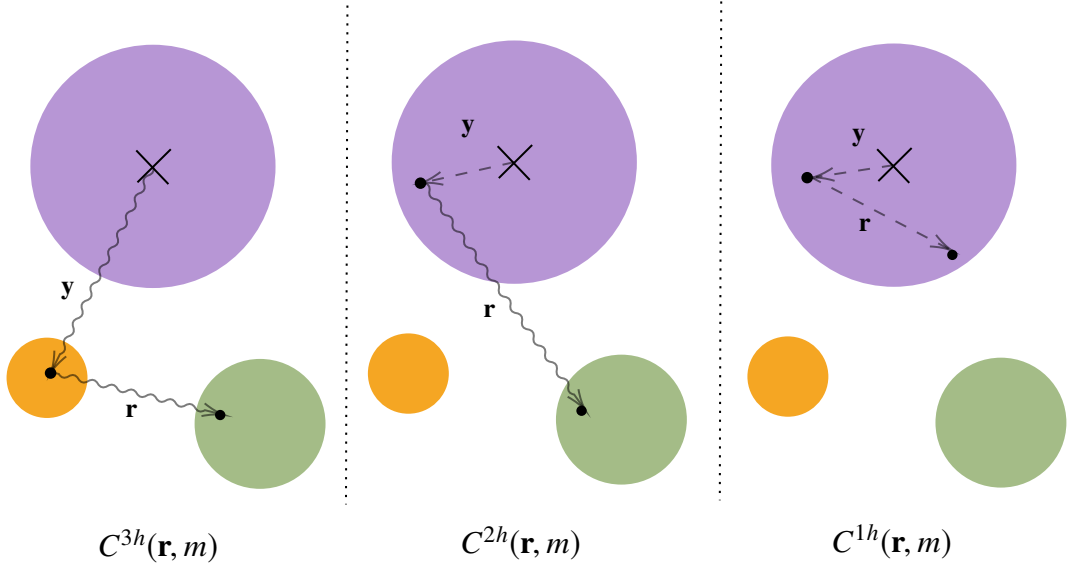
\begin{figure}
    \centering
    \resizebox{0.8\textwidth}{!}{%
    \input{cov-fig}   }%
    \caption{Cartoon showing the various terms in the covariance between halo number counts and the matter power spectrum. The cross indicates the position of the halo used for the number counts (of mass $m$) and the two points indicate matter particles, which are assumed to lie in halos. To form the covariance, we integrate over both $\vec y$ and sum over halos. Dashed and wavy l (and the one-halo term) is sourced by $\sigma(m)$ and correlation function $\xi(\vec x)$ (and non-linear corrections) respectively. Whilst a second two-halo term is possible (with two matter particles in the same halo) this evaluates to zero and is thus excluded from the figure.}\label{fig: cartoon-cov}
  \end{figure}

The intrinsic covariance may be formulated following a similar methodology to the power spectrum derivation (Sec.\,\ref{sec: Pk_derivation}). We begin {by considering the covariance of $N(m)$ and $\xi(\vec r)$}, inserting the definitions (Eqs.\,\ref{eq: N-def}\,\&\,\ref{eq: 2PCF-definition}) into the covariance definition;
\beq\label{eq: cov_initial}
    \operatorname{cov}\left(N(m),\xi(\vec r)\right)^\mathrm{intrinsic} &=& \int \frac{d\vec xd\vec y}{V}\,\left[\av{\hat{n}(m|\vec x)\hat{\delta}_\mathrm{HM}(\vec x+\vec y+\vec r)\hat{\delta}_\mathrm{HM}(\vec x+\vec y)}-\av{\hat{n}(m|\vec x)}\av{\hat{\delta}_\mathrm{HM}(\vec x+\vec y+\vec r)\hat{\delta}_\mathrm{HM}(\vec x+\vec y)}\right]\\\nonumber
    &=& \int \frac{d\vec xd\vec x_1d\vec x_2d\vec y}{V}dm_1dm_2\,\frac{m_1m_2}{\bar\rho^2}u(\vec x+\vec y+\vec r-\vec x_1|m_1)u(\vec x+\vec y-\vec x_2|m_2)\av{\delta\hat{n}(m|\vec x)\delta\hat{n}(m_1|\vec x_1)\delta\hat{n}(m_2|\vec x_2)},
\eeq  
where coordinates 
are chosen such that $\vec x$ is the center of the halo with distances $\vec y$ and $\vec y+\vec r$ to the two matter fields respectively, and we insert the definition of $\hat{\delta}_\mathrm{HM}$ (Eq.\,\ref{eq: deltah-normed}) in the second line.\footnote{Note that $\hat{\delta}_\mathrm{HM}$ is normalized by $\av{n(m)}$, not $\int d\vec x\,\hat{n}(m|\vec x)$, since we assume a fixed mass in the box (for the intrinsic covariance).} The expectation term, henceforth denoted $\av{F}$, may be written as a Poissonian expansion into three-, two- and one-halo terms (cf.\,Eq.\,\ref{eq: poisson_expansion});
\beq
    \av{F} &=& \left[^{}_{}\av{\delta n(m|\vec x)\delta n(m_1|\vec x_1)\delta n(m_2|\vec x_2)}\right]\\\nonumber
    &&\,+\, \left[^{}_{}\delta_D(\vec x-\vec x_1)\delta_D(m-m_1)\av{n(m|\vec x)\delta n(m_2|\vec x_2)} + \delta_D(\vec x-\vec x_2)\delta_D(m-m_2)\av{n(m|\vec x)\delta n(m_1|\vec x_1)}\right.\\\nonumber
    &&\,+\, \left.^{}_{}\delta_D(\vec x_1-\vec x_2)\delta_D(m_1-m_2)\av{n(m|\vec x)n(m_1|\vec x_1)} - \delta_D(\vec x_1-\vec x_2)\delta_D(m_1-m_2)\av{n(m|\vec x)}\av{n(m_1|\vec x_1)}\right]\\\nonumber
    &&\,+\, \left[^{}_{}\delta_D(\vec x-\vec x_1)\delta_D(\vec x_1-\vec x_2)\delta_D(m-m_1)\delta_D(m_1-m_2)\av{n(m|\vec x)}\right],
\eeq
where we do not yet average over the random fields $n(m_i|\vec x_i)$. The bracketed terms above physically correspond to the two density fields and halo of mass $m$ in three, two, and one distinct halos, and are shown schematically in Fig.\,\ref{fig: cartoon-cov}. Writing these in terms of the fluctuation field $\eta(m|\vec x)$ (Eq.\,\ref{eq: eta_definition}) and using $\av{\eta}\equiv 0$, we obtain
\beq
    \av{F} &\equiv& \av{F}^{3h}+\av{F}^{2h}+\av{F}^{1h}\\\nonumber
    \av{F}^{3h} &=& n(m)n(m_1)n(m_2)\av{\eta(m|\vec x)\eta(m_1|\vec x_1)\eta(m_2|\vec x_2)}\\\nonumber
    \av{F}^{2h} &=& 2\delta_D(\vec x-\vec x_1)\delta_D(m-m_1)n(m)n(m_2)\av{\eta(m|\vec x)\eta(m_2|\vec x_2)}\\\nonumber
    &&\,+\, \delta_D(\vec x_1-\vec x_2)\delta_D(m_1-m_2)n(m)n(m_1)\av{\eta(m|\vec x)\eta(m_1|\vec x_1)}\\\nonumber
    \av{F}^{1h} &=& \delta_D(\vec x-\vec x_1)\delta_D(\vec x_1-\vec x_2)\delta_D(m-m_1)\delta_D(m_1-m_2)n(m),
\eeq
where we have combined terms symmetric under $\{\vec x_1,m_1\}\leftrightarrow \{\vec x_2,m_2\}$ for brevity. Inserting these into Eq.\,\ref{eq: cov_initial} and simplifying gives
\beq
    \operatorname{cov}\left(N(m),\xi(\vec r)\right) &\equiv& \mathcal{C}^{3h}(m,\vec r)+\mathcal{C}^{2h}(m,\vec r)+\mathcal{C}^{1h}(m,\vec r)\\\nonumber
    \mathcal{C}^{3h}(m_0,\vec r) &=& n_0\int dm_1dm_2d\vec x_1d\vec x_2\,n_1n_2\frac{m_1m_2}{\bar\rho^2}\left[u_1\ast u_2\right](\vec r+\vec x_2-\vec x_1)\int \frac{d\vec x}{V}\av{\eta_0(\vec x)\eta_1(\vec x_1)\eta_2(\vec x_2)}\\\nonumber
    \mathcal{C}^{2h}(m_0,\vec r) &=& 2n_0\frac{m_0}{\bar\rho}\int dm_2\,n_2\frac{m_2}{\bar\rho}\left[u_0\ast u_2\ast \av{\eta_0\eta_2}\right](\vec r)\\\nonumber
    &&\,+\, n_0\int dm_1\,n_1\frac{m_1^2}{\bar\rho^2}\left[u_1\ast u_1\right](\vec r)\int d\vec x_1\av{\eta_0\eta_1}(\vec x_1)\\\nonumber
    \mathcal{C}^{1h}(m_0,\vec r) &=& n_0\frac{m_0^2}{\bar\rho^2}\left[u_0\ast u_0\right](\vec r),
\eeq
using $X(m_i)\equiv X_i$ for brevity and the convolution operator $\ast$, as before. To proceed, we use the expectations of $\eta$ from Eq.\,\ref{eq: eta-statistics} as well as
\beq
    \av{\eta_0(\vec x)\eta_1(\vec x_1)\eta_2(\vec x_2)} &=& \bias{1}_0\bias{1}_1\bias{1}_2\av{\delta_R(\vec x)\delta_R(\vec x_1)\delta_R(\vec x_2)}\\\nonumber
    &&\,+ \frac{1}{2}\left\{\bias{2}_0\bias{1}_1\bias{1}_2\av{\left(\delta_R^2(\vec x)-\sigma_R^2\right)\delta_R(\vec x_1)\delta_R(\vec x_2)}+\text{2 cyc.}\right\}\\\nonumber
    &&\,+\mathcal{O}(\delta_R^5),
\eeq
where `cyc.' indicates cyclic permuations of the three density fields and masses. In terms of the two-, three- and four-point correlators of $\delta_R$ ($\xi_R$, $\zeta_R$ and $\xi_R^{(4)}$) we find
\beq\label{eq: av-eta2-eta3}
    \av{\eta(m_0|\vec x)\eta(m_1|\vec x_1)} &=& \bias{1}_0\bias{1}_1\xi_R(\vec x-\vec x_1) + \frac{1}{2}\left[\bias{1}_0\bias{2}_1+\bias{2}_0\bias{1}_1\right]\zeta^{(3)}_R(\vec x-\vec x_1,\vec 0)\\\nonumber
    &&\,+\frac{1}{3!}\left[\bias{3}_0\bias{1}_1+\bias{1}_0\bias{3}_1\right]\left[2\sigma_R^2\xi_R(\vec x-\vec x_1)+\xi_R^{(4)}(\vec x-\vec x_1,\vec 0,\vec 0)\right]\\\nonumber
    &&\,+\frac{1}{2}\bias{2}_0\bias{2}_2\left[2\xi_R(\vec x-\vec x_1)\xi_R(\vec x-\vec x_1)+\xi_R^{(4)}(\vec x-\vec x_1,\vec x-\vec x_1,\vec 0)\right]\\\nonumber
    &&\,+\mathcal{O}(\delta_R^5)\\\nonumber
    \av{\eta(m_0|\vec x)\eta(m_1|\vec x_1)\eta(m_2|\vec x_2)} &=& \bias{1}_0\bias{1}_1\bias{1}_2\zeta_R(\vec x-\vec x_1,\vec x-\vec x_2)\\\nonumber
    &&\,+ \frac{1}{2}\left\{\bias{2}_0\bias{1}_1\bias{1}_2\left[2\xi_R(\vec x-\vec x_1)\xi_R(\vec x-\vec x_2)+\xi_R^{(4)}(\vec x-\vec x_1,\vec x-\vec x_2,\vec x_1-\vec x_2)\right]+\text{2 cyc.}\right\}\\\nonumber
    &&\,+\mathcal{O}(\delta_R^5),
\eeq
applying Wick's theorem liberally. Further simplification is achieved by noting that, non-perturbatively (and in the infinite volume limit), 
\beq\label{eq: xi-zeta-zeros}
    \int d\vec x\,\xi_R(\vec x-\vec x_1) &=& 0\\\nonumber
    \int d\vec x\,\zeta_R(\vec x-\vec x_1,\vec x-\vec x_2) &=& 0\\\nonumber
    \int d\vec x\,\xi^{(4)}_R(\vec x-\vec x_1,\vec x-\vec x_2,\vec x_1-\vec x_2) &=& 0,
\eeq
for arbitrary $\vec x_1, \vec x_2$. These relations hold by definition of the correlation functions as over-random probabilities,\footnote{In particular, the 2PCF is the over-random probability of finding particles separated by $\vec r$, thus integrating over all $\vec r$ gives zero by normalization. Further, the probability of finding particles in a triangle described by sides $\vec x$ and $\vec y$ is proportional to $1+\xi(\vec x)+\xi(\vec y)+\xi(\vec x-\vec y)+\zeta(\vec x,\vec y)$. Averaging over $\vec x$ gives $1+\xi(\vec y)+\int d\vec x\,\zeta(\vec x,\vec y)/V = 1+\xi(\vec y)$, since this is just a two-point correlator. A similar argument holds for the 4PCF.} and cancel a large swathe of terms in the two- and three-point covariance.\footnote{In reality, these relations hold only in the $V\rightarrow\infty$ limit, and properly give terms involving $\sigma_\mathrm{box}^2$; the variance of the density field filtered on scales comparable to the survey width. These are however subdominant to the main super-sample covariance terms.}

Inserting these relations and simplifying, keeping only terms up to fourth-order in the random field $\delta_R$, we obtain
\beq
    \mathcal{C}^{3h}(m_0,\vec r) &=& n_0\bias{2}_0\int dm_1dm_2\,n_1n_2\bias{1}_1\bias{1}_2\frac{m_1m_2}{\bar\rho^2}\left[u_1\ast u_2\ast \xi_R\ast \xi_R\right](\vec r)\\\nonumber
    \mathcal{C}^{2h}(m_0,\vec r) &=& 2n_0\frac{m_0}{\bar\rho}\int dm_2\,n_2\frac{m_2}{\bar\rho}\left[u_0\ast u_2\ast \mathcal{F}_{02}\right](\vec r)\\\nonumber
    \mathcal{C}^{3h}(m_0,\vec r) &=& n_0\frac{m_0^2}{\bar\rho^2}\left[u_0\ast u_0\right](\vec r),
\eeq
where we have dropped any terms with higher order biases in $m_1$, $m_2$ (since these are expected to be small via the consistency condition) and define
\beq
    \mathcal{F}_{02}(\vec r) = \bias{1}_2\left[\bias{1}_0+\sigma_R^2\bias{3}_0\right]\xi_R(\vec r) + \frac{1}{2}\bias{2}_0\bias{1}_2\zeta_R(\vec r,\vec 0) + \frac{1}{3!}\resub{\bias{3}_0\bias{1}_3}\xi^{(4)}_R(\vec r,\vec 0,\vec 0).
\eeq

\tikzset{every picture/.style={line width=0.75pt}} 

Shifting to Fourier space, the covariance becomes
\beq
    \operatorname{cov}\left(N(m),P(\vec k)\right)^\mathrm{intrinsic} &\equiv& \mathcal{C}^{3h}(m,\vec k)+ \mathcal{C}^{2h}(m,\vec k)+\mathcal{C}^{1h}(m,\vec k)\\\nonumber
    \mathcal{C}^{3h}(m,\vec k) &=&  n(m)\bias{2}(m)\left[\int dm'\,n(m')\frac{m'}{\bar\rho}\bias{1}_1u(\vec k|m')\right]^2P_\mathrm{L}^2(\vec k)W^4(kR)\\\nonumber
    \mathcal{C}^{2h}(m,\vec k) &=& 2n(m)\frac{m}{\bar\rho} u(\vec k|m)\left[\int dm'n(m')\frac{m'}{\bar\rho}\bias{1}(m')u(\vec k|m')
    \right]\\\nonumber
    &&\times \left[\left(\bias{1}(m)P_\mathrm{NL}(\vec k)+\frac{1}{3}\bias{3}(m)\sigma_R^2P_\mathrm{L}(\vec k)\right)W^2(kR) + \frac{1}{2}\bias{2}(m)\mathcal{B}(\vec k) + \frac{1}{3!}\bias{3}(m)\mathcal{T}(\vec k)\right]\\\nonumber
    \mathcal{C}^{1h}(m,\vec k) &=& n(m)\frac{m^2}{\bar\rho^2}u^2(\vec k|m),
\eeq
where we have noted that the power spectrum of the filtered field $\delta_R$ is just that of $\delta$ multiplied by $W^2(kR)$ and defined the collapsed bispectrum and trispectrum
\beq
    \mathcal{B}(\vec k) &\equiv& \int \frac{d\vec p_1d\vec p_2}{(2\pi)^6}\,W(kR)W(p_1R)W(p_2R)B(\vec k,\vec p_1,\vec p_2)(2\pi)^3\delta_D(\vec k+\vec p_1+\vec p_2)\\\nonumber
    \mathcal{T}(\vec k) &\equiv& \int \frac{d\vec p_1d\vec p_2d\vec p_3}{(2\pi)^9}\,W(kR)W(p_1R)W(p_2R)W(p_3R)T(\vec k,\vec p_1,\vec p_2,\vec p_3)(2\pi)^3\delta_D(\vec k+\vec p_1+\vec p_2+\vec p_3).
\eeq

To obtain the full covariance at one-loop order, one should evaluate the power spectrum at one-loop order (using the EFT corrections described in Sec.\,\ref{subsec: quasi-linear-power-model}), but $P^2$, $\sigma^2P$, $\mathcal{B}$ and $\mathcal{T}$ at tree-level. This can be done via the relations
\beq\label{eq: s2BT-def}
    \sigma_R^2 &=& \int \frac{d\vec p}{(2\pi)^3}W^2(kR)P_\mathrm{L}(\vec k)+...\\\nonumber
    B(\vec k_1,\vec k_2,\vec k_3) &=& \left[2F_2(\vec k_1,\vec k_2)P_\mathrm{L}(\vec k_1)P_\mathrm{L}(\vec k_2) + \text{ cyc.}\right] +...\\\nonumber
    T(\vec k_1,\vec k_2,\vec k_3,\vec k_4) &=& \left[4F_2(\vec k_{12},-\vec k_1)F_2(\vec k_{34},\vec k_4)P_\mathrm{L}(\vec k_1)P_\mathrm{L}(\vec k_{34})P_\mathrm{L}(\vec k_3)+6F_3(\vec k_1,\vec k_2,\vec k_3)P_\mathrm{L}(\vec k_1)P_\mathrm{L}(\vec k_2)P_\mathrm{L}(\vec k_3)+\text{ cyc.}\right]+...
\eeq
\citep{1999MNRAS.308..119S}, where $\vec k_{ij}\equiv \vec k_i+\vec k_j$ and we note that the sum of the arguments of $B$ and $T$ is identically zero. These use the coupling kernels $F_2$, $F_3$ tabulated in \citet{2002PhR...367....1B}. Despite its complexity, the collapsed bispectrum can be significantly simplified
\beq\label{eq: col-B-form}
    \mathcal{B}(\vec k) &=& 2W(kR)\int \frac{d\vec p}{(2\pi)^3}F_2(\vec p,\vec k-\vec p)W(pR)W(|\vec k-\vec p|R)P_\mathrm{L}(\vec p)P_\mathrm{L}(\vec k-\vec p)\\\nonumber
    &&\,+4P_\mathrm{L}(\vec k)\int \frac{d\vec p}{(2\pi)^3}F_2(\vec k,-\vec p)P_\mathrm{L}(\vec p)W^2(pR)W(|\vec p-\vec k|R),
\eeq
which requires integration only over $|\vec p|$ and the angle between $\vec k$ and $\vec p$. The latter term simply evaluates to $\left(68/21\right)\sigma^2P_L(\vec k)$ in the limit of no smoothing, which simply renormalizes the halo bias \citep{2014JCAP...08..056A}, in the same manner as the $\bias{3}(m)\sigma_R^2$ term. In the computations below, we will assume that the collapsed trispectrum makes only a small contribution to the covariance and thus set it to zero. We further assume that the measured $\bias{1}$ is itself renormalized, allowing us to absorb the $\bias{3}$ term and $P_\mathrm{L}(\vec k)$-like $\bias{2}$ term into $\bias{1}$ in $\mathcal{C}^{2h}$ \citep{2014JCAP...08..056A,2020MNRAS.492.1614W}.

To aid comparison with data, we should consider the effects of using mass bins of finite size. In a bin labelled by the index $i$, the number counts are given by
\beq
    \hat{N}_i \equiv \int_{m\in i}dm\,\hat{N}(m).
\eeq
The addition of mass-space binning is thus as simple as integrating our covariances over $m$. Recalling the definitions of the $I_p^q$ functions (Eq.\,\ref{eq: integral-notation}) and introducing
\beq\label{eq: integral-notation-J}
    {}_iJ^q_p(\vec k_1,...,\vec k_p) \equiv \int_{m\in i} dm\,n(m)\bias{q}(m)\left(\frac{m}{\bar\rho}\right)^p n(m)\prod_{i=1}^p u(\vec k_i|m),
\eeq
(with ${}_iJ^p_q=I^p_q$ when $i$ covers the full domain of $m$), we thus obtain a complete formula for the intrinsic covariance at one-loop order, ignoring higher order biases constrained by the consistency condition;
\beq\label{eq: cov_no_SSC_final} 
    \operatorname{cov}\left(N_i,P(\vec k)\right)^\mathrm{intrinsic} &=& \mathcal{C}_i^{3h}(\vec k)+\mathcal{C}_i^{2h}(\vec k)+\mathcal{C}_i^{1h}(\vec k)\\\nonumber
    \mathcal{C}_i^{3h}(\vec k) &=& {}_iJ^2_0 I_1^1(\vec k)I_1^1(\vec k)P_\mathrm{L}^2(\vec k)W^4(kR)\\\nonumber
    \mathcal{C}_i^{2h}(\vec k) &=& 2I_1^1(\vec k)\left[{}_iJ_1^1(\vec k)W^2(kR)P_\mathrm{NL}(\vec k)+\frac{1}{3}{}_iJ^3_1(\vec k)\sigma_R^2W^2(kR)P_\mathrm{L}(\vec k) + \frac{1}{2}{}_iJ^2_1(\vec k)\mathcal{B}(\vec k) + \frac{1}{3!}{}_iJ^3_1(\vec k)\mathcal{T}(\vec k)\right]
    \\\nonumber
    \mathcal{C}_i^{1h}(\vec k) &=& {}_iJ^0_2(\vec k,\vec k).
\eeq
Note that we have presented covariances continuous in $\vec k$-space; these may be simply converted into bandpowers for $P(k)$ by integrating over $k$-bins, via 
\beq
    \operatorname{cov}(N_i,P_a) = \frac{1}{V_a}\int_{\vec k\in a}\operatorname{cov}(N_i,P(\vec k)) \approx \operatorname{cov}(N_i,P(k_a)),
\eeq
where $V_a$ is the volume of bin $a$ centered at $k_a$.

It is interesting to compare our results to previous derivations, which have been performed only at $\mathcal{O}(\delta_R^2)$. At this order (in the limit of zero smoothing), we are in agreement with the results presented in \citet{2014PhRvD..90l3523S} and \citet{2014MNRAS.441.2456T}, however, any quadratic analysis necessarily neglects the three-point covariance that scales as $P_\mathrm{L}^2$. As discussed in Sec.\,\ref{subsec: sims-fixed-mass}, we find this to be the dominant large-scale intrinsic term at high-redshift (though usually hidden behind super-sample effects), thus its inclusion is of significant importance if we wish to obtain accurate covariance matrix predictions.

\subsubsection{Autocovariance of $N(m)$}
{In a similar vein to the above, we can compute the auto-covariance of halo counts in two (differential) bins, starting from the $\hat{N}(m)$ definition (Eq.\,\ref{eq: N-def})
\beq
    \operatorname{cov}(N(m_1),N(m_2))^\mathrm{intrinsic} &=& \int d\vec x_1d\vec x_2\,\av{\delta\hat{n}(m_1|\vec x_1)\delta\hat{n}(m_2|\vec x_2)}.
\eeq
As before, we proceed by performing a Poissonian expansion and expressing in terms of the fluctuation field $\eta$;
\beq\label{eq: cov-NN-intrinsic-fluc}
    \operatorname{cov}(N(m_1),N(m_2))^\mathrm{intrinsic} &=& \int d\vec x_1d\vec x_2\,\left\{_{ }^{ }n_1n_2\av{\eta_1\eta_2}(\vec x_1-\vec x_2) + \delta_D(\vec x_1-\vec x_2)\delta_D(m_1-m_2)n_1\left[1+\av{\eta_1(\vec x_1)}\right]\right\}\\\nonumber
    &=& Vn_1n_2\int d\vec r\av{\eta_1\eta_2}(\vec r) + \delta_D(m_1-m_2)\,n_1V 
\eeq
where we have integrated over $\vec x_2$ in the second line, noting that $\av{\eta_1} = 0$ and using the notation $f_i\equiv f(m_i)$. Next, note that $\av{\eta_1\eta_2}$ integrated over all radius should be small (and vanishing in the limit $V\rightarrow\infty$). This is true via the bias expansion of Eq.\,\ref{eq: av-eta2-eta3} and the fact that infinite integrals over the correlation functions are zero (Eq.\,\ref{eq: xi-zeta-zeros}). The only remaining term in Eq.\,\ref{eq: av-eta2-eta3} scales as a product of second-order biases and the integral of $\xi_R^2$, which is expected to be small. This yields the simple form
\beq
    \operatorname{cov}(\hat{N}(m_1),\hat{N}(m_2))^\mathrm{intrinsic} &=& Vn_1\delta_D(m_1-m_2),
\eeq
or in finite mass bins $i$, $j$,
\beq
    \operatorname{cov}(N_i,N_j)^\mathrm{intrinsic} &=& V{}_iJ_0^0\,\delta_K^{ij}
\eeq
where $\delta_K$ is the Kronecker delta.} \resub{This agrees with the infinite-volume part of the \citet{2004PhRvD..70d3504L} and \citet{2011MNRAS.418..729S} halo count covariances, with the latter derived within a more rigorous framework.}

\subsection{Halo Exclusion Covariance}\label{subsec: exclusion-cov}

\subsubsection{The Excluded Halo Density Function}\label{subsubsec: excluded-halo-fun}
An important effect that we have so far neglected is that of halo exclusion; the fact that two halos cannot co-exist at the same spatial location. In our context, halos are selected via the FoF algorithm, and, for a halo of mass $m$ to exist at spatial location $\vec x$, we require there to be no halos of mass $m'>m$ within some exclusion radius $R_\mathrm{ex}(m,m')$ of $\vec x$.\footnote{The restriction to $m'>m$ is important; if a larger halo is present, the given halo of mass $m$ will simply be part of this halo and not an individual object, whilst any halo smaller than $m$ can be thought of as part of the halo of mass $m$. Considering the largest mass halos first also naturally arises from Press-Schechter theory.} To formulate this mathematically, we first consider the discrete form of the halo number density in some infinitesimal mass bin $\delta m$;
\beq
    \hat{n}_\mathrm{ex}(m|\vec x)\delta m &=& \sum_{\vec x_i\in \hat{S}_m}\delta_D(\vec x-\vec x_i) = \sum_{\vec x_i\in \hat{S}_m^\ast}\delta_D(\vec x-\vec x_i) - \sum_{\vec x_i\in\hat{E}_m}\delta_D(\vec x-\vec x_i),
\eeq
where $\hat{S}_m$ is the set of all halos positions of mass in $[m,m+\delta m]$. In the second equality, we have split this into a set $\hat{S}^\ast_m$, representing the halo positions if there were no exclusion and $\hat{E}_m$, representing the subset of those halos which are excluded by virtue of an adjacent halo of mass $m'>m$. The number of \resub{such halos} within the exclusion radius \resub{around} $\vec x$ is given by
\beq
    \int_m^\infty dm'\int d\vec y\,\left[\frac{1}{\delta m'}\sum_{\vec y_i\in \hat{S}_{m'}}\delta_D(\vec y-\vec y_i)\right]\Theta_\mathrm{ex}(\vec y-\vec x|m,m'),
\eeq
where $\Theta_\mathrm{ex}(\vec r|m,m')$ is unity for $|r|<R_\mathrm{ex}(m,m')$ and zero else. Assuming the exclusion radius is small, we expect this to be unity if there is an excluding halo nearby and zero else. The excluded number density is thus
\beq
    \hat{n}_\mathrm{ex}(m|\vec x)\delta m = \sum_{\vec x_i\in\hat{S}_m^*}\delta_D(\vec x-\vec x_i)\left[1-\int_m^\infty dm'\int d\vec y\,\left[\frac{1}{\delta m'}\sum_{\vec y_i\in \hat{S}_{m'}}\delta_D(\vec y-\vec y_i)\right]\Theta_\mathrm{ex}(\vec y-\vec x|m,m')\right].
\eeq
We now promote this to continuous form, assuming the unclustered halos are distributed according to some function $\hat{\tilde{n}}(m,\vec x)$;
\beq
    \hat{n}_\mathrm{ex}(m|\vec x) \approx \hat{\tilde{n}}(m|\vec x)\left[1-\int_m^\infty dm'\int d\vec y\,\hat{n}_\mathrm{ex}(m'|\vec y)\Theta_\mathrm{ex}(\vec y-\vec x|m,m')\right].
\eeq
The second term will be henceforth known as the `exclusion fraction'. Note that this is strictly a recursive definition; $\hat{n}_\mathrm{ex}(m)$ is defined in terms of itself at a larger mass $m'$, which leads to an infinite hierarchy of exclusion integrals. Here, we opt to truncate the sum at the first term, setting $\hat{n}_\mathrm{ex}(m'|\vec y)$ to $\hat{\tilde{n}}(m'|\vec y)$, which corresponds to working to first order in the halo exclusion fraction. We must further ensure that the halo number density is preserved, i.e. that $\av{\hat{n}_\mathrm{ex}} = \av{\hat{n}}$. Noting that we cannot have Poissonian contractions between $\hat{\tilde{n}}(m,\vec x)$ and $\hat{n}(m'|\vec y)$ as the masses are distinct and assuming $\tilde{n}(m|\vec x) = \tilde{n}(m)(1+\eta(m|\vec x))$, we obtain
\beq\label{eq: av-n-ex}
    \av{\hat{n}_\mathrm{ex}(m)} &\approx& \tilde{n}(m)\left[1 - \int_m^\infty dm'\int \frac{d\vec xd\vec y}{V}\tilde{n}(m)\tilde{n}(m')\resub{\av{\left(1+\eta(m|\vec x)\right)\left(1+\eta(m'|\vec y)\right)}}\Theta_\mathrm{ex}(\vec y-\vec x|m,m')\right]\\\nonumber
    &=& \tilde{n}(m)\left[1-\int_m^\infty dm'\tilde{n}(m')\left(V_\mathrm{ex}(m,m')+\bias{1}(m)\bias{1}(m')S_\mathrm{ex}(m,m')\right)\right] + \text{higher order},
\eeq
where we have defined
\beq
    V_\mathrm{ex}(m,m') &\equiv& \int d\vec y\,\Theta_\mathrm{ex}(\vec y|m,m')\\\nonumber
    S_\mathrm{ex}(m,m') &\equiv& \int d\vec y\,\xi_R(\vec y)\Theta_\mathrm{ex}(\vec y|m,m'),
\eeq
and included only linear biases. Setting $\av{\hat{n}_\mathrm{ex}(m)} = \av{\hat{n}(m)}$ and again working to linear order in the exclusion fraction, we can use Eq.\,\ref{eq: av-n-ex} to replace $\tilde{n}$, obtaining the final ansatz for the excluded number density field;
\beq\label{eq: n-ex-full}
    \hat{n}_\mathrm{ex}(m|\vec x) &\equiv& \hat{n}(m|\vec x)\left\{1 - \int_m^\infty dm'\,\left[\int d\vec y\,\hat{n}(m'|\vec y)\Theta_\mathrm{ex}(\vec y-\vec x|m,m') - n(m')V_\mathrm{ex}(m,m') - n(m')\bias{1}(m)\bias{1}(m')S_\mathrm{ex}(m,m')\right]\right\}\\\nonumber
    &\equiv& \hat{n}(m|\vec x)\left\{1 - \int_m^\infty dm'\,\int d\vec y\,\left[\delta\hat{n}(m'|\vec y)\Theta_\mathrm{ex}(\vec y-\vec x|m,m') - n(m')\bias{1}(m)\bias{1}(m')\frac{S_\mathrm{ex}(m,m')}{V}\right]\right\}.
\eeq
Importantly, this form has the same mean density as $\hat{n}(m,\vec x)$, but now contains two copies of the stochastic field $\hat{n}$ which will lead to the appearance of new covariance matrix terms. It is pertinent to note that this formalism should only be applied to the number density field included in $\hat{N}(m)$, not those in $\hat{\delta}_\mathrm{HM}$. We expect the exclusion effects to be \resub{both a physical effect} and a consequence of our halo-finding algorithms; their prescription thus differs for the continuous fields $\delta_\mathrm{HM}$ (where one always considers all possible mass bins).

It remains to choose the form of the exclusion radius $R_\mathrm{ex}(m,m')$. Given that FoF halos have a variety of geometries (often far from spherical), it is not \textit{a priori} obvious how to set this. Here, we note that the exclusion radius should not be larger than the sum of the two halos' Lagrangian radii $R_L(m) \equiv \resub{\left(3m/4\pi\bar\rho\right)^{1/3}}$ (which would correspond to overlapping halos in Lagrangian space), and thus parametrize as
\beq\label{eq: R-ex-def}
    R_\mathrm{ex}(m,m') \equiv \alpha\left[R_L(m)+R_L(m')\right],
\eeq
where $\alpha \leq 1$ is a free parameter. Physically, we expect the exclusion radius to be at least as large as the sum of the halos Eulerian radii; assuming halos have $\rho_h = 200\bar\rho$, this corresponds to $\alpha \gtrsim 0.17$. Considering our lack of knowledge of halo geometries, it is not possible to place a much stronger prior on $\alpha$.

\subsubsection{Contributions to the Covariance of $N(m)$ and $P(\vec k)$}
We now consider the covariance contributions arising from the exclusion model detailed above. Noting that this model is strictly only an approximation, we shall work to first order in bias and halo exclusion fraction, and consider our model a success if it is able to approximately capture the frequency and mass dependence of the covariance. To begin, we write the covariance in real space (ignoring super-sample effects), as in Eq.\,\ref{eq: cov_initial};
\beq\label{eq: cov-ex-expansion}
    \operatorname{cov}(N_\mathrm{ex}(m),\xi(\vec r)) &=& \int\frac{d\vec x d\vec y}{V}\av{\delta\hat{n}_\mathrm{ex}(m|\vec x)\hat{\delta}_\mathrm{HM}(\vec x+\vec y+\vec r)\hat{\delta}_\mathrm{HM}(\vec x+\vec y)}\\\nonumber
    &=& \int\frac{d\vec x d\vec y}{V}\av{\delta\hat{n}(m|\vec x)\hat{\delta}_\mathrm{HM}(\vec x+\vec y+\vec r)\hat{\delta}_\mathrm{HM}(\vec x+\vec y)}\\\nonumber
    &&\,- \int \frac{d\vec xd\vec yd\vec z}{V}\int_{m}^\infty dm' \av{\hat{n}(m|\vec x)\delta\hat{n}(m'|\vec z)\hat{\delta}_\mathrm{HM}(\vec x+\vec y+\vec r)\hat{\delta}_\mathrm{HM}(\vec x+\vec y)}\Theta_\mathrm{ex}(\vec z - \vec x|m,m')\\\nonumber
    &&\,+ \int \frac{d\vec xd\vec y}{V}\int_{m}^\infty dm' \av{\hat{n}(m|\vec x)\hat{\delta}_\mathrm{HM}(\vec x+\vec y+\vec r)\hat{\delta}_\mathrm{HM}(\vec x+\vec y)}n(m')\bias{1}(m)\bias{1}(m')S_\mathrm{ex}(m,m')\\\nonumber
    &\equiv& \operatorname{cov}(N(m),\xi(\vec r))^\mathrm{intrinsic} + \operatorname{cov}(N(m),\xi(\vec r))^\mathrm{exclusion},
\eeq
where we have separated out the intrinsic covariance in the final line. Including the definitions of $\hat{\delta}_\mathrm{HM}$ and simplifying, we obtain
\beq\label{eq: G-ex-def}
    \operatorname{cov}(N(m_0),\xi(\vec r))^\mathrm{exclusion} &=& \int d\vec s\,dm_1dm_2\frac{m_1m_2}{\bar\rho^2}\left[u_1\ast u_2\right](\vec r+\vec s)\av{\mathcal{G}(\vec s|m_0,m_1,m_2)}\\\nonumber
    \mathcal{G}(\vec s|m_0,m_1,m_2) &\equiv& -\int_{m_0}^\infty dm_3\int \frac{d\vec td\vec y}{V}\delta\hat{n}_1(\vec t)\delta\hat{n}_2(\vec s+\vec t)\hat{n}_0(\vec y)\left[\int d\vec z\,\delta\hat{n}_3(\vec z)\thex(\vec z - \vec y|m_0,m_3) - n_3\bias{1}_0\bias{1}_3\sex(m_0,m_3)\right].
\eeq
again adopting the notation $f_i\equiv f(m_i)$. 

As usual, we proceed by considering the Poisson contractions of stochastic density fields (noting that $\hat{n}_0$ and $\hat{n}_3$ cannot contract since $m_3\neq m_0$) and writing in terms of the fluctuation field $\eta$. For the three-halo term (which requires no contractions) this yields
\beq
    -\mathcal{G}^{3h}(\vec s|m_0,m_1,m_2) &=& \int_{m_0}^\infty dm_3\int \frac{d\vec td\vec y}{V}n_0n_1n_2n_3\eta_1(\vec t)\eta_2(\vec s+\vec t)\left(1+\eta_0(\vec y)\right)\\\nonumber
    &&\,\qquad\qquad\times\left[\int d\vec z\,\eta_3(\vec z)\thex(\vec z-\vec y|m_0,m_3) - \bias{1}_0\bias{1}_3\sex(m_0,m_3)\right]\\\nonumber
    -\av{\mathcal{G}^{3h}(\vec s|m_0,m_1,m_2)} &=& 2n_0n_1n_2\bias{1}_0\bias{1}_1\bias{1}_2\int_{m_0}^\infty dm_3\,n_3\bias{1}_3\left[\xi_R\ast\xi_R\ast\thex\right](\vec s|m_0,m_3),
\eeq
where we have kept only linear bias terms and ignored contributions from higher-point statistics. This has made liberal use of the relations of Eqs.\,\ref{eq: av-eta2-eta3}\,\&\,\ref{eq: xi-zeta-zeros}. For the two-halo term, multiple Poisson contractions are possible; between two density fields (fields 1 and 2), between density and halo fields (0 and 1 or 0 and 2) or between density and exclusion fields (1 and 3 or 2 and 3). The first is zero (at linear order in bias) since we force $\hat{n}_\mathrm{ex}$ and $\hat{n}$ to have the same normalization. From the others;
\beq
    -\mathcal{G}^{2h-I}(\vec s|m_0,m_1,m_2) &=& \int_{m_0}^\infty dm_3\int \frac{d\vec y}{V}n_0n_2n_3\eta_2(\vec s+\vec y)\left(1+\eta_0(\vec y)\right)\\\nonumber
    &&\quad\quad\times\left[\int d\vec z\,\eta_3(\vec z)\thex(\vec z-\vec y|m_0,m_3) - \bias{1}_0\bias{1}_3\sex(m_0,m_3)\right]\delta_D(m_1-m_0)+(m_1\leftrightarrow m_2)\\\nonumber
    -\av{\mathcal{G}^{2h-I}(\vec s|m_0,m_1,m_2)} &=& n_0n_2\bias{1}_2\int_{m_0}^\infty dm_3\,n_3\bias{1}_3\left[\left[\xi_R\ast\thex\right](\vec s|m_0,m_3)-\bias{1}_0\bias{1}_0\sex(m_0,m_3)\xi_R(\vec s)\right]+\,(m_1\leftrightarrow m_2),
\eeq
and
\beq
-\mathcal{G}^{2h-II}(\vec s|m_0,m_1,m_2) &=& \int_{m_0}^\infty dm_3\int \frac{d\vec yd\vec z}{V}n_0n_2n_3\eta_2(\vec s+\vec z)\left(1+\eta_0(\vec y)\right)\\\nonumber
    &&\quad\quad\times\,\left(1+\eta_3(\vec z)\right)\thex(\vec y - \vec z|m_0,m_3)\delta_D(m_1-m_3)+(m_1\leftrightarrow m_2)\\\nonumber
    -\av{\mathcal{G}^{2h-II}(\vec s|m_0,m_1,m_2)} &=& n_0n_2\bias{1}_2\int_{m_0}^\infty dm_3\,\resub{n_3}\left[\bias{1}_3\vex(m_0,m_3)\xi(\vec s)+\bias{1}_0\left[\xi_R\ast\thex\right](\vec s|m_0,m_3)\right]\delta_D(m_1-m_3) + (m_1\leftrightarrow m_2).
\eeq
Finally, for the one-halo term we must consider three types of contraction; between two density fields and the halo field (fields 0, 1 and 2), between two density fields and the exclusion field (1, 2 and 3) or between a density field and the halo field and the other density field and the exclusion field (0 and 1, and 2 and 3, or 0 and 2, and 1 and 3). As before, the first vanishes due to the normalization of $\hat{n}_\mathrm{ex}$; the others give
\beq
    -\mathcal{G}^{1h-I}(\vec s|m_0,m_1,m_2) &=& \int_{m_0}^\infty dm_3\int \frac{d\vec yd\vec z}{V}n_0n_3\left(1+\eta_0(\vec y)\right)\left(1+\eta_3(\vec z)\right)\thex(\vec z-\vec y|m_0,m_3)\\\nonumber
    &&\,\times\,\delta_D(m_1-m_3)\delta_D(m_2-m_3)\delta_D(\vec s)\\\nonumber
    -\av{\mathcal{G}^{1h-I}(\vec s|m_0,m_1,m_2)} &=& n_0 \delta_D(\vec s)\int_{m_0}^\infty dm_3\,n_3\left[\vex(m_0,m_3)+\bias{1}_0\bias{1}_3\sex(m_0,m_3)\right]\delta_D(m_1-m_3)\delta_D(m_2-m_3),
\eeq
and 
\beq
    -\mathcal{G}^{1h-II}(\vec s|m_0,m_1,m_2) &=& \int_{m_0}^\infty dm_3\int \frac{d\vec y}{V}n_0n_3\left(1+\eta_0(\vec y)\right)\left(1+\eta_3(\vec y+\vec s)\right)\thex(\vec s|m_0,m_3)\delta_D(m_0-m_1)\delta_D(m_2-m_3)\\\nonumber
    &&\,+\,(m_1\leftrightarrow m_2)\\\nonumber
    -\av{\mathcal{G}^{1h-II}(\vec s|m_0,m_1,m_2)} &=& n_0\delta_D(m_0-m_1)\int_{m_0}^\infty dm_3\,n_3\left[\thex(\vec s|m_0,m_3)+\bias{1}_0\bias{1}_3\xi_R(\vec s)\thex(\vec s|m_0,m_3)\right]\delta_D(m_2-m_3) + (m_1\leftrightarrow m_2).
\eeq

From the above expansions, we can compute the covariance from Eq.\,\ref{eq: G-ex-def}. Transforming into Fourier space and omitting tedious algebra, we arrive at the final result for the exclusion covariance to first-order in biases and the exclusion fraction;
\beq
    \operatorname{cov}(N(m),P(\vec k))^\mathrm{exclusion} &\equiv& \mathcal{C}^{3h,\mathrm{ex}}(m,\vec k) + \mathcal{C}^{2h,\mathrm{ex}}(m,\vec k) + \mathcal{C}^{1h,\mathrm{ex}}(m,\vec k)\\\nonumber
    \mathcal{C}^{3h,\mathrm{ex}}(m,\vec k) &=& -2n(m)\bias{1}(m)I_1^1(\vec k)I_1^1(\vec k)\left[\int_m^\infty dm'\,n(m')\bias{1}(m')\tilde\Theta_\mathrm{ex}(\vec k|m,m')\right]W^4(kR)P_\mathrm{L}^2(\vec k)\\\nonumber
    \mathcal{C}^{2h,\mathrm{ex}}(m,\vec k) &=& - 2n(m)\frac{m}{\bar\rho}u(\vec k|m)I_1^1(\vec k)\left[\int_m^\infty dm'\,n(m')\bias{1}(m')\tilde\Theta_\mathrm{ex}(\vec k|m,m')\right]W^2(kR)P_\mathrm{NL}(\vec k)\\\nonumber
    &&+2n(m)\bias{1}(m)\bias{1}(m)\frac{m}{\bar\rho}I_1^1\left[\int_m^\infty dm'\,n(m')\bias{1}(m')\sex(m,m')\right]W^2(kR)P_\mathrm{L}(\vec k)\\\nonumber
    &&-2n(m)\bias{1}(m)I_1^1(\vec k)\left[\int_m^\infty dm'\,n(m')\frac{m'}{\bar\rho}u(\vec k|m')\tilde\Theta_\mathrm{ex}(\vec k|m,m')\right]W^2(kR)P_\mathrm{NL}(\vec k)\\\nonumber
    &&-2n(m)I_1^1(\vec k)\left[\int_m^\infty dm'\,n(m')\bias{1}(m')\frac{m'}{\bar\rho}u(\vec k|m')\vex(m,m')\right]W^2(kR)P_\mathrm{NL}(\vec k)\\\nonumber
    \mathcal{C}^{1h,\mathrm{ex}}(m,\vec k) &=& -n(m)\int_m^\infty dm'\,n(m')\frac{m'^2}{\bar\rho^2}u^2(\vec k|m')\vex(m,m')\\\nonumber
    &&- n(m)\bias{1}(m)\int_m^\infty dm'\,n(m')\bias{1}(m')\frac{m'^2}{\bar\rho^2}u^2(\vec k|m')\sex(m,m')\\\nonumber
    &&-2n(m)\frac{m}{\bar\rho}u(\vec k|m)\int_m^\infty dm'\,n(m')\frac{m'}{\bar\rho}u(\vec k|m')\tilde\Theta_\mathrm{ex}(\vec k|m,m')\\\nonumber
    &&-2n(m)\bias{1}(m)\frac{m}{\bar\rho}u(\vec k|m)\int_m^\infty dm'\,n(m')\bias{1}(m')\frac{m'}{\bar\rho}u(\vec k|m')\left[W^2P_\mathrm{NL}\ast\tilde\Theta\right](\vec k|m,m'),
\eeq
using the notation of Eq.\,\ref{eq: integral-notation} where $\tilde\Theta_\mathrm{ex}$ is the Fourier transform of $\Theta_\mathrm{ex}$, i.e. $\tilde\Theta_\mathrm{ex}(\vec k|m,m') = 4\pi R_\mathrm{ex}^2(m,m')j_1\left(|\vec k|R_\mathrm{ex}(m,m')\right)/|\vec k|$. Note that we use a non-linear power spectrum in general (evaluated using EFT; Sec.\,\ref{subsec: quasi-linear-power-model}), though a linear power spectrum for terms involving $P_\mathrm{L}\sex$, to keep the order consistent.

To properly compare this to data, we must integrate over mass bins, as before. This strictly requires a non-separable double integral in $m$ and $m'$; given that our exclusion model is at best approximate, we will assume (without incurring any significant error) the lower limit of the $m'$ integral to be equal to the average mass in the bin, giving the final form;
\beq\label{eq: cov-ex-final}
    \operatorname{cov}(N_i,P(\vec k))^\mathrm{exclusion} &\approx&     \mathcal{C}_i^{3h,\mathrm{ex}}(\vec k) + \mathcal{C}_i^{2h,\mathrm{ex}}(\vec k) + \mathcal{C}_i^{1h,\mathrm{ex}}(\vec k)\\\nonumber
    \mathcal{C}^{3h,\mathrm{ex}}_i(\vec k) &=& -2{}_iJ^1_0I_1^1(\vec k)I_1^1(\vec k){}_iK^1_0\left[\tilde\Theta_\mathrm{ex}\right](\vec k)W^4(kR)P_\mathrm{L}^2(\vec k)\\\nonumber
    \mathcal{C}_i^{2h,\mathrm{ex}}(\vec k) &=&
    - 2{}_iJ_1^0(\vec k)I_1^1(\vec k){}_iK_0^1\left[\tilde\Theta_\mathrm{ex}\right](\vec k)W^2(kR)P_\mathrm{NL}(\vec k) + 2{}_iJ_1^{1,1}(\vec k)I_1^1(\vec k){}_iK_0^1\left[\sex\right](\vec k)W^2(kR)P_\mathrm{L}(\vec k)\\\nonumber
    &&-2{}_iJ_0^1I_1^1(\vec k){}_iK_1^0\left[\tilde\Theta_\mathrm{ex}\right](\vec k)W^2(kR)P_\mathrm{NL}(\vec k) - 2{}_iJ_0^0I_1^1(\vec k){}_iK_1^1\left[\vex\right](\vec k)W^2(kR)P_\mathrm{NL}(\vec k)\\\nonumber
    \mathcal{C}_i^{1h,\mathrm{ex}}(\vec k) &=& -{}_iJ_0^0{}_iK_2^0\left[\vex\right](\vec k)-{}_iJ_0^1{}_iK_2^1\left[\sex\right](\vec k)\\\nonumber
    &&-2{}_iJ_1^0(\vec k){}_iK_1^0\left[\tilde\Theta_\mathrm{ex}\right](\vec k) - 2{}_iJ_1^1(\vec k){}_iK_1^1\left[W^2P_\mathrm{NL}\ast\tilde\Theta_\mathrm{ex}\right](\vec k),
\eeq
using the notation introduced in Eq.\,\ref{eq: integral-notation-J} and the shorthand;
\beq\label{eq: integral-notation-K}
    {}_iK_p^q\left[f\right](\vec k) \equiv \int_{\av{m}_i}^\infty dm \left(\frac{m}{\bar\rho}\right)^pn(m)\bias{q}(m)f(\vec k|\av{m}_i,m)\prod_{i=1}^pu(\vec k|m),
\eeq
where $\av{m}_i$ is the average mass in bin $i$.\footnote{Note also that ${}_iJ_1^{1,1}$ is analogous to ${}_iJ_1^1$, except with two factors of $\bias{1}(m')$ in the integrand.} 

\subsubsection{Contributions to the $N(m)$ Autocovariance}
{We may similarly estimate the exclusion covariance of $N(m_1)$ and $N(m_2)$, as for the intrinsic covariance in Sec.\,\ref{subsec: no-ssc-deriv}. As in the above, we will work to first order in bias and exclusion fraction, starting from the form
\beq
    \mathrm{cov}(N_\mathrm{ex}(m_1),N_\mathrm{ex}(m_2)) &=& \int d\vec x_1d\vec x_2\,\av{\delta\hat{n}_\mathrm{ex}(m_1|\vec x_1)\delta\hat{n}_\mathrm{ex}(m_2|\vec x_2)}\\\nonumber
    &=& \int d\vec x_1d\vec x_2 \,\av{\hat{n}(m_1|\vec x_1)\hat{n}(m_2|\vec x_2)} \\\nonumber
    &&\,-\,\left\{\int d\vec x_1d\vec x_2d\vec x_3\,\int_{m_1}^\infty dm_3\,\av{\hat{n}(m_1|\vec x_1)\delta\hat{n}(m_2|\vec x_2)\delta\hat{n}(m_3|\vec x_3)}\thex(\vec x_3-\vec x_1|m_1,m_3)\right.\\\nonumber
    &&\,+\,\left.\int d\vec x_1d\vec x_2\,\int_{m_1}^\infty dm_3\,\av{\hat{n}(m_1|\vec x_1)\delta\hat{n}(m_2|\vec x_2)} n(m_3)\bias{1}(m_1)\bias{1}(m_3)\sex(m_1,m_3)\right\}\\\nonumber
    &&\,+\,\left(1\leftrightarrow 2\right)\\\nonumber
    &=& \mathrm{cov}(N(m_1),N(m_2))^\mathrm{intrinsic} + \mathrm{cov}(N(m_1),N(m_2))^\mathrm{exclusion},
\eeq
inserting the definition of the exclusion number density (Eq.\,\ref{eq: n-ex-full}), as in Eq.\,\ref{eq: cov-ex-expansion}. Applying the Poissonian expansion (and noting that $m_3>m_1$ thus $\hat{n}(m_1)$ and $\hat{n}(m_3)$ cannot contract), we obtain two- and one-halo terms, given by
\beq
    \mathrm{cov}(N(m_1),N(m_2))^\mathrm{exclusion} &=& \av{\mathcal{H}^{2h}(m_1,m_2)} + \av{\mathcal{H}^{1h}(m_1,m_2)}\\\nonumber
    -\mathcal{H}^{2h}(m_1,m_2) &=& \int d\vec x_1d\vec x_2\int_{m_1}^\infty dm_3\, n_1n_2n_3\left[1+\eta_1(\vec x_1)\right]\eta_2(\vec x_2)\left\{\int d\vec x_3\,\eta_3(\vec x_3)\thex(\vec x_3-\vec x_1|m_1,m_3) - n_3\bias{1}_1\bias{1}_3\sex(m_1,m_3)\right\}\\\nonumber
    &&\,+\,\left(1\leftrightarrow 2\right)\\\nonumber
     -\mathcal{H}^{1h}(m_1,m_2) &=& \int d\vec x_1d\vec x_2\,n_1n_2\left[1+\eta_1(\vec x_1)\right]\left[1+\eta_2(\vec x_2)\right]\thex(\vec x_2-\vec x_1|m_1,m_2)\Theta_\mathrm{H}(m_2-m_1) + \left(1\leftrightarrow 2\right),
\eeq
where we only consider the $\vec x_2 = \vec x_3$ two-halo term, since the $\vec x_1 = \vec x_2$ term vanishes by the normalization of $\hat{n}_\mathrm{ex}$. Note that we have integrated over $m_3$ in the one-halo term, with the Heaviside function, $\Theta_\mathrm{H}$, enforcing $m_2>m_1$. Taking the expectation and setting higher-order biases to zero, we obtain
\beq
    -\av{\mathcal{H}^{2h}(m_1,m_2)} &=& 0\\\nonumber
    -\av{\mathcal{H}^{1h}(m_1,m_2)} &=& Vn_1n_2\Theta_\mathrm{H}(m_2-m_1)\left[\vex(m_1,m_2) + \bias{1}_1\bias{1}_2\sex(m_1,m_2)\right] + \left(1\leftrightarrow 2\right),
\eeq
where the two-halo term vanishes since each term contains an unrestricted spatial integral over a correlation function (Eq.\,\ref{eq: xi-zeta-zeros}). The exclusion covariance in infinitesimal bins is thus
\beq
    \mathrm{cov}(N(m_1),N(m_2))^\mathrm{exclusion} &=& Vn_1n_2\left[\vex(m_1,m_2) + \bias{1}_1\bias{1}_1\sex(m_1,m_2)\right],
\eeq
and in finite bins,
\beq
    \mathrm{cov}(N_i,N_j)^\mathrm{exclusion} &=& \int_{m_1\in i}\int_{m_2 \in j}dm_1dm_2\,n(m_1)n(m_2)\left[\vex(m_1,m_2)+\bias{1}(m_1)\bias{1}(m_2)\sex(m_1,m_2)\right].
\eeq
}

\subsection{Super-Sample Covariance}\label{subsec: ssc-cov}
The final component of our covariance model is the super-sample covariance (hereafter SSC), arising from density perturbations on scales comparable to the survey (or simulation box) width. Practically, these modify the background density of the region, creating additional covariance from the fluctuations in $N(m)$ and $P(\vec k)$ sourced by a background overdensity $\delta_b$. Following the treatment of \citet{2014PhRvD..90l3523S}, we note that the effect of such a perturbation on a general observable $\hat{f}$ is given by a Taylor expansion of its expectation $f=\av{\hat{f}}$;
\beq
    f(\delta_b) \approx f(0) + \left.\frac{\partial f}{\partial\delta_b}\right|_{\delta_b = 0}\delta_b + \frac{1}{2}\left.\frac{\partial^2 f}{\partial\delta_b^2}\right|_{\delta_b = 0}\delta_b^2 + \mathcal{O}\left(\delta_b^3\right).
\eeq
Averaging over $\delta_b$ we obtain
\beq
    \av{f(\delta_b)}_{\delta_b} = f(0) + \frac{1}{2}\left.\frac{\partial^2 f}{\partial\delta_b^2}\right|_{\delta_b = 0}\sigma^2(V)+\mathcal{O}\left(\sigma^4(V)\right),
\eeq
where $\sigma^2(V)$ is the variance of the density field smoothed on the scale of the survey. This can be written
\beq\label{eq: sigma2V-def}
    \sigma^2(V) = \int \frac{d\vec k}{(2\pi)^3}\left|W_\mathrm{survey}(\vec k)\right|^2P_\mathrm{L}(\vec k),
\eeq
where $W_\mathrm{survey}$ is the survey window function and we have assumed that the modes are large enough to be in the linear regime. Whilst the correction to the observable $f$ is expected to be small, it can be shown that this leads to a non-trivial \textit{covariance} between any two observables which have dependence on $\delta_b$. For a pair of observables $\hat{f}$ and $\hat{g}$,
\beq\label{eq: cov-ssc-fg}
    \operatorname{cov}\left(\hat{f},\hat{g}\right) = \av{\operatorname{cov}\left(\hat{f},\hat{g})\right)}_{\delta_b} + \left.\frac{\partial f}{\partial\delta_b}\right|_{\delta_b=0}\left.\frac{\partial g}{\partial\delta_b}\right|_{\delta_b=0}\sigma^2(V) + \mathcal{O}\left(\sigma^4(V)\right),
\eeq
\citep[Appendix A.2]{2014PhRvD..90l3523S}
where the first term is the usual covariance averaged over $\delta_b$, which is equal to the $\delta_b = 0$ covariance plus some negligible correction term. In the case of the $N(m)$ and $P(\vec k)$ covariance, the additional term is hence
\beq
    \operatorname{cov}\left(N(m),P(\vec k)\right)^\mathrm{SSC} &=& \left.\frac{\partial N(m)}{\partial\delta_b}\right|_{\delta_b=0}\left.\frac{\partial P(\vec k)}{\partial\delta_b}\right|_{\delta_b=0}\sigma^2(V).
\eeq
We will drop the `$\delta_b = 0$' subscript henceforth.

To compute the $N(m)$ derivative, we recall the definition (Eq.\,\ref{eq: N-def}), taking the expectation over $\hat{n}(m|\vec x)$;
\beq
    \frac{\partial N(m)}{\partial\delta_b} \equiv \frac{\partial}{\partial\delta_b}\int d\vec x\,n(m) = Vn(m)\bias{1}(m),
\eeq
where we note that the derivative of $n(m)$ with respect to a long mode $\delta_b$ is equal to $\bias{1}(m)n(m)$ by definition. In some mass bin $i$, we have
\beq\label{eq: dN_ddelta-model}
    \frac{\partial N_i}{\partial\delta_b} \equiv V\frac{\partial}{\partial\delta_b}{}_iJ_0^0 =  V{}_iJ_0^1,
\eeq
using the notation of Eq.\,\ref{eq: integral-notation-J}. This agrees with standard results \citep[e.g.][]{2014PhRvD..90l3523S,2016JCAP...08..005L,2018A&A...611A..83L}. Note that we do not include halo exclusion effects here since these do not contribute to the expectation of $\hat{N}(m)$.

For the $P(\vec k)$ derivative, more care is needed. As shown in \citet{2014PhRvD..89h3519L}, variation in the power spectrum due to a long wavelength mode $\delta_b$ is sourced by three effects; the response of small scale modes to a large scale overdensity (`beat-coupling'), the increase in halo number density due to an increased background density (`halo sample variance'), and the coordinate rescaling induced by a $\delta_b$ modifying the local expansion factor $a$ (`linear dilation'). A variety of manners exist in which to model this, including taking the squeezed limit of the matter bispectrum \citep[e.g.][]{2013PhRvD..88h3502P,2014JCAP...05..048C}, considering the impact of long wavelength modes on the matter trispectrum  \citep[e.g.][]{2013PhRvD..87l3504T}, or using separate universe approaches \citep[e.g.][]{2014PhRvD..89h3519L}. 

Here, we adopt the last approach, roughly following the treatment of \citet{2014JCAP...05..048C}. In this approximation, a region with overdensity $\delta_b$ is treated as an isolated region with modified cosmology chosen to give background density $\bar\rho(a)\left[1+\delta_b(a)\right]$ at scale factor $a$. Firstly, we consider linear dilation, which can be shown to modify the power spectrum at fixed time via
\beq
    P(k,t) \rightarrow P(k,t)\left[1-\frac{1}{3}\frac{d\log k^3P(k,t)}{d\log k}\delta_b(t)\right]
\eeq
\citep[Appendix A]{2013PhRvD..88h3502P}. The other effects may be included by considering the change in the power spectrum induced by $\delta_b(a)$ at fixed $k$ (as in \citealt{2013PhRvD..87l3504T}) Writing in terms of the scale factor $a$, we obtain
\beq\label{eq: Pka-transform}
    P(k,a) \rightarrow \left[\left(1+2q\delta_b(a)\right)P(k,a)+\left.\frac{d P(k,a)}{d\delta_b(a)}\right|_{k}\delta_b(a)\right]\times \left[1-\frac{1}{3}\left.\frac{d\log k^3P(k,a)}{d\log k}\right|_{\delta_b = 0}\delta_b(a)\right],
\eeq
at linear order in $\delta_b(a)$. Note that we have introduced the parameter $q$; this accounts for the fact that the density fields can be normalized by the background density of the separate universe ($q = 0$) or the global mean density ($q = 1$). The latter is relevant for weak lensing analyses (where the background density is set by cosmological parameters rather than being measured) and will be assumed henceforth. Writing Eq.\,\ref{eq: Pka-transform} in terms of a derivative (often known as the power spectrum response) we obtain
\beq
    \frac{dP(k,a)}{d\delta_b(a)} &=& \left.\frac{d P(k,a)}{d\delta_b(a)}\right|_k + \left[2 - \frac{1}{3}\left.\frac{d\log k^3P(k,a)}{d\log k}\right|_{\delta_b = 0}\right]P(k,a).
\eeq

We now proceed to evaluate the above derivative in the context of our power spectrum model (Eq.\,\ref{eq: pk-summary}). For the dilation term, we make the assumption that the coordinate rescaling does not impact the halo profiles; this leads to a dilation term
\beq
    \frac{d\log k^3P_\mathrm{HM}(k)}{d\log k} &=& \frac{d\log k^3W^2(kR)P_\mathrm{NL}(k)}{d\log k}.
\eeq
In practice, this is found to be valid. For the other terms, working at fixed $k$ and making the $a$ dependence implicit, we may write
\beq
    \frac{dP_{2h}(k)}{d\delta_b} &=& \left[I_1^1(k)\right]^2\frac{d}{d\delta_b}\left[W^2(kR)P_\mathrm{NL}(k)\right] + 2I_1^1(k)W^2(kR)P_\mathrm{NL}(k)\frac{dI_1^1(k)}{d\delta_b}\\\nonumber
    \frac{dP_{1h}(k)}{d\delta_b} &=& \frac{dI_2^0(k,k)}{d\delta_b}.
\eeq
It is not immediately clear how the smoothing scale $R$ (nor the speed-of-sound counterterm $c_s^2$) should depend on the background overdensity $\delta$. For this reason, we set the relevant derivatives to zero. For the mass integrals, we note that
\beq
    \frac{d}{d\delta_b}\left[n(m)\bias{1}(m)\right] &=& \frac{dn(m)}{d\delta_b}\bias{1}(m) + n(m)\frac{d\bias{1}(m)}{d\delta_b} = n(m)\left(\bias{1}(m)\right)^2 + n(m)\left[\bias{2}(m) - \left(\bias{1}(m)\right)^2\right] = n(m)\bias{2}(m),
\eeq
\resub{(evaluating the $\bias{1}(m)$ derivative using the definition of the bias in Eq.\,\ref{eq: bias-definitions})}, and hence
\beq
    \frac{dI_1^1(k)}{d\delta_b} &=& I_1^2(k), \qquad \frac{dI_2^0(k,k)}{d\delta_b} = I_2^1(k,k).
\eeq
Note that the former term is constrained by the consistency condition (Eq.\,\ref{eq: bias_consistency_relation}) and expected to be small (and should strictly be set to zero given that we have previously ignored $\bias{2}$ terms in the power spectrum derivation). For the non-linear power spectrum, $P_\mathrm{NL}(k)$, we use the approach of \citet{2014JCAP...05..048C}, considering the perturbation to be sourced from a change to the growth factor;
\beq
    \frac{dP_\mathrm{NL}(k)}{d\delta_b} &=& \frac{\partial P_\mathrm{NL}(k)}{\partial D(a)}\frac{d D(a)}{d\delta_b},
\eeq
where $d\log D(a)/d\delta_b = 13/21$ \citep{2011JCAP...10..031B}. Noting that the linear and one-loop terms scale as $D^2(a)$ and $D^4(a)$ respectively, this gives
\beq
    \frac{dP_\mathrm{NL}(k)}{d\delta_b} &=& \frac{26}{21}P_\mathrm{L}(k) + \frac{52}{21}\left(P_\mathrm{SPT}(k)+P_\mathrm{ct}(k)\right)
\eeq
(cf.\,Eq.\,\ref{eq: Pnl-def}). Combining terms, we obtain the final model for the power spectrum derivative
\beq\label{eq: dP_ddelta-model}
    \frac{dP_\mathrm{HM}(k)}{d\delta_b} &=& 
    2I_1^2(k)I_1^1(k)W^2(kR)P_\mathrm{NL}(k) + I_2^1(k,k)\\\nonumber
    &&\,+\left[I_1^1(k)\right]^2W^2(kR)P_\mathrm{NL}(k)\left(\frac{68}{21} + \frac{26}{21}\frac{P_\mathrm{SPT}(k)+P_\mathrm{ct}(k)}{P_\mathrm{NL}(k)}\right)\\\nonumber
    &&\,- \frac{1}{3}\frac{d\log k^3 P_\mathrm{NL}(k)}{d\log k}P_\mathrm{HM}(k),
\eeq
where the three lines correspond to halo sample variance, beat coupling and linear dilation terms respectively. This is compared to simulations in Sec.\,\ref{subsec: sims-DC-mode}.

Our final model for the super-sample covariance is thus
\beq\label{eq: full-SSC-model}
    \operatorname{cov}\left(N_i,P(\vec k)\right)^\mathrm{SSC} &=& V\sigma^2(V){}_iJ_0^1\\\nonumber
    &&\,\times \left\{I_1^1(k)W^2(kR)P_\mathrm{NL}(k)\left[2I_1^2(k)+I_1^1(k)\left(\frac{68}{21} + \frac{26}{21}\frac{P_\mathrm{SPT}(k)+P_\mathrm{ct}(k)}{P_\mathrm{NL}(k)}\right)\right]+I_2^1(k,k) - \frac{1}{3}\frac{d\log k^3 P_\mathrm{NL}(k)}{d\log k}P_\mathrm{HM}(k)\right\}.
\eeq
Notably, this scales as $V\sigma^2(V)$, unlike the volume-independent intrinsic and exclusion terms. 

{For the number count auto-covariance, the result is straightforward;
\beq
    \mathrm{cov}(N_i,N_j)^\mathrm{SSC} &=& V^2\sigma^2(V){}_iJ_0^1{}_jJ_0^1,
\eeq
which is equal to the linear-part of the three-halo term in Eq.\,\ref{eq: cov-NN-intrinsic-fluc}, had we considered the 2PCF integral to be performed across the survey volume, rather than infinite space.} \resub{In combination with the intrinsic covariance, this matches the result of \citet{2004PhRvD..70d3504L} and \citet{2011MNRAS.418..729S}.}

\section{Halo Count Covariances: Comparison to Simulations}\label{sec: sims-cov}

In Sec.\,\ref{sec: sims}, we have shown our model for $P(k)$ to be robust {and} 
capable of producing predictions of percent-level accuracy up to $k\sim 1\hMpc$. It remains to test the other main prediction of this paper; the {cluster count covariances} (Sec.\,\ref{sec: cov_N_Pk_derivation}). In principle, our method to do this is straightforward; take a large set of $N$-body simulations, measure the matter power spectrum and halo number counts in each, then compute the associated covariances. {These are defined by the standard estimators} for mass bins $i$ and $j$
\beq
    \operatorname{cov}(N_i,P(k)) &=& \frac{1}{N_\mathrm{sim}-1}\sum_{n=0}^{N_\mathrm{sim}}\left[\hat{N}_i^{(n)}\hat{P}^{(n)}(k)-\overline{N}_i\overline{P}(k)\right]\\\nonumber
    \operatorname{cov}(N_i,N_j) &=& \frac{1}{N_\mathrm{sim}-1}\sum_{n=0}^{N_\mathrm{sim}}\left[\hat{N}_i^{(n)}\hat{N}_j^{(n)}-\overline{N}_i\overline{N}_j\right],
\eeq
where the superscript $(n)$ indicates the value obtained for the $n$-th simulation, and an overbar indicates averaging over the $N_\mathrm{sim}$ realizations.

In practice, robustly testing the covariance matrix model is more difficult, since the three contributions are heavily entangled with different dependencies on the survey volume and redshift. Here, we adopt a hybrid method, first testing the intrinsic and exclusion covariances using $N$-body simulations of fixed total mass (which do not have super-sample effects, since the mean box density is fixed to the cosmological average $\bar\rho$), then using separate universe simulations to constrain the necessary super-sample derivatives. Finally, we turn to subbox simulations (with varying total mass) to constrain the full model. In addition, we can separate intrinsic and exclusion effects by their redshift dependence; at high-$z$, the fraction of mass in large halos is small, thus exclusion effects are subdominant.

\subsection{Simulations with Fixed Total Mass: Intrinsic and Exclusion Covariances}\label{subsec: sims-fixed-mass}

To assess the validity of our intrinsic and exclusion covariance model, we make use of the 15,000 standard-resolution \texttt{Quijote} simulations of fixed cosmology (Sec.\,\ref{subsec: quijote-and-abacus-sims}). {Since these have a fixed total mass of dark matter, the simulation mean density is equal to the cosmological value $\bar\rho$, thus they do not include super-sample effects.} For each simulation, we count the number of halos in bins of width $\Delta\log_{10}\left(M/h^{-1}M_\odot\right) = 0.2\,\mathrm{dex}$ with $M_\mathrm{min} = 10^{13.1}h^{-1}M_\odot$, $M_\mathrm{max} = 10^{14.5}h^{-1}M_\odot$. These are chosen such that all halos are relatively abundant and have masses considerably above the mass resolution ($M_\mathrm{min} \approx 20M_\mathrm{res}$). Since the exclusion covariance is expected to be a strong function of $z$, we compute the sample covariance at four redshifts; $z \in \{0,0.5,1,2\}$. The theory models are computed in a similar manner as for the power spectrum (Sec.\,\ref{subsec: practical-evaluation}), requiring evaluation of various $I_p^q$, ${}_iJ_p^q$ and ${}_iK_p^q$ integrals (Eqs.\,\ref{eq: integral-notation},\,\ref{eq: integral-notation-J}\,\&\,\ref{eq: integral-notation-K}), which can simply be approximated by finely binned numerical quadrature. Only the convolution terms are non-trivial (i.e. $\left[W^2P_\mathrm{NL}\ast\tilde{\Theta}_\mathrm{ex}\right](\vec k)$ in Eq.\,\ref{eq: cov-ex-final}); these are performed via a simple application of the FFTLog algorithm \citep{2018JCAP...04..030S}.

\subsubsection{Covariance of $N(m)$ and $P(\vec k)$}
\begin{figure}
    \centering
    \includegraphics[width=0.9\textwidth]{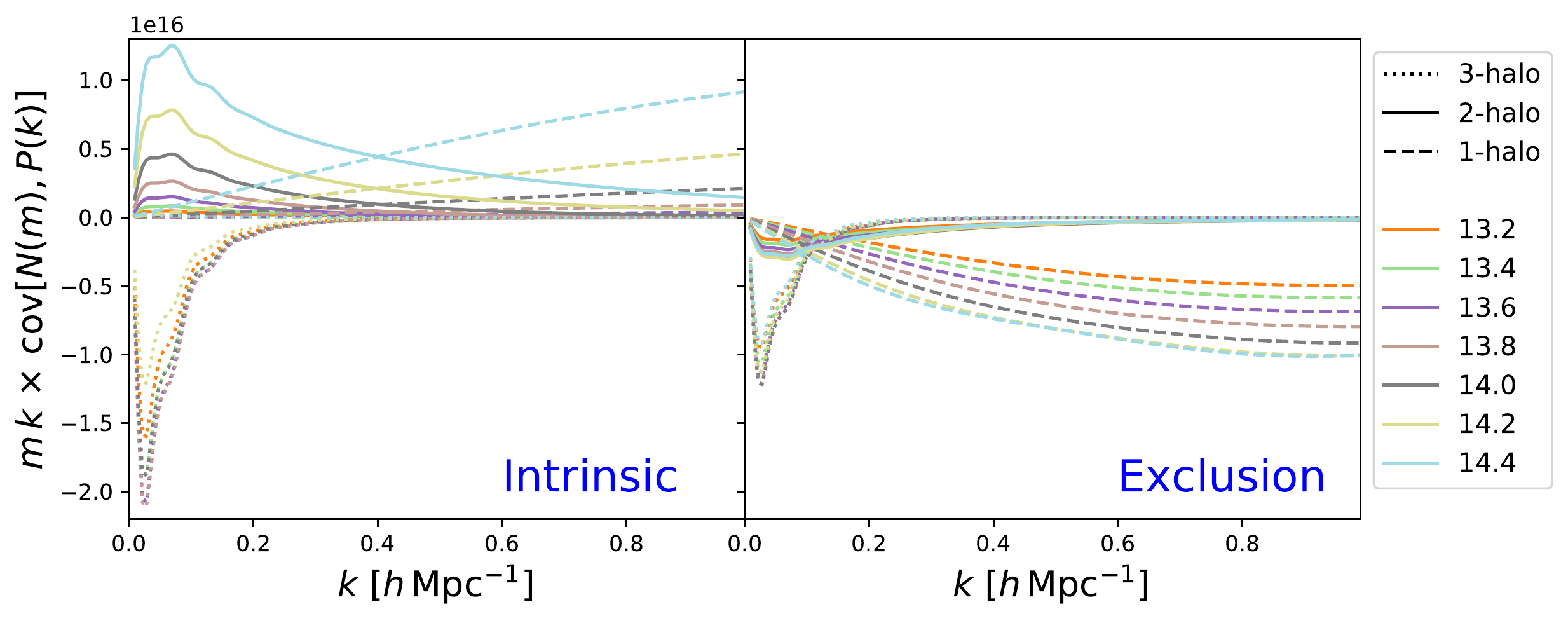}
    \caption{Contributions to the covariance between halo number counts $N(m)$ and the matter power spectrum $P(k)$ at $z = 0$ from the intrinsic (left) and halo exclusion (right) models discussed in Secs.\,\ref{subsec: no-ssc-deriv}\,\&\,\ref{subsec: exclusion-cov} using the {effective halo model}. Covariances are plotted as a function of wavenumber $k$ for seven mass bins (plotted in different colors), with the central value of $\log_{10}\left(M/h^{-1}M_\odot\right)$ in each bin indicated by the caption. Line textures show the contributions from different types of terms, indicating the positions of matter particles and halos in the theory model. The covariances are multiplied by $m\times k$ for visibility, and assume the parameter set $(c_s^2 = 9.25h^{-2}\mathrm{Mpc}^2, R = 1.86h^{-1}\mathrm{Mpc}, \alpha = 0.5)$. Note that we do not include super-sample terms here, {which will dominate for small survey volumes.}}
    \label{fig: cov-breakdown}
\end{figure}

{We begin by considering the cross-covariance between the halo counts and the power spectrum.} Before comparing our models to data, it is instructive to plot the various terms as a function of mass and scale, as shown in Fig.\,\ref{fig: cov-breakdown}. Working at redshift zero (where all contributions are non-negligible), we note large contributions from both intrinsic and exclusion covariances. In the former case, the one-halo term is seen to dominate at high $k$ as expected, with a substantial positive contribution from the two-halo term on larger scales. Importantly, we observe a large negative contribution from the three-halo term, scaling as $n(m)\bias{2}(m)P_\mathrm{L}^2(k)$ for halo mass $m$. Whilst this contributes only for $k\lesssim 0.2\hMpc$ it is nonetheless non-negligible, and of particular interest since it has not been included in previous analyses, which worked only to first order in bias and $P_\mathrm{L}(\vec k)$. The negative sign is attributed to the negative sign of the second order bias at redshift zero. In addition, we note that the various terms have substantially different mass dependencies. This is to be expected since they are sourced by different integrals of combinations of $n(m)$, $m$, $\bias{1}(m)$ and $\bias{2}(m)$.

Turning to the exclusion covariance, we similarly note that we become one-halo dominated on relatively large scales, here at $k\sim 0.2\hMpc$. In this case, the two-halo term is seen to be subdominant to the three-halo term, which, due to its $P_\mathrm{L}^2$ scaling, is sharply peaked at low-$k$, as for the intrinsic covariance. Due to the large number of contributors to our exclusion model, the magnitude of the various terms is not \textit{a priori} clear, but we note all are negative definite. This is to be expected since, whilst the model for $\delta\hat{n}_\mathrm{ex}$ (Sec.\,\ref{subsubsec: excluded-halo-fun}) has zero mean, the probability of halo exclusion in a particular spatial region is strongly correlated with the overdensity therein, giving a negative $\hat{n}_\mathrm{ex}\delta$ expectation.

\begin{figure}
    \centering
    \includegraphics[width=0.9\textwidth]{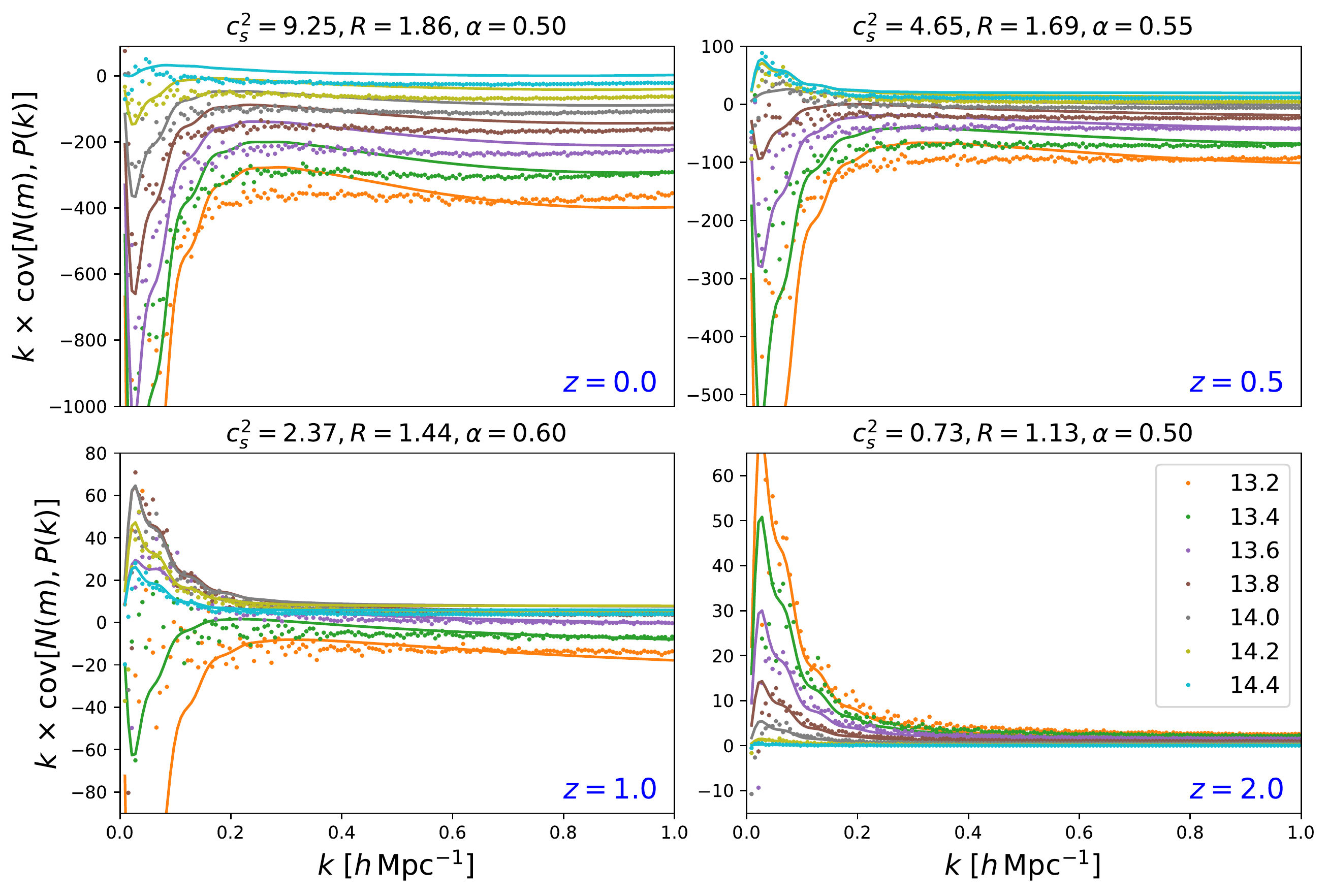}
    \caption{Covariance between halo number counts, $N(m)$, and the matter power spectrum, $P(k)$, computed from 15,000 \texttt{Quijote} simulations at a variety of redshifts. For each redshift, we plot the measured covariances (points) and the theory model from Sec.\,\ref{sec: cov_N_Pk_derivation} (lines) as a function of scale and mass, as in Fig.\,\ref{fig: cov-breakdown}, but without the normalization factor of $m$. Note that these simulations do not include super-sample effects. For each simulation, the free parameters $c_s^2$ and $R$ (specified in the title in $h^{-2}\mathrm{Mpc}^2$ and $h^{-1}\mathrm{Mpc}$ units respectively) are set by fitting the power spectrum model (Sec.\,\ref{sec: Pk_derivation}) at the specified redshift with $\alpha$ (the dimensionless ratio of halo exclusion radius to Lagrangian radius) chosen to give an approximate fit to the measured covariance. We note that the covariances are dominated by intrinsic (exclusion) effects at high (low) redshift.}
    \label{fig: all-z-exclusion-cov}
\end{figure}

In Fig.\,\ref{fig: all-z-exclusion-cov} we plot the measured covariance alongside the theory model for intrinsic and exclusion covariances. Notably, we observe strong negative covariances on all scales for low and intermediate mass halos at low redshift. Since the high-$k$ behavior of the intrinsic covariance model is set by the one-halo term, which is necessarily positive (since it has no dependence on halo bias), any theoretical model which does not account for halo exclusion cannot provide an accurate model of the low-redshift covariance of fixed mass simulations. For modest volumes, super-sample covariance usually dominates, explaining why these effects have not previously been noted.

From the figure, we note relatively good agreement between simulated and theoretical covariance at all redshifts. In particular, the $z = 2$ comparison implies that our intrinsic covariance model appears to work well, and we note the existence of a sharp peak in at low-$k$, which is well modeled by the (previously unmodeled) three-point covariance term. Indeed, it can be shown that the model is significantly deficient if the higher-order squared power spectrum and collapsed bispectrum terms are not included. At lower redshift, exclusion effects dominate and we see that the model of Sec.\,\ref{subsec: exclusion-cov} is able to provide a fair model of the covariance shape. Whilst the fit is not perfect, the general shape and amplitude dependence are certainly captured. As previously mentioned, we do not expect our somewhat rudimentary exclusion model to provide a perfect model for the covariance; it is however clear that the functional form is correct, and addition of higher order terms in exclusion fraction and bias are likely to improve this fit. We further note that the covariance model is strongly affected by our choice of bias parameters; whilst the exact modeling of these is of limited importance for the power spectrum (since biases are constrained by the consistency condition), this is not the case here. In general, PBS biases are known only to be accurate to $\sim 10\%$ \citep[e.g.][]{2016JCAP...02..018L}, thus the amplitudes of the various terms (both intrinsic and exclusion) are $\sim 10\%$ certain. Whilst it is certainly possible to measure the linear and quadratic biases from simulations or data \citep[e.g.][]{2012PhRvD..86h3540B} this is non-trivial, and goes against our philosophy of using as little simulation-based information as possible. Our model (and indeed any such treatment) should therefore be viewed with some caution, noting that deficiencies in the bias parameters are degenerate with those in the exclusion model.

The covariance model plotted in Fig.\,\ref{fig: all-z-exclusion-cov} depends on three free parameters; the effective-sound-speed $c_s^2$, the smoothing radius $R$ and the ratio of halo exclusion to Lagrangian radius $\alpha$. In this case, we choose $c_s^2$ and $R$ by fitting the power spectrum model of Sec.\,\ref{sec: Pk_derivation} to the measured matter spectra; only $\alpha$ remains as a free parameter. We find relatively good agreement between observed and model covariances for $\alpha \sim 0.5$, with slight variation seen between redshifts. (Note that the value of $\alpha$ is arbitrary for $z = 2$ since the exclusion terms are subdominant.) \textit{A priori}, one may not expect $\alpha$ to be redshift-dependent; however, since our exclusion model is fairly rudimentary and does not encapsulate additional effects such as higher-order biases or bispectrum terms, some residual dependence is unsurprising.\footnote{{We further note that the optimal value of $\alpha$ may depend on our choice of halo-finding algorithm.}}

\subsubsection{Autocovariance of $N(m)$}
{We can apply a similar methodology to test our theory model for the covariance of halo counts in different bins. Notably, the theoretical covariance has only minor dependence on the free parameters $c_s^2$ and $R$ (since these appear only in the non-linear part of the $\sex$ exclusion term at second order). The exclusion parameter $\alpha$ is significantly more important, however, controlling the magnitude of the (negative) exclusion terms. To ensure that our estimates are compatible with those obtained from the cross-covariance analysis above, we choose to fix this to the (redshift-dependent) value used in Fig.\,\ref{fig: all-z-exclusion-cov}, rather than fitting it from scratch.}

\begin{figure}
    \centering
    \includegraphics[width=0.8\textwidth]{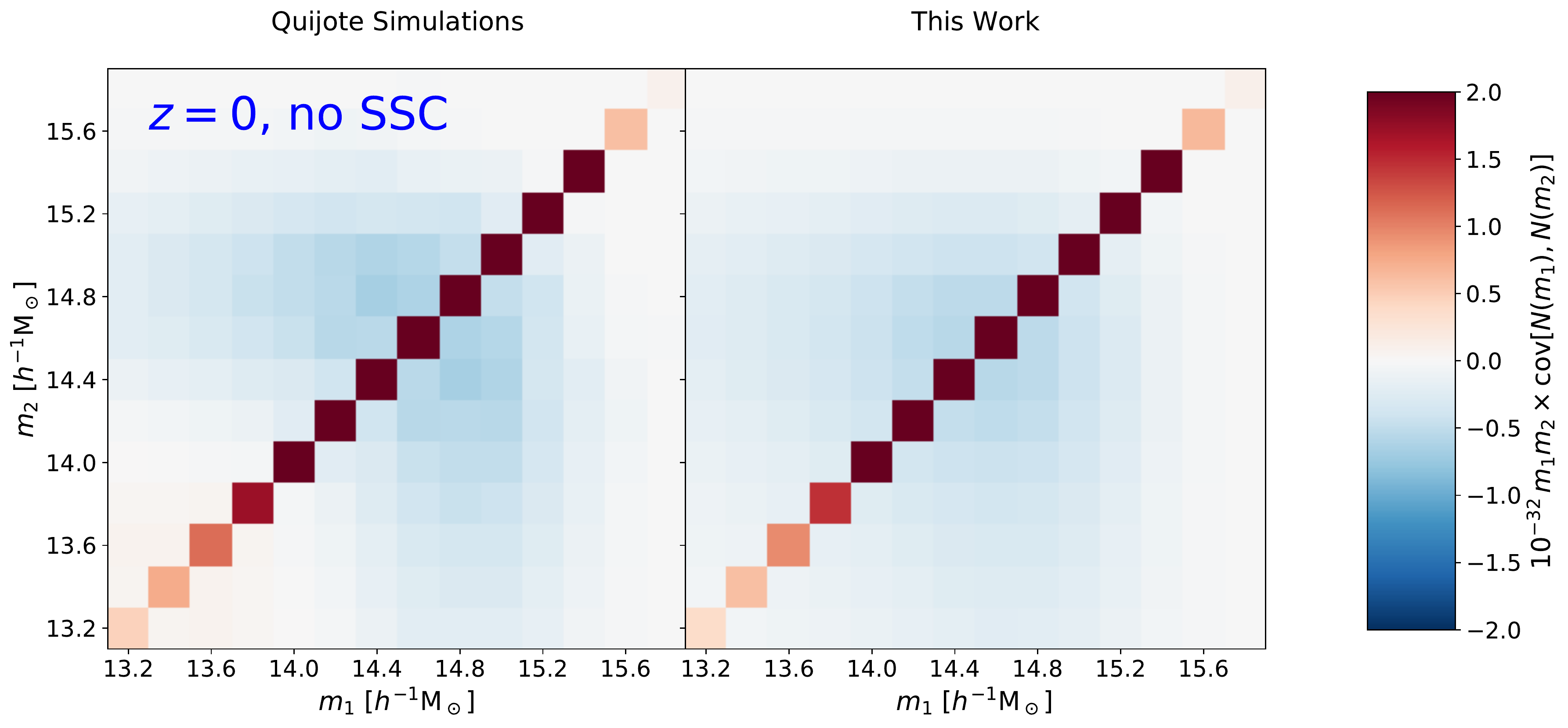}
    \includegraphics[width=0.8\textwidth]{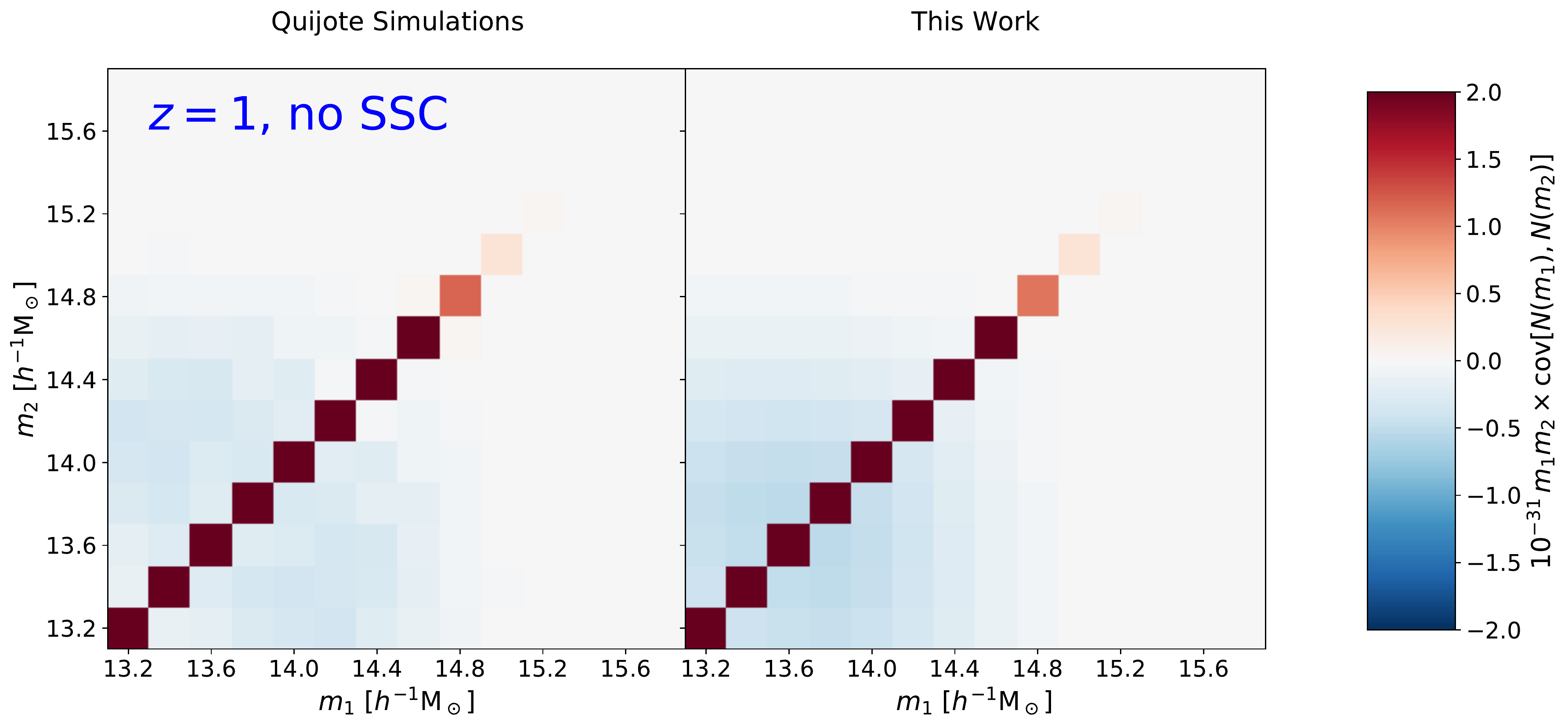}
    \caption{{Covariance matrices for cluster counts $N(m)$ in two mass bins $m_1$ and $m_2$. Results are obtained for $z = 0$ (upper figure) and $z = 1$ (lower figure) comparing the sample covariances from 15,000 $N$-body simulations (left) to the theory model of Sec.\,\ref{sec: cov_N_Pk_derivation} (right). The free parameters for each model are fixed to those used in Fig.\,\ref{fig: all-z-exclusion-cov}, and we note that super-sample effects are not included. Matrices are multiplied by $m_1$ and $m_2$ for visualization, with the colorbar chosen such that the off-diagonal exclusion terms are clear. For comparison, note that the maximum amplitude of the diagonal in these figures is $\sim 6$ in the plotted units for both figures.}}
    \label{fig: fullbox-Nm}
\end{figure}

{In Fig.\,\ref{fig: fullbox-Nm}, we plot the $N(m)$ covariance matrices for two redshifts, $z = 0$ and $z = 1$. To obtain increased diagnostic power, a greater range of masses is used than in previous sections, setting $M_\mathrm{max} = 10^{15.9}M_\odot$, which gives a total of 14 mass bins. Plotting the model covariances alongside those obtained from the \texttt{Quijote} simulations, we see good overall agreement, with the covariance dominated by the diagonal one-halo term in all cases (as expected). Since these are simply equal to the mean halo counts, they are well modelled and usually the only covariance term included. Of particular interest are the off-diagonal terms; our model clearly captures the heuristic trend, especially at $z = 1$, with significant negative contributions ($\sim 30\%$ of the diagonal power) for all but the largest masses. As expected, the effects of exclusion become smaller at early times, and are concentrated towards lower masses. Considering the amplitude of exclusion, whilst the $z = 1$ predictions appear to be appropriate, the exclusion effect at $z = 0$ is somewhat underestimated. This is likely a consequence of the simplicity of our exclusion model (for example in the restriction to first-order exclusion) and from not refitting the $\alpha$ parameter despite the mass range changing. An additional discrepancy between the simulations and theory appears at low mass for $z = 0$, where the simulated off-diagonal covariance becomes positive. This is likely caused by the perturbative two-halo intrinsic term (which scales as the integral of $\av{\eta_1\eta_2}$ over the survey). Whilst this is negligible in the infinite-volume limit, the finite $(1000\Mpch$) size likely causes this to be important here. However, the restriction to fixed mass in the box places global constraints on $\xi(\vec r)$ (since the 2PCF integrated over the box is equal to the mass variance, and hence zero), thus it is difficult to model. Practically, this is not seen to be important unless very small halo masses are used. Noting that previous models did not include any halo exclusion, we conclude that the model presented in this work is a significant upgrade.}

\subsection{Separate Universe Simulations: Super Sample Effects}\label{subsec: sims-DC-mode}
Before comparing our full covariance model to simulations, it is important to test our expressions for the response of $N(m)$ and $P(\vec k)$ to a long wavelength perturbation (Eqs.\,\ref{eq: dN_ddelta-model}\,\&\,\ref{eq: dP_ddelta-model}) since these underlie the SSC models. Practically, this is achieved via the separate universe \texttt{Quijote} simulations introduced in Sec.\,\ref{subsec: quijote-and-abacus-sims}. Whilst the details of these are non-trivial (and discussed at length in \citealt{2014PhRvD..89h3519L}) it is sufficient here to say that the simulations can be regarded as boxes with fiducial cosmology but with background overdensities $\delta_b = \pm \delta_0$ for $\delta_0 = 0.035$. For an observable $X$ (either $N(m)$ or $P(k)$), we approximate the derivative via
\beq
    \frac{dX}{d\delta_b} \approx \frac{X(\delta_0) - X(-\delta_0)}{2\delta_0}.
\eeq
and average over the 100 simulations with each choice of $\delta_b$. In practice, one must proceed with caution since the Hubble parameter $h$ differs between the simulations (due to the separate universe assumptions), thus we work in physical $M_\odot$ units to define the halo counts. Furthermore, the power spectra are evaluated relative to their \textit{local} mean density rather than the global mean, thus we must add a term $-2P(k)$ to the power spectrum derivative model (Eq.\,\ref{eq: dP_ddelta-model}).

\begin{figure}
\centering
\begin{minipage}[t]{.48\textwidth}
  \centering
  \includegraphics[width=\textwidth]{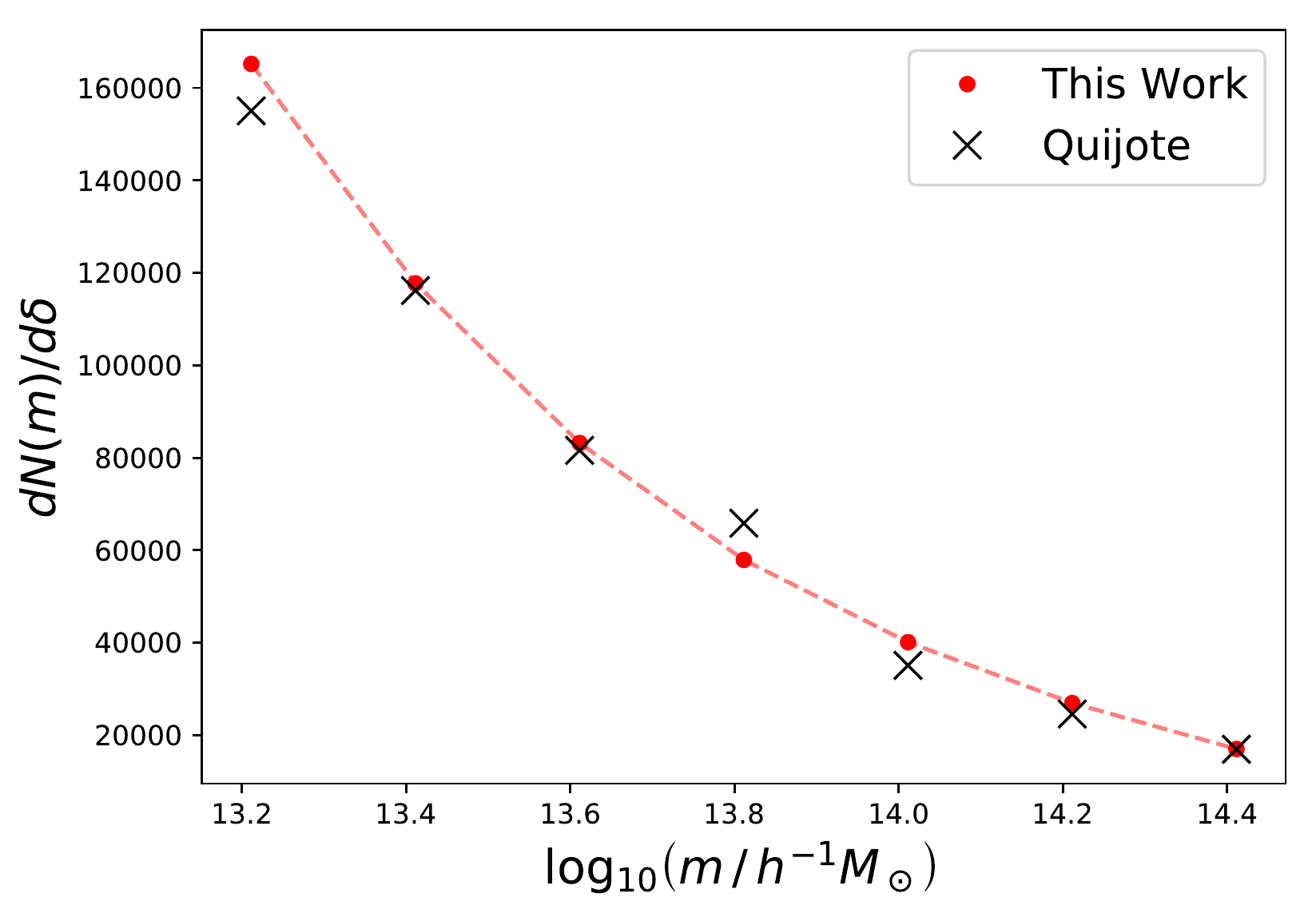}
  \captionof{figure}{Response of the halo number counts, $N(m)$, to a long wavelength perturbation $\delta_b$. The derivative is approximated numerically using 100 pairs of {standard-resolution}
  separate universe simulations from the \texttt{Quijote} suite, across seven mass bins with width $\Delta\log_{10}\left(m\,/h^{-1}M_\odot\right) = 0.2$ dex. The theoretical model (red) is computed from Eq.\,\ref{eq: dN_ddelta-model} and seen to be in fair agreement.}
  \label{fig: Nm-deriv-DC}
\end{minipage}%
\quad
\begin{minipage}[t]{.48\textwidth}
  \centering
  \includegraphics[width=\textwidth]{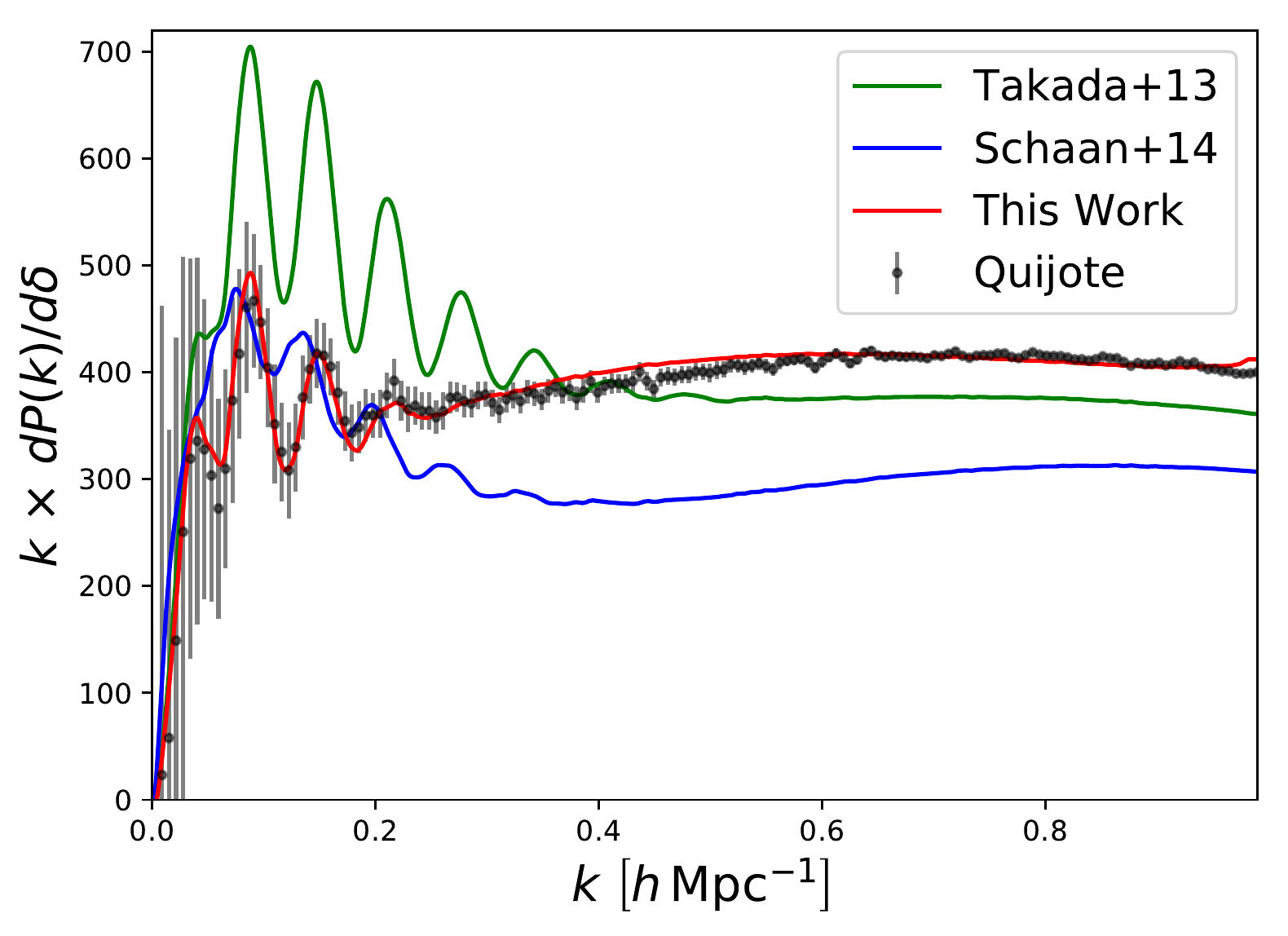}
  \captionof{figure}{Response of the matter power spectrum, $P(k)$, to a long wavelength perturbation $\delta_b$. As in Fig.\,\ref{fig: Nm-deriv-DC}, the derivative is approximated numerically via separate universe simulations, and we plot a variety of physical models alongside, from \citet{2013PhRvD..87l3504T}, \citet{2014PhRvD..90l3523S} and this work, corresponding to Eqs.\,\ref{eq: dPk-Takada},\,\ref{eq: dPk-Schaan}\,\&\,\ref{eq: dP_ddelta-model} respectively. Note that the simulated power spectra are computed using the local mean value of the average density rather than value at $\delta_b = 0$, and a factor $-2P(k)$ has been subtracted from each model to account for this, as discussed in the text. The derivatives used to compute the super-sample covariance in Fig.\,\ref{fig: subbox-cov} do not include this factor and it enhances the differences between models. 
  \resub{Whilst the model depends on two free parameters, $c_s^2$ and $R$, these are fit using the power spectrum only (and not the response).}
  }
  \label{fig: Pk-deriv-DC}
\end{minipage}
\end{figure}

In Fig.\,\ref{fig: Nm-deriv-DC}, we plot the measured halo number count derivative alongside the model of Sec.\,\ref{subsec: ssc-cov}. This is holistically in agreement, though does not precisely capture the fine details of the mass dependence (though this is difficult to probe without finer mass bins, requiring more simulations). Note that our model heavily relies on the measured linear biases $\bias{1}(m)$, which are estimated from the PBS formalism and not expected to be highly accurate. Nevertheless, it is seen to capture the leading mass dependence, and thus suitable for the SSC analysis.

Fig.\,\ref{fig: Pk-deriv-DC} shows the analogous results for the power spectrum response. The model introduced in this work \resub{(with $c_s^2$ and $R$ parameters obtained by fitting $P(k)$ data, as before)} is found to give an excellent fit across all $k$ scales probed (though with a slight excess of power at $k\sim 0.4\hMpc$ likely due to the uncalibrated bias parameters used). For comparison, we show two other models from the literature; firstly, the linear model presented in \citet{2013PhRvD..87l3504T} (and corrected in \citealt{2014PhRvD..89h3519L}), which gives
\beq\label{eq: dPk-Takada}
    \frac{dP_\mathrm{Takada}(k)}{d\delta_b} &=& 
    \left(\frac{68}{21}-\frac{1}{3}\frac{d\log k^3\left[I_1^1(k)\right]^2P_\mathrm{L}(k)}{d\log k}\right)\left[I_1^1(k)\right]^2P_\mathrm{L}(k) + I_2^1(k,k).
\eeq
Aside from the assumption of linearity, this differs in that the dilation derivative is applied to (and multiplied by) the full two-halo term and the subdominant $I_1^2(k)$ term is omitted. The second model shown is from \citet{2014PhRvD..90l3523S} and considers only halo sample variance;
\beq\label{eq: dPk-Schaan}
    \frac{dP_\mathrm{Schaan}(k)}{d\delta_b} &=& 
    I_1^1(k)I_1^{1,1}(k)P_\mathrm{L}(k) + I_2^1(k,k).
\eeq
To derive this, the authors make the (potentially unjustified) assumption that only $n(m)$ is affected by the super-sample mode. We further note that the $I_{1}^{1,1}(k)$ term is cancelled if the bias is also allowed to vary. 

It is important to note that the power spectrum derivative in Fig.\,\ref{fig: Pk-deriv-DC} differs from that for a weak-lensing type analysis by $-2P(k)$. {This was noted in Sec.\,\ref{subsec: ssc-cov}, and arises from the density field normalization by $\bar\rho(1+\delta_b)$ rather than $\bar\rho$, giving $P(k) \rightarrow \av{P(k)/(1+\delta_b)^2} \approx P(k)\left[1 - 2 \sigma^2(V)\right]$.} Since this acts to suppress the derivative, and $P(k)$ is reasonably well modeled by both linear and non-linear models, the differences between the models become more significant. From the figure, we note that both our model and that of \citet{2013PhRvD..87l3504T} are able to predict the positions of the power spectrum wiggles at low-$k$ (though the latter suffers from increased amplitude due to the lack of IR resummation) whilst that of \citet{2014PhRvD..90l3523S} suffers from a slight offset, since it neglects the beat-coupling effects. Furthermore, although all models agree on scales $\lesssim 0.05\hMpc$, the response is over-(under-)estimated by \citealt{2013PhRvD..87l3504T} (\citealt{2014PhRvD..90l3523S}), as a consequence of not including non-linear effects. Perhaps counter-intuitively, the models differ at large $k$, in the one-halo regime. Whilst all models have a common factor $I_2^1(k,k)$, only the model of this work accurately captures the derivative. We attribute this to the prefactor of $P_\mathrm{HM}(k)$ multiplying the linear dilation term, as opposed to $P_\mathrm{2h}(k)$ in the \citet{2013PhRvD..87l3504T} model.

\subsection{Subbox Simulations: Full Covariance}\label{subsec: sims-subbox}
The analysis of the preceding sections affords us confidence that our models capture the dominant contributions to the covariance of a survey in the infinite volume limit and can capture the necessary super-sample mode derivatives. It remains, therefore, to compute the full covariance including all terms; intrinsic, exclusion and super-sample. As previously noted, we cannot generate a covariance matrix including super-sample effects from the standard \texttt{Quijote} simulations since they contain only modes on scales less than the box-size. Furthermore, whilst the separate universe simulations used in Sec.\,\ref{subsec: sims-DC-mode} are useful for constraining overdensity derivatives, they are computed for only two choices of overdensity, and thus do not provide an appropriate testing ground for the full covariance model. 

To obtain an accurate sample covariance including all physical effects (including hitherto unconsidered tidal super-sample effects), we split each of the 100 high-resolution \texttt{Quijote} simulations into a set of $3^3$ disjunct \textit{subboxes}, each with $1/27$ of the total volume and side-length $L = (1000/3) \Mpch$. Though the total background density, summed over all subboxes, is fixed, this is \textit{not} true for the individual subboxes, as required. To reduce correlations between subboxes, we use only the nine diagonal subboxes from each main box, giving a total of $N_\mathrm{subbox}=900$ subboxes, which we assume to be independent. In each subbox, the matter power spectrum is estimated as in Sec.\,\ref{subsec: quijote-and-abacus-sims}, using a grid-size of $N_\mathrm{grid} = 512$ cells per dimension.\footnote{Note that this implicitly assumes the subboxes to have periodic boundaries; a false assumption in this case. The resulting spectra were compared to those computed by embedding the subbox in a larger empty region and using the estimator of \citet{1994ApJ...426...23F}, and found to be highly consistent.}  Though this is lower than the $N_\mathrm{grid}=2048$ used previously it is of limited importance since the Nyquist frequency scales as $N_\mathrm{grid}/L$ and $L$ has been reduced by a factor of three. When computing the overdensity fields $\delta$ of the subboxes, we normalize by the fullbox mean density $\bar\rho$ rather than that of the subbox, such that our covariances are applicable to weak-lensing analyses which do not depend on the density normalizations. We additionally work at $z = 0$, since this is where the non-SSC covariances are strongest.

Following this, the model of Sec.\,\ref{sec: cov_N_Pk_derivation} is computed as in the previous sections. There are two points to note; firstly, we must be aware of power spectrum effects arising from the finite volume of the box. As discussed in Appendix \ref{appen: Pk-trunc}, the non-linear power spectrum of the subbox is \textit{not} simply a function of the full linear power spectrum drawn from some cosmological Boltzmann code. Due to the fixed size of the region, modes larger than the subbox cannot affect the power spectrum (except as a super-sample covariance), which warrants using a linear power spectrum with zero power for $k \le k_\mathrm{min} = 2\pi/L_\mathrm{subbox}$ to compute the non-linear corrections. For the full $L=1000\Mpch$ boxes this is of limited importance, though it starts to become important on the subbox scale. Secondly, the computation of the super-sample mode variance $\sigma^2(V)$ is non-trivial due to the cubic geometry of the box. Starting from the definition (Eq.\,\ref{eq: sigma2V-def}), we may write
\beq
    \sigma^2(V) &\equiv& \int \frac{d\vec k}{(2\pi)^3}\left|W_\mathrm{survey}(\vec k)\right|^2P_\mathrm{L}(\vec k)\\\nonumber
    &=& \int_{k_0}^\infty \frac{k^2dk}{2\pi^2}\left[W_\mathrm{survey}^2\right]_0(k)P_\mathrm{L}(k),
\eeq
where we have integrated over the angular part of $\vec k$ in the second line, noting that, since $P_\mathrm{L}$ is isotropic, only the monopole part of $W^2_\mathrm{survey}(\vec k)$ can give a non-zero contribution. Note that we include only modes with $k>k_0 = 2\pi/L$ since longer wavelength modes are not present in the $L = 1000\Mpch$ box from which the subboxes are drawn. The monopole of the squared window can be computed via a spherical Fourier transform
\beq
    \left[W^2_\mathrm{survey}\right]^2(k) = 4\pi \int_0^\infty r^2dr\,j_0(kr)\left[W^2_\mathrm{survey}\right]_0(r),
\eeq
where $\left[W^2_\mathrm{survey}\right]_0(r)$ is the real-space monopole of the square survey window function, simply computed via pair-counting. This gives $\sigma^2(V) = 4.72\times 10^{-4}$ for the subbox, which may be compared to $4.82\times 10^{-4}$ if we use a spherical window function of equivalent volume rather than the monopole of the cubic function. This is further in percent-level agreement with the sample variance of mass in the 900 subboxes.

\begin{figure}
    \centering
    \includegraphics[width=0.95\textwidth]{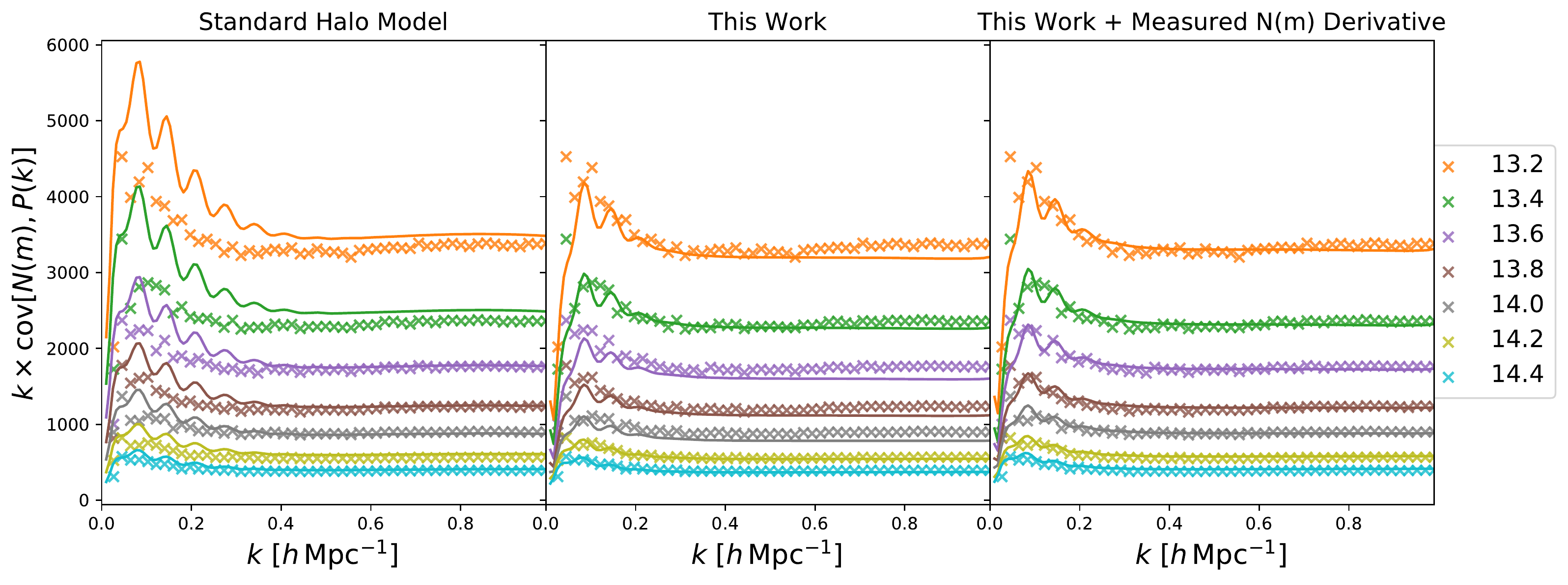}
    \caption{Covariance between the halo number counts $N(m)$ and matter power spectrum $P(k)$ computed from 900 $z = 0$ subboxes of the high-resolution \texttt{Quijote} simulation suite, each of which have side-length \resub{$3.7\times 10^{7}h^{-3}\mathrm{Mpc}^3$}. 
    These are of the same format as Fig.\,\ref{fig: all-z-exclusion-cov} but include super-sample effects from variations in the mean density of the simulation box. Simulation results (points) are plotted for seven mass bins (as in previous figures) and we plot the standard halo model covariance (left panel, {using \citet{2013PhRvD..87l3504T} super-sample covariances}) in addition to the new model introduced in this work (central panel). Free parameters for our model are taken from fitting the power spectrum model and fixed-mass simulations, and are not optimized to produce this figure. The right panel shows our model but with the $N(m)$ super-sample derivative replaced by its measured value from the subbox simulations.}
    \label{fig: subbox-cov}
\end{figure}

\subsubsection{Covariance of $N(m)$ and $P(\vec k)$}
In Fig.\,\ref{fig: subbox-cov} we compare the measured and modeled subbox covariances {of $N(m)$ and $P(\vec k)$}. Our first observation is that super-sample effects greatly dominate, with small-scale amplitudes in excess of ten times those found for the (volume-independent) fixed-mass covariance in Fig.\,\ref{fig: all-z-exclusion-cov}. This effect is enhanced by the small volume of the region (since intrinsic and exclusion covariances are volume independent, yet SSC terms scale as $V\sigma^2(V)$), though usually expected to be present. Comparing the sample covariance to that predicted by the standard halo model (using the model of \citealt{2014PhRvD..90l3523S} and \citealt{2014MNRAS.441.2456T}, coupled with an SSC term from \citealt{2013PhRvD..87l3504T}), we see a severe over-prediction of power on large scales, with a slight overestimate for low-mass halos on small scales. We attribute this to two effects; (1) the vanilla model over-predicts the low-$k$ power spectrum response as in Fig.\,\ref{fig: Pk-deriv-DC}, and (2) this does not include a model for halo exclusion, leading the low-mass covariance to be overestimated on small scales. Indeed, Fig.\,\ref{fig: Pk-deriv-DC} also indicates that the vanilla model under-predicts $dP(k)/d\delta_b$ at high-$k$, thus the exclusion effects are somewhat stronger than indicated.

The central panel of Fig.\,\ref{fig: subbox-cov} plots the covariance model obtained in this work, as detailed in  Sec.\,\ref{sec: cov_N_Pk_derivation}. The free parameters $c_s^2$ and $R$ are obtained by fitting the power spectrum (as in Sec.\,\ref{sec: sims}) with the halo exclusion parameter $\alpha=0.5$ taken from the $L = 1000\Mpch$ fixed-mass covariances of Fig.\,\ref{fig: subbox-cov}. Our model is thus not recalibrated for this visualization. Our approach provides an adequate model of the sample covariance, particularly regarding its wavenumber dependence, as a consequence of our improved model of the power spectrum response term. The mass dependence is roughly correct, though there are some discrepancies possibly relating to inaccuracies in determining the bias parameters, which are important for the $N(m)$ response and exclusion terms.  It can be shown that the model severely over-predicts the low-mass covariance if the exclusion terms are not included, highlighting the importance of this effect.

Given the good agreement between simulations and theory in the previous sections, it is worth asking the question of where the dominant source of error in our model lies. From the figure, it is clear that the greatest uncertainty is in the mass dependence; this is likely to be sourced by incomplete knowledge of the bias parameters and could be significantly improved by (a) using measured bias parameters rather than those from the PBS formalism or (b) allowing the exclusion parameter $\alpha$ to vary freely (rather than being set by the fixed-mass covariances), which has a large mass-dependent effect. In the right panel of Fig.\,\ref{fig: subbox-cov}, we plot the same covariance, but replacing the theory $dN(m)/d\delta_b$ derivative with that taken from the subbox simulations (via $dN(m)/d\delta_b = \operatorname{cov}\left(N(m),\delta_b\right)/\sigma^2(V)$).\footnote{It is not immediately obvious whether this should be identical to the derivative measured from the separate universe simulations. One effect that is present only in the latter is the restriction to fixed total mass, which may affect the halo counts.} Notably, our model is now seen to capture both the mass and $k$ dependence with high precision (and would be significantly deficient at low mass without halo exclusion). Since our model for $dN(m)/d\delta_b$ is deceptively simple, it stands to reason that the model inaccuracies are primarily sourced by a lack of knowledge of the exact bias parameters. We therefore conclude that our covariance matrix model captures all the dominant physical effects and is accurate, modulo a lack of analytic knowledge of halo bias.

\subsubsection{Autocovariance of $N(m)$}
\begin{figure}
    \centering
    \includegraphics[width=0.95\textwidth]{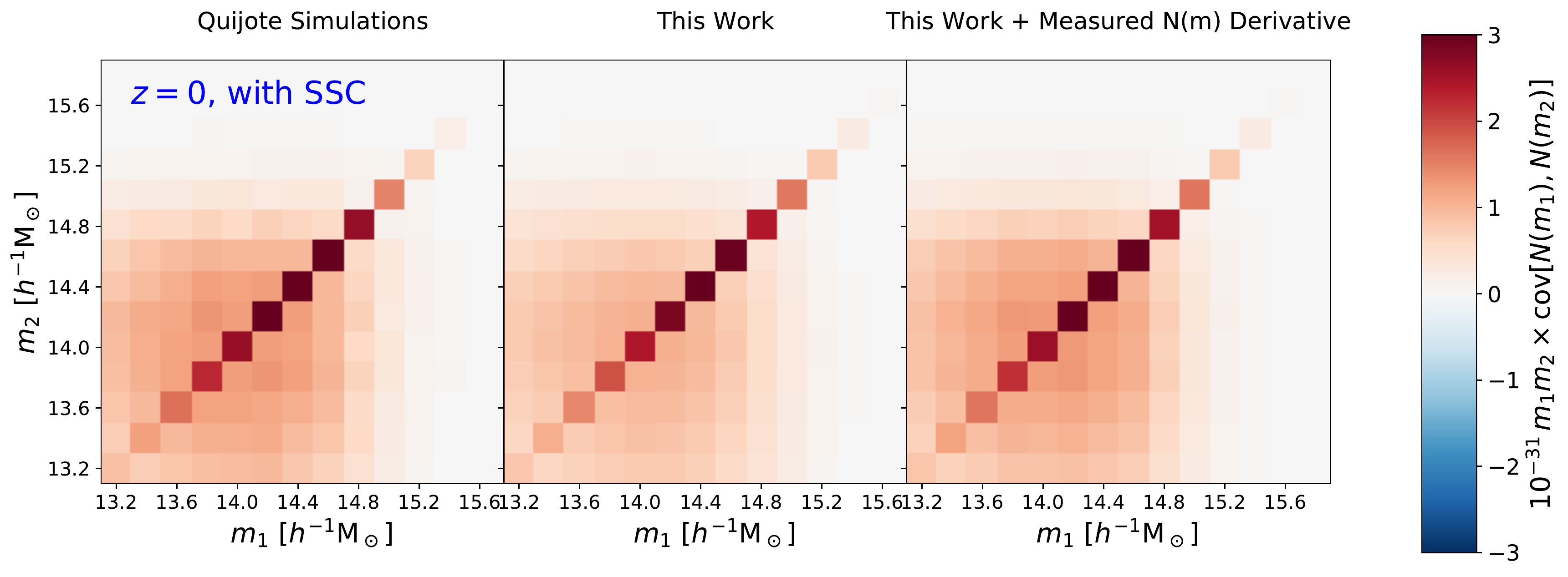}
    \caption{{Covariance matrices for cluster counts $N(m)$ in two bins $m_1$ and $m_2$, including super-sample effects. This uses the same data-set as in Fig.\,\ref{fig: subbox-cov}, plotting in the format of Fig.\,\ref{fig: fullbox-Nm}. The left panel shows the results from \texttt{Quijote} subbox simulations, whilst the central panel displays the theory model of Sec.\,\ref{sec: cov_N_Pk_derivation}. In the right panel, the derivative term $dN(m)/d\delta_b$ is replaced by its observational value, as in the right panel of Fig.\,\ref{fig: subbox-cov}. In all cases, the off-diagonal contributions are dominated by super-sample effects, though the contributions from halo exclusion are $\sim 15\%$.}}
    \label{fig: subbox-Nm}
\end{figure}

{Our final task is to compare the predictions of our $N(m)$ covariance model (including super-sample effects) to simulations. This is shown in the left and center panels of Fig.\,\ref{fig: subbox-Nm} in the same format as Fig.\,\ref{fig: fullbox-Nm}. Whilst the diagonal elements are still primarily controlled by the one-halo term, we note strong off-diagonal contributions at low-mass, with correlation coefficients up to $70\%$ found at low mass. These are sourced by the SSC terms, which greatly dominate at this volume. Indeed, halo exclusion is found to have only a minor effect here, though its importance would grow with the survey volume (as $V\sigma^2(V)$ shrinks). Comparing our model to data, we observe that it is able to provide a good heuristic fit, though, as before, its amplitude is slightly underestimated at low-mass. Whilst one might be tempted to relate this to the low-mass deficit seen in the full-box power at $z = 0$ in Fig.\,\ref{fig: fullbox-Nm}, this is unlikely to be the case, since the former was expected to arise from a neglected higher-order super-sample effect, which will be swamped by our SSC covariance terms. Instead, we posit that the underestimate is caused by inaccuracies in the modelling of $dN(m)/d\delta_b$, as proposed in the previous subsection. To test this, we follow a similar method to before, replacing the theoretical model for the $N(m)$ response with the observed value $\operatorname{cov}(N(m),\delta_b)/\sigma^2(V)$, (using the modeled $\sigma^2(V)$). This is shown in the right panel of Fig.\,\ref{fig: subbox-Nm}, and we note good agreement between theory and simulations (though a large overestimate if exclusion is not accounted for). From this, we attribute the discrepancy in our model to poor understanding of the $N(m)$ response. At higher masses however, these effects are negligible, and the covariance becomes almost diagonal.}

\section{Summary}\label{sec: conclusion}

In this paper, we have developed a new formalism for computing cosmological statistics across a broad range of scales, formulated in the context of the cosmological halo model and perturbation theory. Our approach, {named the `Effective Halo Model'}, relies on two physical assumptions: (1) halos are Poisson distributed with a distribution function that depends on the non-linear density field $\delta$ smoothed on an unknown scale $R$; (2) the statistics of $\delta$ can be adequately modeled using the Effective Field Theory of Large Scale Structure (EFT). Starting from these, we have derived a simple model for the matter power spectrum at one-loop order that depends on two free parameters, $R$ and the EFT effective sound-speed $c_s^2$. Practically, our model is just the usual halo model power spectrum, but with an additional smoothing window applied to the two halo term and the linear power spectrum replaced with a non-linear one, featuring infra-red resummation and a (Pade-resummed) ultraviolet counterterm. {Whilst our model does carry free parameters, these are inherent in \textit{any} perturbative model of the matter power spectrum, due to the existence of counterterms. It is further the case that these encode not only non-linearities in the underlying density field but additionally unmodeled physics on the shell-crossing scale. It is an important conclusion of this work that both effects are degenerate in the power spectrum, and thus can be fully modelled using simply the perturbative free parameters.}   A Python package incorporating our power spectrum model has been publicly released.\footnote{\href{https://EffectiveHalos.rtfd.io}{EffectiveHalos.readthedocs.io}}.

The second half of this paper extends this model to compute the {covariance matrices of halo number counts, both alone and in combination with the matter power spectrum}. These are key underlying statistics for weak lensing and thermal Sunyaev-Zel'dovich (tSZ) analyses. In particular, we provide new models for the intrinsic (infinite volume) and super-sample covariances, which, unlike previous analyses, include second order terms in $P_\mathrm{L}(k)$. Furthermore, it has been shown that to obtain an adequate model of the covariance, one must include halo exclusion effects, arising from the impossibilitity of having two halos co-located. An approximate method for including this has been developed, and all contributions rigorously tested with $N$-body simulations. This represents a dramatic improvement in our modeling of such covariances, though we are still limited by the lack of precision in analytic halo bias models.   The new free parameter in this extension is a characteristic scale for halo exclusion, an effect that is important at low redshift.

\resub{Given that the principal application of matter power spectrum models is to integrated statistics such as weak lensing, it is important to make a comparison between the model introduced in this work and standard techniques. In many analyses, semi-analytic models are used, including the `HaloFit' \citep{2003MNRAS.341.1311S,2012ApJ...761..152T}  and `CosmicEmu' \citep{2010ascl.soft10030L} emulators, as well as `HMCode' \citep{2015MNRAS.454.1958M}, a fitted halo model. The first difference lies in accuracy; as the Effective Halo Model is based around EFT it naturally produces highly accurate models on quasi-linear scales (including all physical effects such as BAO wiggles) allowing for robust parameter inference. In contrast, emulators are usually fit to a wide range of cosmologies, and typically obtain $\sim 5\%$ accuracy. On the other hand, since our model contains redshift-dependent free parameters ($c_s^2$ and $R$), performing the projection integrals necessary to compute statistics such as weak lensing convergence spectra will be non-trivial, though likely possible with some assumed redshift dependence \citep[cf.][]{2016JCAP...04..033F}. This complication is not present for emulator-based models, and should be discussed in future work.}

\resub{A further point of distinction concerns scale dependence. In this work, we have tested the Effective Halo Model only for $k < 1\hMpc$, whilst most emulators claim a good fit to $N$-body simulations up to much larger $k$. Since the standard halo model is known to provide a good fit to data on small scales, our model, which uses the same one-halo term, is expected to be similarly accurate. However, we caution that the effects of baryon feedback are significant beyond $k\approx 1\hMpc$ \citep[e.g.,][]{2019OJAp....2E...4C} and thus, since neither our model nor most emulators encapsulate this, we find it somewhat academic to extend the statistics to larger $k$, as dark-matter-only spectra are not cosmologically relevant. Given that baryonic effects are poorly understood, it is desirable to perform some kind of baryon marginalization, for example with the free feedback parameters of \citet{2015MNRAS.454.1958M} or \citet{2014MNRAS.445.3382M}, though this suffers the same projection integral complexities as our model.\footnote{Baryonic effects on large scales are expected to be absorbed by the model $R$ and $c_s^2$ parameters; a further motivation for keeping them free.} With proper consideration of baryonic effects, our model can thus be extended to larger $k$. This is an important next step for the Effective Halo Model.}

There are a number of additional ways in which the above analysis can be extended. In particular;
\begin{itemize}
    \item \textbf{Exclusion Modeling}: Whilst we have demonstrated the importance of including halo exclusion effects in a model of the halo count versus power spectrum covariance, our model for it is somewhat rudimentary. More work is needed to accurately model this effect, including its dependence on redshift as well as working to higher order in bias parameters, exclusion fraction and perturbation theory.
    \item \textbf{Large-Scale Effects}: A fundamental problem with the halo model of the matter power spectrum is that it is inaccurate on the largest scales. Although not evident in this work (since we are limited to $k\gtrsim 10^{-2}\hMpc$), the conventional one-halo term tends to a constant on large scales rather than having the $k^4$ dependence required by mass and momentum conservation. This {dominates} at $k\lesssim 10^{-3}\hMpc$ and could be ameliorated by \resub{halo exclusion \citep{2007PhRvD..75f3512S, 2011PhRvD..83d3526S}}, modifying the halo profile \citep{2020PhRvD.101j3522C} {or including a stochastic perturbation field \citep{2016PhRvD..93f3512S}}. 
    \item \textbf{Number Count Responses}: {As shown in Sec.\,\ref{sec: sims-cov}, the largest uncertainty in our modelling of the cluster count covariances lies in the determination of the response of the halo number counts to a long mode. Developing a more robust model for this will improve predictions of the super-sample covariance model.} 
    \item {\textbf{Other Statistics}: Whilst this work has concentrated on the matter power spectrum, there are a proliferation of halo models available for other statistics, including galaxy \citep[e.g.][]{2002PhR...372....1C} and void \citep{2014PhRvL.112d1304H,2020arXiv200306411V} statistics. Our method can simply be applied to these models, and would be expected to improve their accuracy. This is more complex than for the matter power spectrum however, since the precision achievable depends strongly on our knowledge of the halo occupation distributions and bias parameters, since these are no longer constrained by consistency conditions.}
     \item \textbf{Likelihood Analyses:}  This analytical model offers a potential route towards developing an analytical model for the joint likelihood function for weak lensing measurements and cluster counts (e.g., through tSZ measurements).  By developing a rigorous theory that includes higher order statistics and extends to smaller physical scales, we have the potential to extract significantly more information from our large investments in large-scale structure observations, weak lensing measurements, and maps of the microwave background. 
\end{itemize}

\bibliographystyle{mnras}
\bibliography{adslib} 

\section*{Acknowledgements}
We thank Jo Dunkley, Yin Li, Andrina Nicola, {Fabian Schmidt}, Marko Simonovi\'c and Matias Zaldarriaga for useful discussions. We additionally thank Colin Hill, Mikhail Ivanov, Leonardo Senatore, Emmanuel Schaan, Marcel Schmittfull, Uro\v{s} Seljak, Masahiro Takada and Ben Wandelt for comments on a draft of this paper. \resub{Furthermore, the authors acknowledge insightful and detailed feedback from the anonymous referee.} OHEP and FAVN acknowledge funding from the WFIRST program through NNG26PJ30C and NNN12AA01C.
The Flatiron Institute is supported by the Simons Foundation.



\appendix

\section{Derivation of the $P(k)$ Covariance}\label{appen: cov-Pk}
We present a brief derivation of the one-loop covariance of the matter power spectrum model introduced in this work. Note that a similar derivation is presented in \citet{2017MNRAS.466..780M}, though not including halo model effects. We begin in configuration space (ignoring super-sample covariance), using Eq.\,\ref{eq: 2PCF-definition} to write
\beq
    \operatorname{cov}\left(\xi(\vec r),\xi(\vec s)\right) &=& \av{\hat{\xi}(\vec r)\hat{\xi}(\vec s)}-\av{\hat{\xi}(\vec r)}\av{\hat{\xi}(\vec s)}\\\nonumber
    &=&\prod_{j=1}^4\left[\int d\vec x_jn_j\frac{m_j}{\bar\rho}\right]\int \frac{d\vec xd\vec y}{V^2}u_1(\vec x-\vec x_1)u_2(\vec x+\vec r-\vec x_2)u_3(\vec y-\vec x_3)u_4(\vec y+\vec s-\vec x_4)\left\{\av{\delta \hat{n}_1\delta \hat{n}_2\delta \hat{n}_3\delta \hat{n}_4}-\av{\delta \hat{n}_1\delta \hat{n}_2}\av{\delta \hat{n}_3\delta \hat{n}_4}\right\},
\eeq
where $u_j(\vec x)\equiv u(\vec x|m_j)$ and $\delta \hat{n}_i\equiv \hat{n}(m_i|\vec x_i) - n(m_i)$. Note that the $\vec x$ ($\vec y$) integral simply simply gives a convolution of the $u_1$ and $u_2$ ($u_3$ and $u_4$) density profiles. To proceed we can expand the term in braces into four-, three-, two- and one-halo terms;
\beq\label{eq: initial-covariance-Pk}
    \left\{\av{\delta \hat{n}_1\delta \hat{n}_2\delta \hat{n}_3\delta \hat{n}_4}-\av{\delta \hat{n}_1\delta \hat{n}_2}\av{\delta \hat{n}_3\delta \hat{n}_4}\right\} &=& n_1n_2n_3n_4\left[\av{\eta_1\eta_2\eta_3\eta_4}_c+2\av{\eta_1\eta_3}\av{\eta_2\eta_4}\right]\\\nonumber
    &&\,+\, 4\delta_D(1-3)n_1n_2n_4\av{(1+\eta_1)\eta_2\eta_4}+\delta_D(1-2)n_1n_3n_4\av{\eta_1\eta_3\eta_4}+\delta_D(3-4)n_1n_2n_3\av{\eta_1\eta_2\eta_3}\\\nonumber
    &&\,+\,2\delta_D(1-3)\delta_D(2-4)n_1n_2\left[1+\av{\eta_1\eta_2}\right]+\delta_D(1-2)\delta_D(3-4)n_1n_3\av{\eta_1\eta_3}\\\nonumber
    &&\,+\, \delta_D(1-3)\left[\delta_D(1-4)+\delta_D(3-4)\right]n_1n_2\av{\eta_1\eta_2}\delta_D(1-3)\left[\delta_D(1-2)+\delta_D(2-3)\right]n_3n_4\av{\eta_3\eta_4}\\\nonumber
    &&\,+\, \delta_D(1-2)\delta_D(2-3)\delta_D(3-4)n_1,
\eeq
where we have assumed Poissonian statistics and included the definition of $\eta$ (Eq.\,\ref{eq: eta_definition}), writing $n_j\equiv n(m_j)$,  $\eta_j \equiv \eta(\vec x_j|m_j)$ and $\delta_D(i-j)\equiv\delta_D(m_i-m_j)\delta_D(\vec x_i-\vec x_j)$. This has additionally used the Wick expansion of the random field $\eta$, $\av{\eta_i}\equiv0$ and grouped symmetric terms. To evaluate these terms we require the expectation of up to four (connected) products of $\eta$. In real space, and working up to terms of order $\left(P_L(k)\right)^2$, these may be written
\beq
    \av{\eta_1(\vec x_1)\eta_2(\vec x_2)} &=&\bias{1}_1\bias{1}_2\av{\delta_R(\vec x_1)\delta_R(\vec x_2)} + \left\{\frac{1}{2}\bias{1}_1\bias{2}\av{\delta^2_R(\vec x_1)\delta_R(\vec x_2)}+\text{ 1 sym.}\right\}\\\nonumber
    &&\,+\, \frac{1}{4}\bias{2}_1\bias{2}_2\left[\av{\delta^2_R(\vec x_1)\delta^2_R(\vec x_2)}-\av{\delta^2_R}^2\right]+\left\{\frac{1}{6}\bias{1}_1\bias{3}_2\av{\delta_R^3(\vec x_1)\delta_R(\vec x_2)}+\text{ 1 sym.}\right\}\\\nonumber
    \av{\eta_1(\vec x_1)\eta_2(\vec x_2)\eta_3(\vec x_3)} &=& \bias{1}_1\bias{1}_2\bias{1}_3\av{\delta_R(\vec x_1)\delta_R(\vec x_2)\delta_R(\vec x_3)}+\left\{\frac{1}{2}\bias{2}_1\bias{1}_2\bias{1}_3\left[\av{\delta^2_R(\vec x_1)\delta_R(\vec x_2)\delta_R(\vec x_3)}-\av{\delta_R^2}\av{\delta_R(\vec x_2)\delta_R(\vec x_3)}\right]+\text{ 2 sym.}\right\}\\\nonumber
    \av{\eta_1(\vec x_1)\eta_2(\vec x_2)\eta_3(\vec x_3)\eta_4(\vec x_4)}_c &=& \bias{1}_1\bias{1}_2\bias{1}_3\bias{1}_4\av{\delta_R(\vec x_1)\delta_R(\vec x_2)\delta_R(\vec x_3)\delta_R(\vec x_4)}_c,
\eeq
Following some algebra, these may be written in Fourier space as
\beq\label{eq: full-eta-statistics}
    \av{\eta_1\eta_2}(\vec k) &=& \bias{1}_1\bias{1}_2P_R(\vec k)+\frac{1}{2}\left(\bias{1}_1\bias{2}_2+\bias{2}_1\bias{1}_2\right)\int \frac{d\vec p}{(2\pi)^3}B_R(\vec p,\vec k)\\\nonumber
    &&\,+\,\frac{1}{4}\bias{2}_1\bias{2}_2\left[2\int\frac{d\vec p}{(2\pi)^3}P_R(\vec p)P_R(\vec k-\vec p)+\int\frac{d\vec p_1}{(2\pi)^3}\frac{\resub{d}\vec p_2}{(2\pi)^3}T_R(\vec p_1,\vec p_2,\vec k-\vec p_1)\right]\\\nonumber
    &&\,+\, \frac{1}{6}\left(\bias{1}_1\bias{3}_2+\bias{3}_1\bias{1}_2\right)\left[3P_R(\vec k)\sigma^2_R+\int\frac{d\vec p_1}{(2\pi)^3}\frac{d\vec p_2}{(2\pi)^3}T_R(\vec p_1,\vec p_2,\vec k)\right]\\\nonumber
    \av{\eta_1\eta_2\eta_3}(\vec k_1,\vec k_2,\vec k_3)\delta_D(\vec k_{123}) &=& \bias{1}_1\bias{1}_2\bias{1}_3B_R(\vec k_1,\vec k_2)+\left\{\frac{1}{2}\bias{2}_1\bias{1}_2\bias{1}_3\left[2P(\vec k_2)P(\vec k_3)+\int \frac{d\vec p}{(2\pi)^3}T_R(\vec p,\vec k_1-\vec p,\vec k_2)\right]+\text{ 2 sym.}\right\}\\\nonumber
    \av{\eta_1\eta_2\eta_3\eta_4}(\vec k_1,\vec k_2,\vec k_3,\vec k_4)\delta_D(\vec k_{1234}) &=& \bias{1}_1\bias{1}_2\bias{1}_3\bias{1}_4T_R(\vec k_1,\vec k_2,\vec k_3),
\eeq
where $\delta_D(\vec k_{1..n})\equiv \delta_D(\vec k_1+...\vec k_n)$. $\sigma_R$, $P_R$, $B_R$ and $T_R$ are the variance, power spectrum, bispectrum and trispectrum of the \textit{smoothed} field $\delta_R$, defined in terms of the unsmoothed matter correlators in Eq.\,\ref{eq: s2BT-def}. Here the bispectrum and trispectrum should be evaluated at tree-level, whilst we require a one-loop power spectrum (or linear-order for $P^2$ terms). 

Inserting Eq.\,\ref{eq: full-eta-statistics} into Eq.\,\ref{eq: initial-covariance-Pk}, and following a straightforward, yet extremely laborious, computation, we obtain an expression for the Fourier-space covariance at one-loop order in our halo model using the notation of Eq.\,\ref{eq: integral-notation};
\beq\label{eq: full-covariance-Pk}
    \operatorname{cov}\left(P(\vec k),P(\vec k')\right)^\mathrm{intrinsic} &\equiv& \mathcal{C}^{4h}(\vec k,\vec k')+\mathcal{C}^{3h}(\vec k,\vec k')+\mathcal{C}^{2h}(\vec k,\vec k')+\mathcal{C}^{1h}(\vec k,\vec k')\\\nonumber
    V\mathcal{C}^{4h}(\vec k,\vec k') &=& 2\left[I_1^1(\vec k)\right]^4P^2_R(\vec k)\delta_D(\vec k+\vec k')+\left[I_1^1(\vec k)I_1^1(\vec k')\right]^2T_R(\vec k,-\vec k,\vec k',-\vec k')\\\nonumber
    V\mathcal{C}^{3h}(\vec k,\vec k') &=& 4I_2^0(\vec k,\vec k)\left[I_1^1(\vec k)\right]^2P_R(\vec k)\delta_D(\vec k+\vec k')\\\nonumber
    &&\,+\,\left\{I_2^1(\vec k,\vec k)\left[I_1^1(\vec k')\right]^2B_R(\vec 0,\vec k',-\vec k')+I_2^2(\vec k,\vec k)\left[I_1^1(\vec k')\right]^2\left(P_R^2(\vec k')+\frac{1}{2}\int_{\vec p}T_R(\vec p,-\vec p,\vec k')\right) +\text{ sym.}\right\}\\\nonumber
    &&\,+\,4I_1^1(\vec k)I_1^1(\vec k)\left[I_2^1(\vec k,\vec k')B_R(\vec k,\vec k')+I_2^2(\vec k,\vec k')\left(P(\vec k)P(\vec k')+\frac{1}{2}\int_{\vec p}T_R(\vec p,\vec k,\vec k')\right)\right]\\\nonumber
    V\mathcal{C}^{2h}(\vec k,\vec k') &=& 2\left[I_2^0(\vec k,\vec k)\right]^2\delta_D(\vec k+\vec k')+2\left[I_2^1(\vec k,\vec k')\right]^2P_R(\vec k+\vec k')+2I_2^1(\vec k,\vec k')I_2^2(\vec k,\vec k')\int_{\vec p}B_R(\vec p,\vec k+\vec k'-\vec p)\\\nonumber
    &&\,+\,\frac{1}{2}\left[I_2^2(\vec k,\vec k')\right]^2\left[2\int_{\vec p}P_R(\vec p)P_R(\vec k+\vec k'-\vec p)+\int_{\vec p_1\vec p_2}T_R(\vec p_1,\vec p_2,\vec k+\vec k'-\vec p_1)\right]\\\nonumber
    &&\,+\, \frac{1}{3}I_2^1(\vec k,\vec k')I_2^3(\vec k,\vec k')\left[3P_R(\vec k+\vec k')\sigma_R^2+\int_{\vec p_1\vec p_2}T_R(\vec p_1,\vec p_2,\vec k+\vec k')\right] + I_2^1(\vec k,\vec k)I_2^1(\vec k',
    \vec k')\sigma_R^2\\\nonumber
    &&\,+\,\frac{1}{2}\left\{I_2^1(\vec k,\vec k)I_2^2(\vec k',\vec k')B_R^0+\text{ sym.}\right\} + \frac{1}{4}I_2^2(\vec k,\vec k)I_2^2(\vec k',\vec k')\left[2\sigma_R^4+T_R^0\right]\\\nonumber
    &&\,+\, \frac{1}{6}\left\{I_2^1(\vec k,\vec k)I_2^3(\vec k',\vec k')\left[3\sigma_R^4+T_R^0\right]+\text{ sym.}\right\}\\\nonumber
    &&\,+\,\left\{2I_3^1(\vec k,\vec k,\vec k')I_1^1(\vec k')P_R(\vec k')+I_3^2(\vec k,\vec k,\vec k')I_1^1(\vec k')\int_{\vec p}B_R(\vec p,\vec k')+\text{ sym.}\right\}\\\nonumber
    &&\,+\,\left\{\frac{1}{3}I_3^3(\vec k,\vec k,\vec k')I_1^1(\vec k')\left[3P_R(\vec k')\sigma_R^2+\int_{\vec p_1\vec p_2}T_R(\vec p_1,\vec p_2,\vec k')\right] + \text{ sym.}\right\}\\\nonumber
    V\mathcal{C}^{1h}(\vec k,\vec k') &=& I_4^0(\vec k,\vec k,\vec k',\vec k'),
\eeq
where we use $\int_{\vec p}\equiv \int \frac{d\vec p}{(2\pi)^3}$ for brevity and omitted the final argument of $B_R$ and $T_R$ (which follows by enforcing that the sum of momenta is zero). We additionally define
\beq
    B_R^0 = \int_{\vec p_1\vec p_2}B_R(\vec p_1,\vec p_2), \quad T_R^0 = \int_{\vec p_1\vec p_2\vec p_3}T_R(\vec p_1,\vec p_2,\vec p_3).
\eeq
Note that we have made extensive use of the bias consistency relations (Eq.\,\ref{eq: bias_consistency_relation}) to remove any terms including $I_1^q(\vec k)$ for $q>1$ (since these are negligible at small $k$ and subdominant at large $k$) and ignored zero-momentum terms in this derivation. We further note that the first pieces of the four-, three- and two-halo covariance are simply $2V^{-1}\left[P^{2h}(\vec k)+P^{1h}(\vec k)\right]^2\delta_D(\vec k+\vec k')$, as in the standard Gaussian matter covariance.

To obtain an accurate model of the power-spectrum covariance, we must additionally consider super-sample effects, as in Sec.\,\ref{subsec: ssc-cov}. From Eq.\,\ref{eq: cov-ssc-fg}, we can write
\beq\label{eq: cov-Pk-SSC}
    \operatorname{cov}\left(P(\vec k),P(\vec k')\right)^\mathrm{SSC} = \sigma^2(V)\left.\frac{dP_\mathrm{HM}(\vec k)}{d\delta_b}\right|_{\delta_b = 0}\left.\frac{dP_\mathrm{HM}(\vec k')}{d\delta_b}\right|_{\delta_b = 0},
\eeq
where the derivative is given by Eq.\,\ref{eq: dP_ddelta-model}. Finally, we must consider $k$-space binning. In practical contexts, the power in $k$-bin $a$ is an integral over $\vec k$;
\beq
    P_a \equiv \frac{1}{V_a}\int_{\vec k\in a}d\vec k\,P(\vec k)\approx P(k_a),
\eeq
where $V_a$ is the volume of the bin centered at $k_a$. For the covariance $\operatorname{cov}\left[P(\vec k),P(\vec k')\right]$, we must carefully consider the case $\vec k=\vec k'$;
\beq
    \operatorname{cov}\left[P_a,P_b\right] &\equiv& \frac{1}{V_aV_b}\int_{\vec k\in a}\int_{\vec k'\in b}\left(\operatorname{cov}\left[P(\vec k),P(\vec k)\right]\delta_D(\vec k-\vec k')+\left.\operatorname{cov}\left[P(\vec k),P(\vec k')\right]\right|_{\vec k\neq \vec k'}\right)\\\nonumber
    &=& \frac{\delta^K_{ab}}{V_a}\operatorname{cov}\left[P(k_a),P(k_a)\right]+\operatorname{cov}\left[P(k_a),P(k_b)\right].
\eeq
Noting that $V_aV$ is equal to the number of modes in bin $a$, this recovers the familiar form for the Gaussian diagonal covariance; $\operatorname{cov}[P_a,P_b]^\mathrm{Gaussian,diag} = 2P_a^2\delta^K_{ab}/N_\mathrm{modes}(k_a)$. Combining both SSC and non-SSC covariances, we thus obtain a model for the power spectrum covariance, which may be straightforwardly computed from a set of $I_p^q$ mass-function integrals and the perturbation-theory correlators, as for the power spectrum itself.

\section{The Effective Halo Model using the Zel`dovich Approximation}\label{appen: zel-Pk}
{In \citet{2014MNRAS.445.3382M} and \citet{2015PhRvD..91l3516S}, an alternative halo model is proposed, based on the Zel`dovich Approximation (ZA) rather than Effective Field Theory (and with a significantly different treatment of non-perturbative physics). To place our results into context, it is thus useful to examine the extent to which our results depend on the choice of perturbation theory. To this end, recall that the perturbation theory enters when considering the statistics of the halo density field $n(m|\vec x)$ (Eq.\,\ref{eq: n_m_expansion}) and the corresponding statistics of the smoothed density field $\delta_R(\vec x)$. If we assume these to be modeled by ZA instead of EFT, the one-halo term is unchanged, but the two-halo term becomes:
\beq\label{eq: ZA-2h}
    P^\mathrm{2h}_\mathrm{Z}(k) = \left[I_1^1(k)\right]^2 W^2(kR) P_\mathrm{Z}(k),
\eeq
where $P_\mathrm{Z}(k)$ is the non-linear Zel`dovich power spectrum. Note that this does not include the counterterm $c_s^2$, since this is not a standard ingredient in ZA.} 

\begin{figure}
    \centering
    \includegraphics[width=\textwidth]{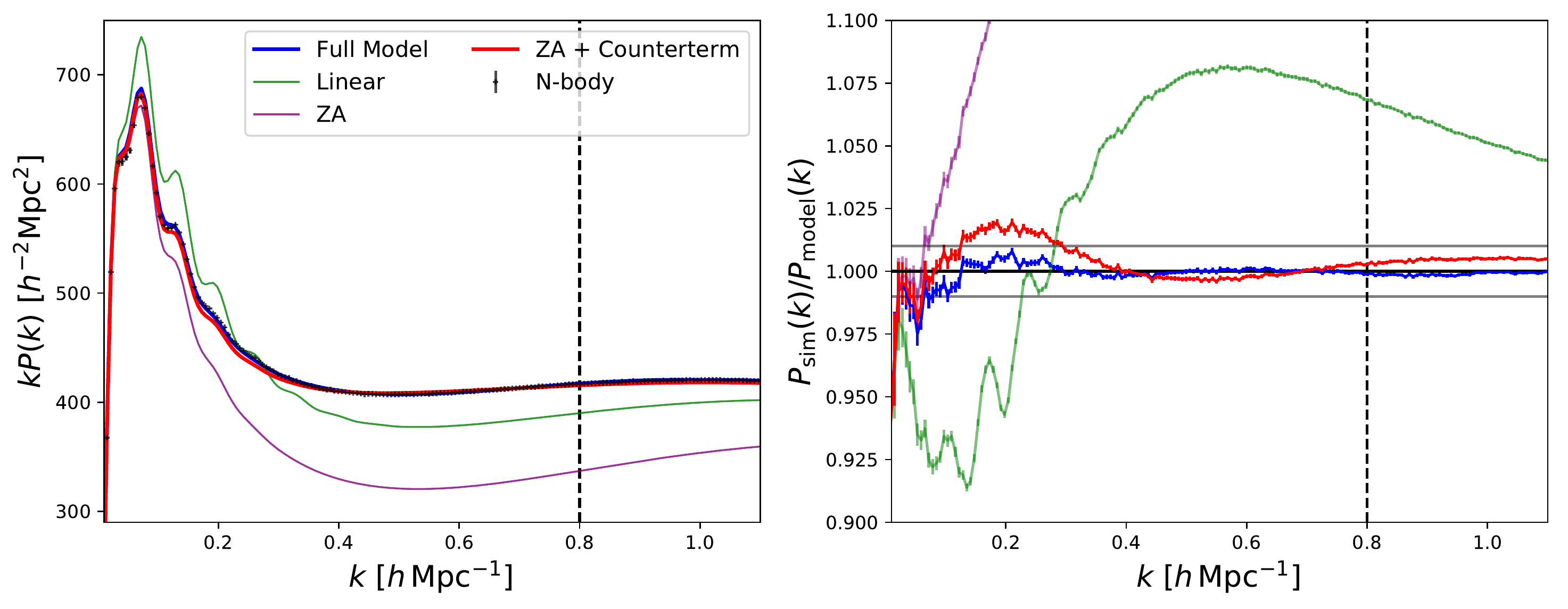}
    \caption{{Comparison of the halo model power spectra using the full model introduced in this work (Eq.\,\ref{eq: pk-summary}, blue), linear theory (green), a halo model based on the Zel`dovich approximation (ZA, purple), and a similar model with an additional counterterm (red). This figure has the same format as Fig.\,\ref{fig: pk_comparison}, except switching to linear axes for the right panel to highlight the differences between models at moderate $k$. The ZA spectra are calculated using \texttt{nbodykit} \citep{2019ascl.soft04027H}.}}
    \label{fig: ZA_Pk}
\end{figure}

{In Fig.\,\ref{fig: ZA_Pk}, we show the halo model power spectrum obtained using Eq.\,\ref{eq: ZA-2h} alongside the effective and standard halo model predictions. Despite the presence of a large one-halo term, the ZA-based model (purple curve) clearly underestimates the power for all $k \gtrsim 0.1\hMpc$. This is as expected, as ZA is known to underestimate power on quasi-linear scales, due to missing perturbative kernels. Whilst Eq.\,\ref{eq: ZA-2h} allows for a smoothing window, its optimal value is simply zero here, since ZA is already an underestimate. Note however that ZA provides a good estimate of the BAO wiggles, since it does not assume the Lagrangian displacement vectors to be small.}

{Clearly an ingredient is missing. The obvious solution is to add a counterterm, as in the EFT model. Perhaps the simplest choice would be to proceed by analogy to EFT and use
\beq
    P_\mathrm{Z}(k) \rightarrow P_\mathrm{Z}(k) - c_s^2k^2P_\mathrm{Z}(k)
\eeq
(cf.\,Eq.\,\ref{eq: counterterm_final}), which is bounded for large $k$.\footnote{Note that using a counterterm $-c_s^2k^2/(1+(k/\hat{k})^2)\times P_\mathrm{L}(k)$ yields similar results.} When this is inserted into the two-halo term, we observe \textit{much} improved agreement between model and simulations, with sub-percent accuracy obtained for $k > 0.3\hMpc$. Unlike for EFT, the optimal value of $c_s^2$ is negative, since ZA underestimates the true power. Notably, the ZA-plus-counterterm model slightly underestimates the power spectrum ($P_\mathrm{sim}/P_\mathrm{model}>1$) at $k\sim 0.2\hMpc$, an effect not seen in the EFT model. This indicates that a more sophisticated model for the correction terms is needed.}

{Given that one-loop EFT and ZA require similar computation time, the above shows a slight preference for EFT due to the improved fit on mildly non-linear scales. More importantly, the counterterm is a key component of EFT (and arises by considering the stress-tensor of the smoothed fluid equations); in ZA, there is not a clear physical motivation. For this reason, the EFT model is preferred and uniformly adopted in this work.}

\section{The Non-Linear Power Spectrum of Small Volumes}\label{appen: Pk-trunc}
In the below, we discuss the corrections required to model the quasi-linear power spectrum in a small regions of space embedded in some larger region. Considering a subbox of volume $L_\mathrm{subbox}$, the internal density field contains contributions from Fourier modes with wavelengths both smaller and larger than $L_\mathrm{subbox}$. However, when the overdensity of the box is computed, the effects of modes larger than the subbox are removed (since these contribute only to a rescaling of the mass in the box), implying that the subbox density field $\delta_\mathrm{subbox}(\vec k)$ contains no power on scales $|\vec k|<k_\mathrm{min}$, where $k_\mathrm{min}$ is the fundamental mode of the box; $k_\mathrm{min} = 2\pi/L_\mathrm{subbox}$. 

In perturbation theory, the non-linear density field is modeled as a functional of the linear density fields;
\beq
    \delta^\mathrm{NL}(\vec k) = \delta^{(1)}(\vec k)+\delta^{(2)}(\vec k)+\delta^{(3)}(\vec k)+\delta^{(ct)}(\vec k)+...
\eeq
where $\delta^{(1)}$ is the linear density field, $\delta^{(ct)}$ is a counterterm (from EFT) and $\delta^{(n)}$ is the $n$-th order field, depending on $n$ copies of the linear field $\delta^{(1)}$ through
\beq
    \delta^{(n)}(\vec k) = \int\left[\prod_{i=1}^n \frac{d\vec q_i}{(2\pi)^3}\delta^{(1)}(\vec q_i)\right]F_{n}(\vec q_1,...,\vec q_n)\delta_D(\vec k - \sum_i\vec q_i),
\eeq
for kernel function $F_n(\vec q_1,..,\vec q_n)$ given in \citet{2002PhR...367....1B}. Imposing that $\delta_\mathrm{subbox}$ contains no power on scales below $k_\mathrm{min}$ thus implies that the $n$-th order field is modified to
\beq
    \delta^{(n)}_\mathrm{subbox}(\vec k) = \int\left[\prod_{i=1}^n \frac{d\vec q_i}{(2\pi)^3}\delta^{(1)}(\vec q_i)\Theta(|\vec q_i|-k_\mathrm{min})\right]F_{n}(\vec q_1,...,\vec q_n)\delta_D(\vec k - \sum_i\vec q_i),
\eeq
where $\Theta$ is a Heaviside function, i.e. we filter the linear field via $\delta^{(1)}(\vec q)\rightarrow \Theta(|\vec q|-k_\mathrm{min})\delta^{(1)}(\vec q)$. (Note that the counterterm $\delta^{(ct)}(\vec k) = -\tfrac{1}{2}c_s^2k^2\delta^{(1)}(\vec k)$ is similarly modified.) With this assumption, we can compute the EFT power spectrum (cf.\,Eq.\,\ref{eq: 1-loop-SPT}) as
\beq
    P_\mathrm{subbox}(\vec k) &\equiv& \av{\left|\delta^\mathrm{NL}_\mathrm{subbox}(\vec k)\right|^2} = P_\mathrm{L}(\vec k)\Theta(|\vec k|-k_\mathrm{min})+P_{22}'(\vec k) + 2P_{13}'(\vec k)-2c_s^2k^2P_\mathrm{L}(\vec k)\Theta(|\vec k|-k_\mathrm{min})\\\nonumber
    P_\mathrm{22}'(\vec k) &=& \int \frac{d\vec q}{(2\pi)^3}P_\mathrm{L}(\vec q)P_\mathrm{L}(|\vec k-\vec q|)|F_2(\vec q,\vec k-\vec q)|^2\Theta(|\vec k-\vec q|-k_\mathrm{min})\Theta(|\vec q|-\vec k_\mathrm{min})\\\nonumber
    P_\mathrm{13}'(\vec k) &=& 3P_\mathrm{L}(\vec k)\Theta(|\vec k|-k_\mathrm{min})\int \frac{d\vec q}{(2\pi)^3}P_\mathrm{L}(\vec q)F_3(\vec k,\vec q,-\vec q)\Theta(|\vec q|-k_\mathrm{min}).
\eeq
Practically, this is simply computed by inserting the filtered linear power spectrum $P_\mathrm{L}(k)\Theta(k-k_\mathrm{min})$ into the \texttt{FAST-PT} code rather than $P_\mathrm{L}(k)$. A similar line of reasoning follows for IR resummation; since the density field of the subbox is affected only by modes below the box frequency, the damping scale $\Sigma$ (Eq.\,\ref{eq: Sigma2-def}) should be computed by integrating only over modes with $k>k_\mathrm{min}$. Whilst we do not expect the one-halo term to be modified by this prescription (since it is independent of $P_\mathrm{L}$, the speed-of-sound parameter $c_s^2$ may be expected to change between large and small subboxes, since it encodes the interplay of short and long modes.

\bsp	
\label{lastpage}
\end{document}

%% file: cov-fig.tex
\begin{tikzpicture}[x=0.75pt,y=0.75pt,yscale=-1,xscale=1]

\draw  [draw opacity=0][fill={rgb, 255:red, 112; green, 46; blue, 171 }  ,fill opacity=0.5 ] (95.53,90.17) .. controls (95.53,56.44) and (122.87,29.09) .. (156.6,29.09) .. controls (190.33,29.09) and (217.67,56.44) .. (217.67,90.17) .. controls (217.67,123.9) and (190.33,151.24) .. (156.6,151.24) .. controls (122.87,151.24) and (95.53,123.9) .. (95.53,90.17) -- cycle ;
\draw  [draw opacity=0][fill={rgb, 255:red, 245; green, 166; blue, 35 }  ,fill opacity=1 ] (83.67,196.3) .. controls (83.67,185.17) and (92.69,176.14) .. (103.83,176.14) .. controls (114.96,176.14) and (123.99,185.17) .. (123.99,196.3) .. controls (123.99,207.44) and (114.96,216.47) .. (103.83,216.47) .. controls (92.69,216.47) and (83.67,207.44) .. (83.67,196.3) -- cycle ;
\draw  [draw opacity=0][fill={rgb, 255:red, 65; green, 117; blue, 5 }  ,fill opacity=0.48 ] (177.35,219.43) .. controls (177.35,201.42) and (191.95,186.82) .. (209.96,186.82) .. controls (227.97,186.82) and (242.58,201.42) .. (242.58,219.43) .. controls (242.58,237.44) and (227.97,252.04) .. (209.96,252.04) .. controls (191.95,252.04) and (177.35,237.44) .. (177.35,219.43) -- cycle ;
\draw    (148.89,98.17) -- (165.49,81.57) ;
\draw    (148.3,82.16) -- (164.31,99.06) ;
\draw  [fill={rgb, 255:red, 0; green, 0; blue, 0 }  ,fill opacity=1 ] (188.02,214.98) .. controls (188.02,213.84) and (188.95,212.91) .. (190.1,212.91) .. controls (191.25,212.91) and (192.18,213.84) .. (192.18,214.98) .. controls (192.18,216.13) and (191.25,217.06) .. (190.1,217.06) .. controls (188.95,217.06) and (188.02,216.13) .. (188.02,214.98) -- cycle ;
\draw  [fill={rgb, 255:red, 0; green, 0; blue, 0 }  ,fill opacity=1 ] (103.23,193.93) .. controls (103.23,192.62) and (104.3,191.56) .. (105.61,191.56) .. controls (106.92,191.56) and (107.98,192.62) .. (107.98,193.93) .. controls (107.98,195.24) and (106.92,196.3) .. (105.61,196.3) .. controls (104.3,196.3) and (103.23,195.24) .. (103.23,193.93) -- cycle ;
\draw [color={rgb, 255:red, 0; green, 0; blue, 0 }  ,draw opacity=0.52 ]   (156.6,90.17) .. controls (157.36,92.4) and (156.62,93.9) .. (154.39,94.66) .. controls (152.16,95.42) and (151.43,96.91) .. (152.19,99.14) .. controls (152.95,101.37) and (152.21,102.87) .. (149.98,103.63) .. controls (147.75,104.4) and (147.02,105.89) .. (147.78,108.12) .. controls (148.54,110.35) and (147.8,111.85) .. (145.57,112.61) .. controls (143.34,113.37) and (142.61,114.86) .. (143.37,117.09) .. controls (144.13,119.32) and (143.39,120.82) .. (141.16,121.58) .. controls (138.93,122.35) and (138.2,123.84) .. (138.96,126.07) .. controls (139.72,128.3) and (138.98,129.8) .. (136.75,130.55) .. controls (134.52,131.32) and (133.79,132.81) .. (134.55,135.04) .. controls (135.31,137.27) and (134.57,138.77) .. (132.34,139.53) .. controls (130.11,140.3) and (129.38,141.79) .. (130.14,144.02) .. controls (130.9,146.25) and (130.16,147.75) .. (127.93,148.5) .. controls (125.7,149.27) and (124.97,150.76) .. (125.73,152.99) .. controls (126.49,155.22) and (125.75,156.72) .. (123.52,157.48) .. controls (121.29,158.24) and (120.55,159.74) .. (121.31,161.97) .. controls (122.07,164.2) and (121.34,165.69) .. (119.11,166.45) .. controls (116.88,167.21) and (116.14,168.71) .. (116.9,170.94) .. controls (117.66,173.17) and (116.93,174.66) .. (114.7,175.43) .. controls (112.47,176.19) and (111.73,177.69) .. (112.49,179.92) .. controls (113.25,182.15) and (112.52,183.64) .. (110.29,184.4) -- (110.02,184.96) -- (106.49,192.14) ;
\draw [shift={(105.61,193.93)}, rotate = 296.17] [color={rgb, 255:red, 0; green, 0; blue, 0 }  ,draw opacity=0.52 ][line width=0.75]    (10.93,-3.29) .. controls (6.95,-1.4) and (3.31,-0.3) .. (0,0) .. controls (3.31,0.3) and (6.95,1.4) .. (10.93,3.29)   ;
\draw [color={rgb, 255:red, 0; green, 0; blue, 0 }  ,draw opacity=0.52 ]   (190.24,214.5) -- (182.47,212.56) .. controls (180.45,213.77) and (178.83,213.36) .. (177.62,211.34) .. controls (176.41,209.32) and (174.79,208.92) .. (172.77,210.13) .. controls (170.75,211.34) and (169.13,210.94) .. (167.92,208.92) .. controls (166.71,206.9) and (165.09,206.5) .. (163.07,207.71) .. controls (161.05,208.92) and (159.43,208.51) .. (158.22,206.49) .. controls (157.01,204.47) and (155.39,204.07) .. (153.37,205.28) .. controls (151.35,206.49) and (149.73,206.09) .. (148.52,204.07) .. controls (147.31,202.05) and (145.69,201.65) .. (143.67,202.86) .. controls (141.65,204.07) and (140.03,203.66) .. (138.82,201.64) .. controls (137.61,199.62) and (135.99,199.22) .. (133.97,200.43) .. controls (131.95,201.64) and (130.33,201.24) .. (129.12,199.22) .. controls (127.91,197.2) and (126.29,196.8) .. (124.27,198.01) .. controls (122.24,199.22) and (120.62,198.81) .. (119.41,196.79) .. controls (118.2,194.77) and (116.58,194.37) .. (114.56,195.58) .. controls (112.54,196.79) and (110.92,196.39) .. (109.71,194.37) -- (107.98,193.93) -- (107.98,193.93) ;
\draw [shift={(192.18,214.98)}, rotate = 194.04] [color={rgb, 255:red, 0; green, 0; blue, 0 }  ,draw opacity=0.52 ][line width=0.75]    (10.93,-3.29) .. controls (6.95,-1.4) and (3.31,-0.3) .. (0,0) .. controls (3.31,0.3) and (6.95,1.4) .. (10.93,3.29)   ;
\draw  [draw opacity=0][fill={rgb, 255:red, 112; green, 46; blue, 171 }  ,fill opacity=0.5 ] (275.54,88.03) .. controls (275.54,54.3) and (302.89,26.96) .. (336.62,26.96) .. controls (370.35,26.96) and (397.69,54.3) .. (397.69,88.03) .. controls (397.69,121.76) and (370.35,149.11) .. (336.62,149.11) .. controls (302.89,149.11) and (275.54,121.76) .. (275.54,88.03) -- cycle ;
\draw  [draw opacity=0][fill={rgb, 255:red, 245; green, 166; blue, 35 }  ,fill opacity=1 ] (263.68,194.17) .. controls (263.68,183.04) and (272.71,174.01) .. (283.84,174.01) .. controls (294.98,174.01) and (304,183.04) .. (304,194.17) .. controls (304,205.3) and (294.98,214.33) .. (283.84,214.33) .. controls (272.71,214.33) and (263.68,205.3) .. (263.68,194.17) -- cycle ;
\draw  [draw opacity=0][fill={rgb, 255:red, 65; green, 117; blue, 5 }  ,fill opacity=0.48 ] (357.37,217.3) .. controls (357.37,199.28) and (371.97,184.68) .. (389.98,184.68) .. controls (407.99,184.68) and (422.59,199.28) .. (422.59,217.3) .. controls (422.59,235.31) and (407.99,249.91) .. (389.98,249.91) .. controls (371.97,249.91) and (357.37,235.31) .. (357.37,217.3) -- cycle ;
\draw    (328.91,96.04) -- (345.51,79.44) ;
\draw    (328.32,80.03) -- (344.32,96.93) ;
\draw  [fill={rgb, 255:red, 0; green, 0; blue, 0 }  ,fill opacity=1 ] (368.04,212.85) .. controls (368.04,211.7) and (368.97,210.77) .. (370.12,210.77) .. controls (371.26,210.77) and (372.19,211.7) .. (372.19,212.85) .. controls (372.19,213.99) and (371.26,214.92) .. (370.12,214.92) .. controls (368.97,214.92) and (368.04,213.99) .. (368.04,212.85) -- cycle ;
\draw  [fill={rgb, 255:red, 0; green, 0; blue, 0 }  ,fill opacity=1 ] (287.52,98.59) .. controls (287.52,97.28) and (288.58,96.22) .. (289.89,96.22) .. controls (291.2,96.22) and (292.26,97.28) .. (292.26,98.59) .. controls (292.26,99.9) and (291.2,100.96) .. (289.89,100.96) .. controls (288.58,100.96) and (287.52,99.9) .. (287.52,98.59) -- cycle ;
\draw [color={rgb, 255:red, 0; green, 0; blue, 0 }  ,draw opacity=0.52 ] [dash pattern={on 4.5pt off 4.5pt}]  (336.62,88.03) -- (291.84,98.15) ;
\draw [shift={(289.89,98.59)}, rotate = 347.27] [color={rgb, 255:red, 0; green, 0; blue, 0 }  ,draw opacity=0.52 ][line width=0.75]    (10.93,-3.29) .. controls (6.95,-1.4) and (3.31,-0.3) .. (0,0) .. controls (3.31,0.3) and (6.95,1.4) .. (10.93,3.29)   ;
\draw [color={rgb, 255:red, 0; green, 0; blue, 0 }  ,draw opacity=0.52 ]   (371.04,211.21) -- (366.43,204.68) .. controls (364.1,204.28) and (363.14,202.92) .. (363.54,200.59) .. controls (363.94,198.27) and (362.98,196.91) .. (360.66,196.51) .. controls (358.34,196.1) and (357.38,194.74) .. (357.78,192.42) .. controls (358.18,190.1) and (357.22,188.74) .. (354.9,188.34) .. controls (352.57,187.94) and (351.61,186.58) .. (352.01,184.25) .. controls (352.41,181.93) and (351.45,180.57) .. (349.13,180.17) .. controls (346.81,179.76) and (345.85,178.4) .. (346.25,176.08) .. controls (346.65,173.76) and (345.69,172.4) .. (343.37,172) .. controls (341.04,171.6) and (340.08,170.24) .. (340.48,167.91) .. controls (340.88,165.59) and (339.92,164.23) .. (337.6,163.82) .. controls (335.28,163.42) and (334.32,162.06) .. (334.72,159.74) .. controls (335.12,157.41) and (334.16,156.05) .. (331.83,155.65) .. controls (329.51,155.25) and (328.55,153.89) .. (328.95,151.57) .. controls (329.35,149.25) and (328.39,147.89) .. (326.07,147.48) .. controls (323.75,147.08) and (322.79,145.72) .. (323.19,143.4) .. controls (323.59,141.07) and (322.63,139.71) .. (320.3,139.31) .. controls (317.98,138.91) and (317.02,137.55) .. (317.42,135.23) .. controls (317.82,132.91) and (316.86,131.55) .. (314.54,131.14) .. controls (312.21,130.75) and (311.25,129.39) .. (311.65,127.06) .. controls (312.05,124.74) and (311.09,123.38) .. (308.77,122.97) .. controls (306.45,122.57) and (305.49,121.21) .. (305.89,118.89) .. controls (306.29,116.57) and (305.33,115.21) .. (303.01,114.8) .. controls (300.68,114.41) and (299.72,113.05) .. (300.12,110.72) .. controls (300.52,108.4) and (299.56,107.04) .. (297.24,106.63) .. controls (294.92,106.23) and (293.96,104.87) .. (294.36,102.55) .. controls (294.76,100.23) and (293.8,98.87) .. (291.48,98.46) -- (289.89,96.22) -- (289.89,96.22) ;
\draw [shift={(372.19,212.85)}, rotate = 234.79] [color={rgb, 255:red, 0; green, 0; blue, 0 }  ,draw opacity=0.52 ][line width=0.75]    (10.93,-3.29) .. controls (6.95,-1.4) and (3.31,-0.3) .. (0,0) .. controls (3.31,0.3) and (6.95,1.4) .. (10.93,3.29)   ;
\draw  [draw opacity=0][fill={rgb, 255:red, 112; green, 46; blue, 171 }  ,fill opacity=0.5 ] (464.62,89.54) .. controls (464.62,55.81) and (491.96,28.47) .. (525.69,28.47) .. controls (559.42,28.47) and (586.76,55.81) .. (586.76,89.54) .. controls (586.76,123.27) and (559.42,150.62) .. (525.69,150.62) .. controls (491.96,150.62) and (464.62,123.27) .. (464.62,89.54) -- cycle ;
\draw  [draw opacity=0][fill={rgb, 255:red, 245; green, 166; blue, 35 }  ,fill opacity=1 ] (452.76,195.68) .. controls (452.76,184.55) and (461.78,175.52) .. (472.92,175.52) .. controls (484.05,175.52) and (493.08,184.55) .. (493.08,195.68) .. controls (493.08,206.81) and (484.05,215.84) .. (472.92,215.84) .. controls (461.78,215.84) and (452.76,206.81) .. (452.76,195.68) -- cycle ;
\draw  [draw opacity=0][fill={rgb, 255:red, 65; green, 117; blue, 5 }  ,fill opacity=0.48 ] (546.44,218.81) .. controls (546.44,200.79) and (561.04,186.19) .. (579.05,186.19) .. controls (597.07,186.19) and (611.67,200.79) .. (611.67,218.81) .. controls (611.67,236.82) and (597.07,251.42) .. (579.05,251.42) .. controls (561.04,251.42) and (546.44,236.82) .. (546.44,218.81) -- cycle ;
\draw    (517.98,97.55) -- (534.58,80.95) ;
\draw    (517.39,81.54) -- (533.4,98.44) ;
\draw  [fill={rgb, 255:red, 0; green, 0; blue, 0 }  ,fill opacity=1 ] (552.79,134.61) .. controls (552.79,133.46) and (553.72,132.53) .. (554.86,132.53) .. controls (556.01,132.53) and (556.94,133.46) .. (556.94,134.61) .. controls (556.94,135.75) and (556.01,136.68) .. (554.86,136.68) .. controls (553.72,136.68) and (552.79,135.75) .. (552.79,134.61) -- cycle ;
\draw  [fill={rgb, 255:red, 0; green, 0; blue, 0 }  ,fill opacity=1 ] (480.15,95.83) .. controls (480.15,94.52) and (481.21,93.46) .. (482.52,93.46) .. controls (483.83,93.46) and (484.9,94.52) .. (484.9,95.83) .. controls (484.9,97.14) and (483.83,98.2) .. (482.52,98.2) .. controls (481.21,98.2) and (480.15,97.14) .. (480.15,95.83) -- cycle ;
\draw [color={rgb, 255:red, 0; green, 0; blue, 0 }  ,draw opacity=0.52 ] [dash pattern={on 4.5pt off 4.5pt}]  (525.69,89.54) -- (486.87,95.52) ;
\draw [shift={(484.9,95.83)}, rotate = 351.24] [color={rgb, 255:red, 0; green, 0; blue, 0 }  ,draw opacity=0.52 ][line width=0.75]    (10.93,-3.29) .. controls (6.95,-1.4) and (3.31,-0.3) .. (0,0) .. controls (3.31,0.3) and (6.95,1.4) .. (10.93,3.29)   ;
\draw [color={rgb, 255:red, 0; green, 0; blue, 0 }  ,draw opacity=0.52 ] [dash pattern={on 4.5pt off 4.5pt}]  (553.09,131.6) -- (484.9,95.83) ;
\draw [shift={(554.86,132.53)}, rotate = 207.68] [color={rgb, 255:red, 0; green, 0; blue, 0 }  ,draw opacity=0.52 ][line width=0.75]    (10.93,-3.29) .. controls (6.95,-1.4) and (3.31,-0.3) .. (0,0) .. controls (3.31,0.3) and (6.95,1.4) .. (10.93,3.29)   ;
\draw  [dash pattern={on 0.84pt off 2.51pt}]  (253.49,9.41) -- (252.77,259.87) ;
\draw  [dash pattern={on 0.84pt off 2.51pt}]  (439.71,7.36) -- (439,257.82) ;

\draw (114.18,152.49) node  [font=\large]  {$\mathbf{y}$};
\draw (145.83,212.67) node  [font=\large]  {$\mathbf{r}$};
\draw (313.61,71.05) node  [font=\large]  {$\mathbf{y}$};
\draw (324.28,162.74) node  [font=\large]  {$\mathbf{r}$};
\draw (508.26,75.47) node  [font=\large]  {$\mathbf{y}$};
\draw (511.22,119.54) node  [font=\large]  {$\mathbf{r}$};
\draw (159.21,280.5) node  [font=\Large]  {$C^{3h}(\mathbf{r} ,m)$};
\draw (339.91,279.05) node  [font=\Large]  {$C^{2h}(\mathbf{r} ,m)$};
\draw (524.71,278.43) node  [font=\Large]  {$C^{1h}(\mathbf{r} ,m)$};

\end{tikzpicture}